     \documentstyle[12pt,epsfig,epsf]{article}
     \newlength{\dinwidth}
     \newlength{\dinmargin}
\def\Journal#1#2#3#4{{#1} {\bf #2}, #3 (#4)}

\def\NPB{{\em Nucl. Phys.}\bf B}
\def\PLB{{\em Phys. Lett.} \bf  B}

\def\PRD{{\em Phys. Rev.}\bf  D}
\def\ZPC{{\em Z. Phys.}\bf  C}
\def\lsim{\mathrel{\rlap{\lower4pt\hbox{\hskip1pt$\sim$}}
    \raise1pt\hbox{$<$}}}                
\def\gsim{\mathrel{\rlap{\lower4pt\hbox{\hskip1pt$\sim$}}
    \raise1pt\hbox{$>$}}}                
\begin{document}
\begin{flushright}
 TAUP 2574 - 99\\
 {  \today} \\
{\tt hep-ph/9905381}
\end{flushright}

\vspace*{10mm}
\begin{center}  \begin{Large} \begin{bf}
Low $\mathbf{x}$ news for
Monte Carlo\\
  \end{bf}  \end{Large}
  \vspace*{5mm}
  \begin{large}
Eugene Levin $^a$,\\
  \end{large}
\end{center}
$^a$ HEP Department,  School of Physics, Tel Aviv University,
Ramat
Aviv,
69978,\,ISRAEL\\

~
\centerline{\em Invited talk at Monte Carlo Workshop, DESY, 1999.} 

~

\begin{quotation}
\noindent
{\bf Abstract:} This talk is a review of news in   low $x$ physics
which, I think, would be useful for writing of the Monte Carlo codes.
The following topics are discussed here: (i)  the next-to-leading order
BFKLPomeron; (ii) two indications for shadowing corrections (SC) in DIS
from  HERA data;
(iii) matching of ``soft" and ``hard" photon - proton interactions;
and (iv) survival probability for large rapidity gap ( LRG ) processes
in hadron-hadron scattering and DIS. I hope, that our current
understanding of these topics will allow us  to narrow the gap between the
MC
codes and our microscopic theory  - QCD.

\end{quotation}

\section{Introduction}

~

{\bf Goals:}  The main goal of this talk is to share with you my
understanding of the current situation in low $x$ physics looking at this
subject from  the angle of possible improvement of existing Monte Carlo
codes.  Everyone knows that the typical MC code for deep inelastic
scattering ( DIS ) contains three parts: the perturbative QCD cascade, the
hadronization  stage and the matching between ``soft" anf ``hard"
processes. Since we have no solid theoretical understanding of
the  hadronization stage and the  ``soft" processes, there is a danger to
write
a MC
code that  will describe the experimental data,  but in which the clear
connection with our microscopic theory - QCD, will be broken.  This is the
reason that   MC experts should follow all theoretical and
experimental
news in implementing  in MC everything that we have learned both
theoretically
and experimentally. My   goal is to provide them with a guide to  the
recent
achievements in low $x$ physics,  which is a meeting point for all the 
complicated problems of perturbative and non-perturbative QCD.

{\bf Topics:} The choice of the topics, that I would like to cover here,
is dictated by my personal interests as well as by the  considerable
progress
that has recently  been made  in their understanding. They are:
\begin{enumerate}
\item\,\,\,The next-to-leading order BFKL Pomeron;

\item\,\,\, Two indications for shadowing corrections (SC) in DIS from
HERA data;

\item\,\,\, Matching of ``soft" and ``hard" photon - proton interactions;

\item\,\,\,  Survival probability for large rapidity gap ( LRG ) processes
in hadron-hadron scattering and DIS.

\end{enumerate}
You can see that all above topics are closely related to our main problem-
the matching between long distance ( non-perturbative ) physics and short
distance ( perturbative ) approaches.

{\bf Sources of information:} This talk is a report on what I learned at:
\begin{enumerate}
\item\,\,\, Theory Institute on Deep-Inelastic Diffraction - ANL, Sept.14
- 16,1998;

\item\,\,\, Workshop ``Small $x$ and Diffraction Physics" - Fermilab,
Sept. 17 - 20,1998;

\item\,\,\,3-rd UK Phenomenology Workshop on HERA Physics - Durham, Sept.
21-25,1998;

\item\,\,\,During animated  discussions with J. Bartels, M.Braun, W.
Buchm\"{u}ller, M. Ciafaloni, E. Gotsman,  A. Kaidalov, Yu. Kovchegov, A.
Kovner, J.
Kwiecinski, L. Lipatov, U. Maor,  A. Mueller, A. Martin, L. McLerran, D.
Ross, M.
Ryskin, G. Salam, M. W\"{u}sthoff  and many others, who shared with me
their points of view on the topics that I am going to discuss;

\item\,\,\, My own  thinking on the subject, which has only partly 
been
reflected in the published papers.
\end{enumerate}

\section{The next-to-leading order  BFKL Pomeron}

~

{\bf Why do we love the BFKL Pomeron?}   I think, we can list here three
main reasons for our love to the BFKL Pomeron:
\begin{enumerate}
\item\,\,\, The BFKL Pomeron \cite{BFKL} is  a high energy asymptotic in
a perturbative QCD approach which we call leading log (1/x ) approximation
( LL(x)A ). In the LL(x)A  we select the diagrams using $\alpha_S
\,\ll\,1$,
$\alpha_S\,\ln Q^2 \,\ll\,1$ but $\alpha_S\,\ln(1/x) \,\approx\,1$. The
LL(x)A is quite different from the leading log ($Q^2$) approximation (
LL($Q^2$)A , $ \alpha_S \,\ll\,1,
\,\,\alpha_S\,\ln(1/x)\,\ll\,1,\,\,\alpha_S\,\ln Q^2 \,\approx\,1$ ) in
which  the DGLAP evolution equations \cite{DGLAP} were derived.
We believe, that the BFKL Pomeron gives:
\begin{enumerate}
\item\,\,\, A guide for the matching of perturbative QCD (pQCD) with
non-perturbative one (npQCD), at least for high enegy scattering amplitude
in DIS;

\item\,\,\,A possibility to get the ``soft" Pomeron contribution by  
taking into account  the
npQCD corrections in the BFKL Pomeron.
\end{enumerate}
We have several examples which support our hopes
\cite{LEBFKL},\cite{2BFKL}.

\item\,\,\,The BFKL Pomeron is the only theoretical way to prove that
at sufficiently high energy ( low $x$ ) the density of partons becomes
large. In other words, the BFKL Pomeron shows that the mathching of the
perturbative QCD parton cascade with the non-perurbative high energy
asymptotics,  goes through a new non-perturbative QCD stage i.e. a high
parton
density QCD ( see, for example, review \cite{LALEREV} for details ).
My personal opinion is that we do not need  any theoretical arguments for
a high parton density QCD, since the HERA experimental data show that the
gluon density is high \cite{CSDDREV} \cite{ACREV}. However, the same
experimental data we can be used  as the experimental confirmation of
the BFKL Pomeron;

\item\,\,\, The new HERA data shows that the BFKL Pomeron is needed to
describe the forward jet production ( see Fig.1). These data change the
entire attitude to the BFKL Pomeron, which from a theoretical toy, becomes
   a tool for the description of the experimental data and, therefore, a
part of DIS phenomenology which should be included in the MC codes.
The question still remains could the experimental data of Fig.1 be
described without the BFKL Pomeron? You will answer this question better
than me. However, Fig. 1 looks so impressive that it is difficult to
believe that the agreement with the BFKL prediction is just a lucky
coincidence.

\begin{figure}
\begin{center}
\epsfig{file=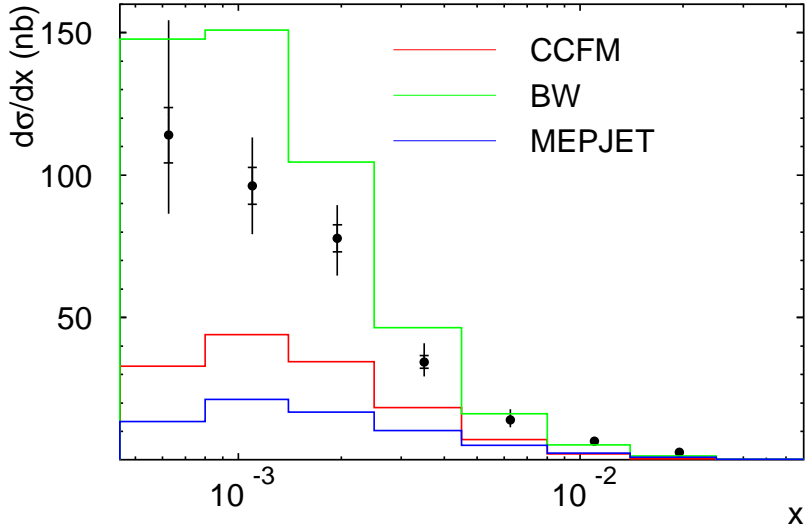,width=120mm}
\end{center}
\vspace*{2mm}
\caption[fig1]{{\it
The ZEUS'98  experimental data \protect\cite{ZEUSFJ} for forward jet
production  and their description in MEPJET MC,
in CCFM evolution equation \protect\cite{CCFM} and in the BFKL approach
( BW )\protect\cite{BW}.
  }}
\end{figure}

\end{enumerate}

{\bf The map of disaster:} It is well known that in the leading order the
BFKL Pomeron leads to Regge-like asymptotics:
\begin{equation} \label{BFKLLO}
\sigma_{tot}(\,BFKL\,\,\, LO\,)\,\,\,\,\,\propto\,\,\,\,
e^{\omega^{LO}_L\,( y - y_0 )\,\,-\,\,\frac{( r -  r_0 )^2}{4 D^{LO} ( y -
y_0 )}}\,\,,
\end{equation}
where $ y - y_0 \,=\,\ln(s/s_0)$ and $r - r_o \,=\,\ln(Q^2/Q^2_0$
( see Fig.2-a for notations ).

This simple asymptotic in the leading order is based on:
\begin{enumerate}
\item\,\,\, The time structure of the parton cascade (see Fig. 2-b );

\item\,\,\, The separation between the longitudinal and transverse degrees
of freedom;

\item\,\,\,The fact that gluon emission leads to $
\sigma\,\,\propto\,\,\frac{1}{x^{\omega_L}}$;

\item\,\,\, The diffusion in log of transverse momenta.
\end{enumerate}

\begin{figure}
\begin{tabular}{l l}
\epsfig{file=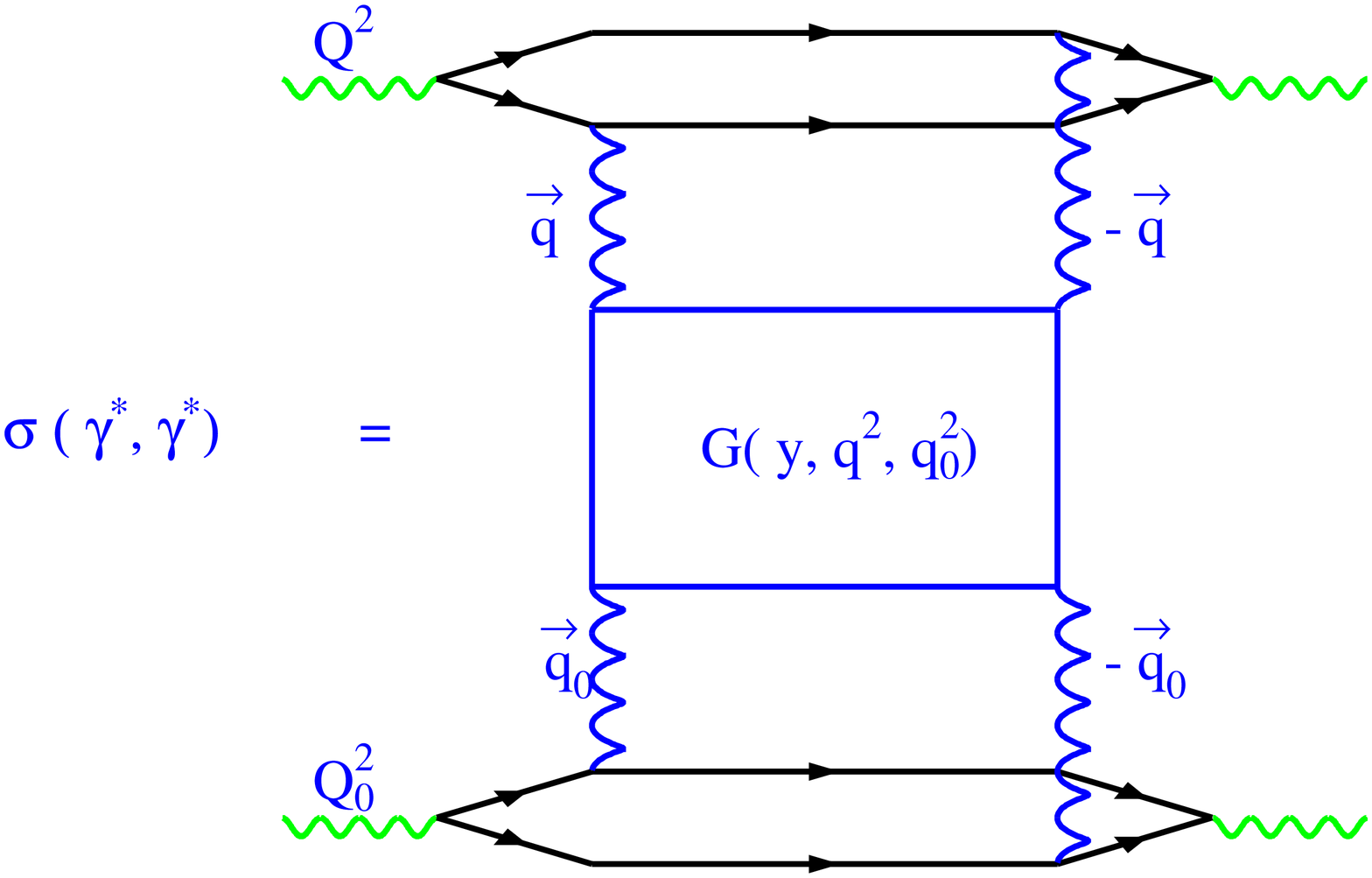,width=8cm,height=4cm} &
\epsfig{file=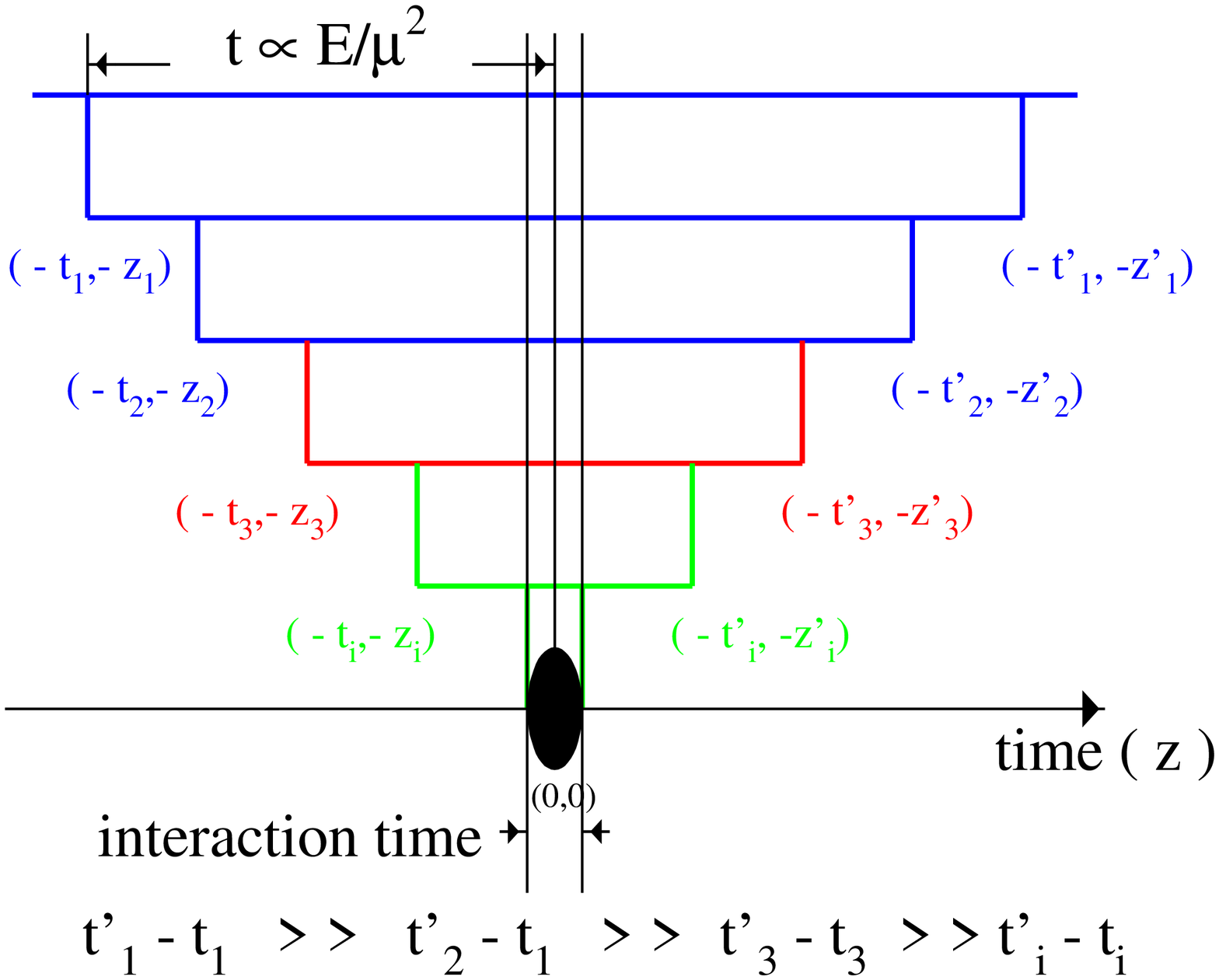,width=9cm,height=4cm}\\
Figure 2-a &
Figure 2-b\\
\end{tabular}
 \vspace*{2mm}
\caption[fig2]{{\it
$\gamma^* \gamma^*$  - interaction in the BFKL approach ( a ) and the
time structure of the BFKL Pomeron ( b ).
  }}
\end{figure}

Recently, the next-to-leading order corrections to the kernel of the
BFKL equation has been calculated
\cite{LF} \cite{CC} and it turns out that these corrections are so large
that they  could change the entire understanding of the structure of the
parton
cascade in pQCD. Indeed, in the NLO:
\begin{itemize}
\item\,\,\,$\omega_L\,\,<\,\,0$ at $q^2_0 \,\lsim\,30\,GeV^2$
\cite{LF}\cite{CC}\cite{BV} ( see Fig.2-a ). It means that in a large 
kinematic region,  the BFKL emission is suppressed, and only the Born term
contributes;

\item\,\,\, For $q^2_0 \,\lsim\,3\, 10^{6}\,GeV^2$  $D\,<\,0$
\cite{DR}. It means that there is no diffusion in log of transverse
momenta in the NLO BFKL. It also turns out that $D \,<\,0$ leads to
oscillation of the total cross section \cite{DR} \cite{LEBFKL} which shows
that the NLO BFKL has a serious pathology;

\item\,\,\,In the NLO, the Regge-like asymptotics of Eq.(~ \ref{BFKLLO} )
is
valid
only in the limited range of energy $y - y_0 = \ln(s/q q_0) \,\leq \,(
\alpha_S )^{- \frac{5}{3}}$ \cite{LEBFKL} \cite{KM}. For large values of
$y - y_0$ an  extra term appears  in exponent of Eq.( \ref{BFKLLO} ) which
is
proportional to $\alpha_s^5\,( y - y_0 )^3$ \cite{KM} \cite{LEBFKL}.
Therefore,  we cannot hope that in NLO BFKL we will get  Regge-like
asymptotics \cite{KM} \cite{LEBFKL}\cite{ABB}.

\end{itemize}

Even a slight glance at the difficulties, listed above, shows that the
main
properties of the LO BFKL are  broken in the NLO.

{\bf A possible way out - the main idea of resummation:} The BFKL equation
in the NLO can be written in the form:

\begin{equation} \label{BFKLNLO}
\frac{\partial G (y, \vec{q},\vec{q}_0)}{\partial y}\,\,=\,\,\delta^{(2)}(
\vec{q} - \vec{q}_0   )\,\,+\,\,\int\,d^2
q'\,K(\vec{q},\vec{q} ')\,G(y, \vec{q} ',\vec{q}_0)\,\,,
\end{equation}
where the NLO kernel $K(\vec{q},\vec{q} ')$ can be written as
\begin{equation} \label{NLOKER}
K(\vec{q},\vec{q} ')\,\,=
\,\,\alpha_S(q^2)\,\,\{\,\,K^{LO}(\vec{q},\vec{q}
')\,\,+
\,\,K^{NLO}(\vec{q},\vec{q}')\,\,\}\,\,=
\end{equation}
$$
\,\,\frac{r_0}{r}\,
K_{conf}(\vec{q},\vec{q}')\,\,=\,\,\frac{r_0}{r}\,\{\,\alpha_S(q_0)\,K^{LO}\,\,+\,\,\alpha^2_S(q_0)\,K^{NLO}\,\}\,\,.
$$

In Ref. \cite{LF} it was found  that the largest  contributions to
$K^{NLO}$ with the
Fadin-Lipatov choice of the energy variable $y - y_0\, =\,  \ln(s/q q_0)$
is $K^{NLO}\,\,\propto\,\,\alpha^2_S \,\,\ln^2(q^2/q'^2)$. One can see
that  such a term appears as a change of the energy scale in the  normal
double
log contribution. Indeed, term  $\alpha^2_S
\,\ln(q^2/q^2_0)\,\ln^2(s/q\,q_0)$ leads to
$\alpha^2_S\,\ln^2(q^2/q'^2)\,\ln(s/q^2_0)$ -contribution if we cange the
energy scale from $s_0 = q \,q_0$ to $s_0 = q^2_0$.

The above example shows    that in the NLO we lost the
very important
property of the LO BFKL Pomeron, namely, matching the BFKL approach with
the DGLAP one,   for both $q^2 \,\,\gg\,\,q^2_0$ and $q^2_0
\,\,\gg\,\,q^2$.
The main idea , suggested by G.Salam \cite{S} and by M.Ciafaloni \cite{C},
is to replace the NLO kernel of the BFKL equation by the new kernel in
which subleading $\alpha^n_S$ -  corrections to the BFKL equation have
been   resummed in a such way that
\begin{equation} \label{RESUM}
K^{BFKL}_{resummed}\,\,\,\,\longrightarrow\,\,\,\,K^{DGLAP}\,\,,
\end{equation}
for $q^2 \,\,\gg\,\,q^2_0$ and for $q^2_0 \,\,\gg\,\,q^2$.

{\bf A possible way out -  a toy model:} Actually, the influence of the
energy scale on the BFKL equation has been studied by Andersson,Gustafson
and Samuelson \cite{AGS} in their Linked Dipole Chain Model which
reproduces the DGLAP double logs $\ln(1/x) \, \ln(q^2/q^2_0)$
for $q^2 \,\,\gg\,\,q^2_0$ and for $q^2_0 \,\,\gg\,\,q^2$.

The key equation of this model
\begin{equation} \label{TOYEQ}
\frac{\partial G (y, \vec{q},\vec{q}_0)}{\partial
y}\,\,=\,\,\frac{\alpha_S N_c}{\pi}\,\,\int\,\frac{d^2
l}{\pi\,l^2}\,\,\{\,\,
G(x',q'^2)\,\,-\,\,G(x',q^2) \,\Theta (q - l )\,\,\}
\end{equation}

with  $ \vec{l}\,\,=\,\vec{q}\,\,-\,\,\vec{q}'$ and
$$  y \,=\,\ln(1/x)\,\,;\,\,\,\,\,\,\, x\,=\,
\frac{q^2}{s}\,\,;\,\,\,\,\,x'\,\,=\,\,{ \rm max}(x, x\,\frac{q'^2}{q^2}
)\,\,.
$$
This equation is the BFKL one if $x' = x$. With a new definition of the
energy variable $x'$,  the equation sums some of the next-to-leading order
corrections to the BFKL kernel.

Fig.3-a shows the Mellin image of the
kernel for the BFKL case  ( $x = x'$ , $\chi^{(0)}$ ), for the
next-to-leading order correction to the BFKL equation ( $ \chi^{( 0 + 1
)}$)  and for the kernel of Eq.(~\ref{TOYEQ} ) ( all orders). One can see,
that the NLO gives a minimum at $\nu $ = 0 which corresponds to
oscillation in the total cross section. However, the resummed kernel (all
orders) has the same qualitatively behaviour as the LO BFKL kernel.
In Fig.3-b the BFKL intercept is plotted for the solution of
Eq.(~\ref{TOYEQ} ). One can see that the negative intercept for
$\alpha_S\,\,> $ 0.2 in the NLO is irrelevant for  the intercept of 
Eq.(~\ref{TOYEQ} ), which is about twice smaller than the intercept of the
LO BFKL equation,  but is definately positive.
\begin{figure}[h]
\begin{tabular}{l l}
\epsfig{file=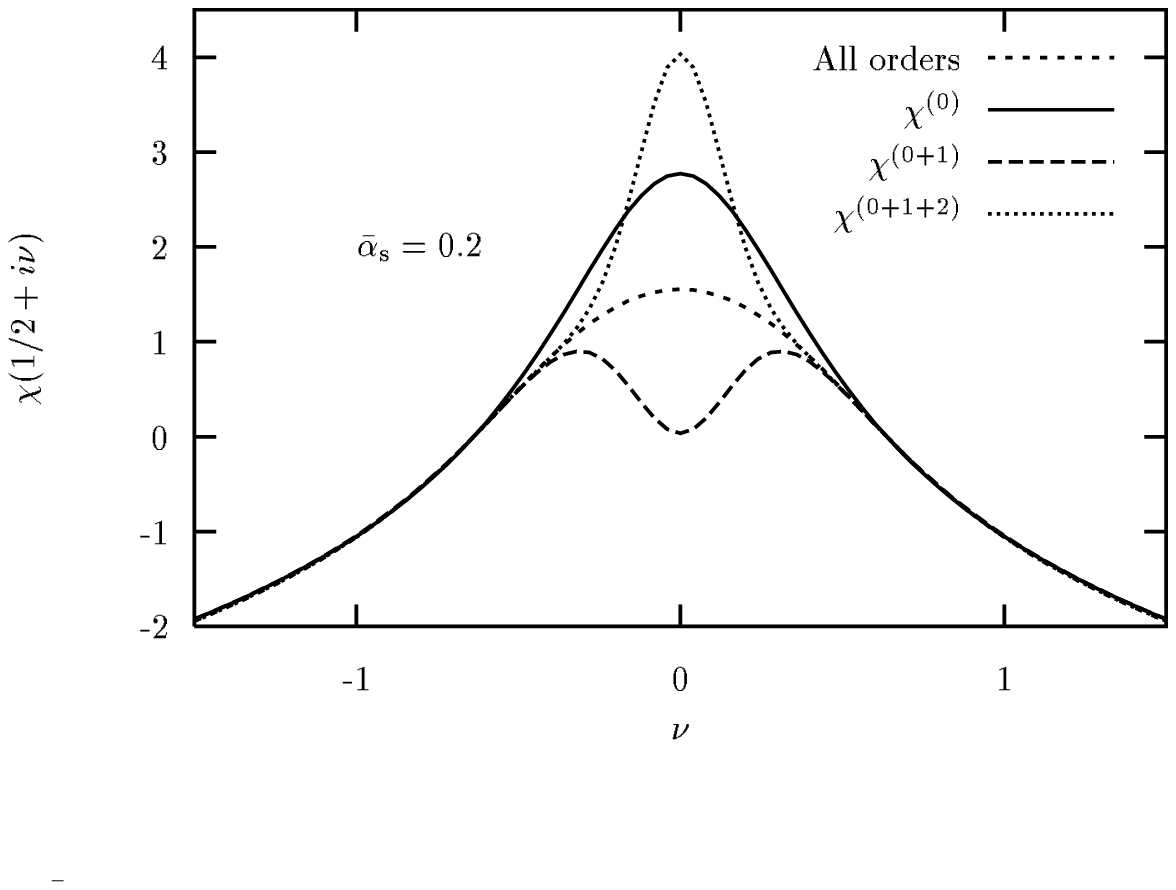,width=8.5cm,height=6.2cm} &
\epsfig{file=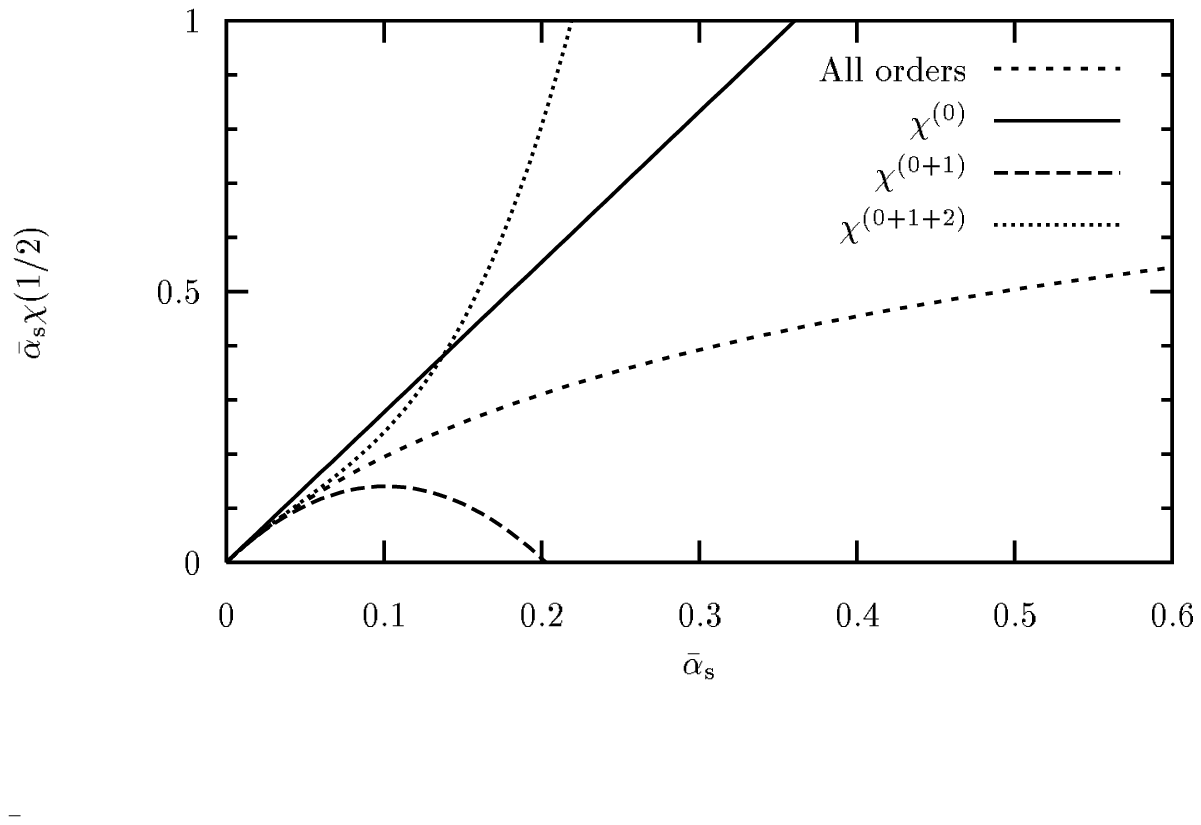,width=8.5cm,height=6.2cm}\\
Figure 3-a &
Figure 3-b\\
\end{tabular}
 \vspace*{2mm}
\caption[fig3]{{\it
The Mellin image of the toy kernel ( a) and the intercept of the
toy-Pomeron (b ). See text for detail description.
  }}
\end{figure}

Therefore, we can conclude from this toy model that the principle,
formulated in the previous subsection ( see Eq.( ~\ref{RESUM}) ), works
and the resummation has a
chance to heal our difficulties in the NLO BFKL equation.

{\bf A possible way out -  an example of resummation:}

G.Salam \cite{S}, using the principle of Eq.( ~\ref{RESUM}~),  suggested
how to deal with unphysical logs of $\alpha^2_S \,\ln^2(q^2/q^2_0)$ -
type. He demonstrated how to extend ( resum ) the NLO BFKL kernel so as to
guarantee the concellation of unphysical logs to all orders and to provide
the
matching of the BFKL Pomeron with the DGLAP evolution equation.
Unfortunately, the prescription for resummation is ambiguous and Salam
suggested four realizations ( schemes) of his ideas. It should be
mentioned that other schemes were advocated \cite{CC} as well as the
arguments for Salam's  schemes 3 and 4 were suggested \cite{CC}\cite{FRS}
\cite{SCH}\cite{FST}. In Refs. \cite{FRS} \cite{SCH} the criteria was
suggested following the idea of Lipatov: the independence of the resummed
result on the energy cutoff. This criteria is in the same  spirit as the
independence of the result on the factorization scale and it can be a good
first step in understanding of the accuracy of the resummed BFKL Pomeron.

Fig.4  shows the behaviour of the BFKL intercept ( $\omega_L =
\bar \alpha_S \,\chi(\frac{1}{2}) )$
in Eq.(~\ref{BFKLLO} )) and  the width of the BFKL Pomeron ( $D = \bar
\alpha_S \,\chi''(\frac{1}{2}) $ ). One can see that the BFKL Pomeron
intercept is positive in contrast with the NLO approximation. $D$ is also
positive,  but only for schemes 3 and 4. The second important conclusions
that we can derive from Fig.4, is the similarity of the resummed NLO BFKL
kernel with  the toy model which has been discussed.

\begin{figure}[h]
\begin{center}
\epsfig{file=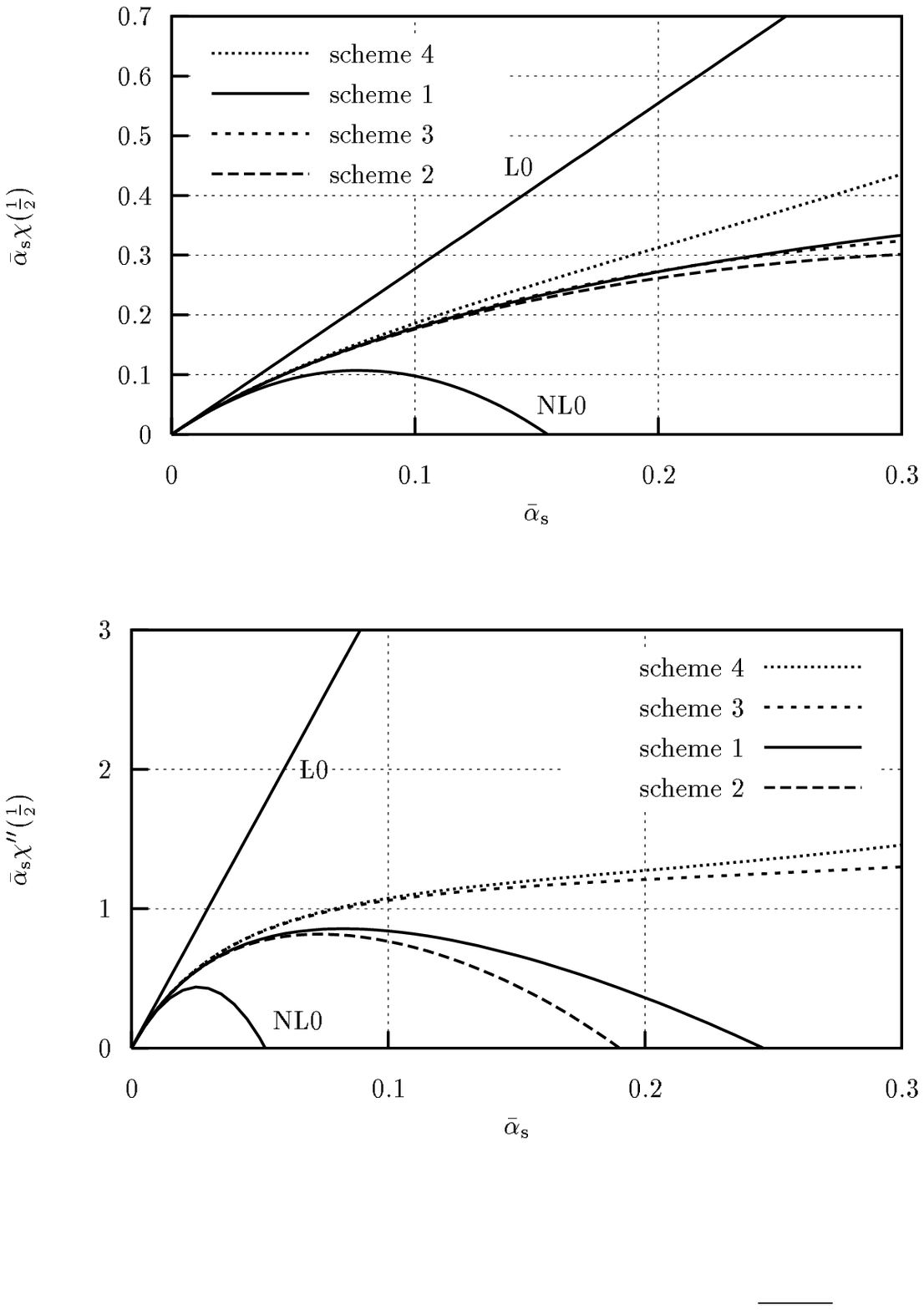,width=12cm}
\end{center}
 \vspace*{2mm}
\caption[fig4]{{\it
The BFKL Pomeron  intercept ($\omega_L\, = \,
\bar \alpha_S \,\chi(\frac{1}{2})$
 and  the width of the BFKL Pomeron ( $D\, = \,\bar
\alpha_S \,\chi''(\frac{1}{2}) $). }}
\end{figure}

{\bf Resume and recommendations:} Let me share with you 
 to-day ( 20.04.1999 ) my  understanding of  the BFKL Pomeron standing.
\begin{enumerate}
\item\,\,\, Large NLO corrections to the BFKL kernel left us without
any theoretical predictions for both  the BFKL Pomeron intercept
( $\omega_L$ in  Eq.(~\ref{BFKLLO} ) ) and the BFKL Pomeron width ( $D$ in
 Eq.(~\ref{BFKLLO} )  ). We can introduce both of them as phenomenological
parameters in a MC code,,  which should be extracted from the experimental
data.  The resummed BFKL kernel can be used to check how reasonable 
 these  extracted values of $\omega_L$ and $D$ will be;

\item\,\,\, For the running QCD coupling constant ($\alpha_S$ ) the
Regge-like
asymptotic of Eq.(~\ref{BFKLLO} ) can be justified only in a limited range
of energies. Therefore, it is better to use the numerical  solution of the
BFKL equation than the analytical expression of Eq.(~\ref{BFKLLO} );

\item\,\,\,The main qualitative properties of the parton BFKL cascade
remain the same as in LO BFKL,  for the NLO resummed BFKL equations. This
is
the very important conclusion for MC, which is the most striking
theoretical result;

\item\,\,\, It is better to use the Linked  Dipole  Chain ( LDC ) Monte
Carlo \cite{KL} than the BFKL equation itself. However, we first  need to
check
 that the LDC MC    describes the LDC model \cite{AGS}\footnote{
I am very thankful to L.Lonblad for discussions with me regarding the LDC
MC. I
realised that the MC experts have a problem in understanding  why the LDC
MC
describes the experimental data worse than other approaches. I am very
certain that as a result of such understanding will be a MC code which
will include  more from the BFKL cascade.}.
\end{enumerate}

\section{Two indications for shadowing corrections from HERA data}

~
{\bf HERA puzzle:} The wide spread opinion is that HERA experimental
data for  $Q^2 \,\gsim\,1\,GeV^2$  can be described using only the DGLAP
evolution equations,   without any
other ingredients such as shadowing corrections, high twist contributions
and so on  ( see, for example reviews
\cite{CSDDREV} \cite{ACREV} ). On the other hand, the most important HERA
discovery is the fact that the density of gluons ( gluon structure
function ) becomes large in HERA kinematic region ( see Fig.5 ).

\begin{figure}[h]
\begin{center}
\epsfig{file=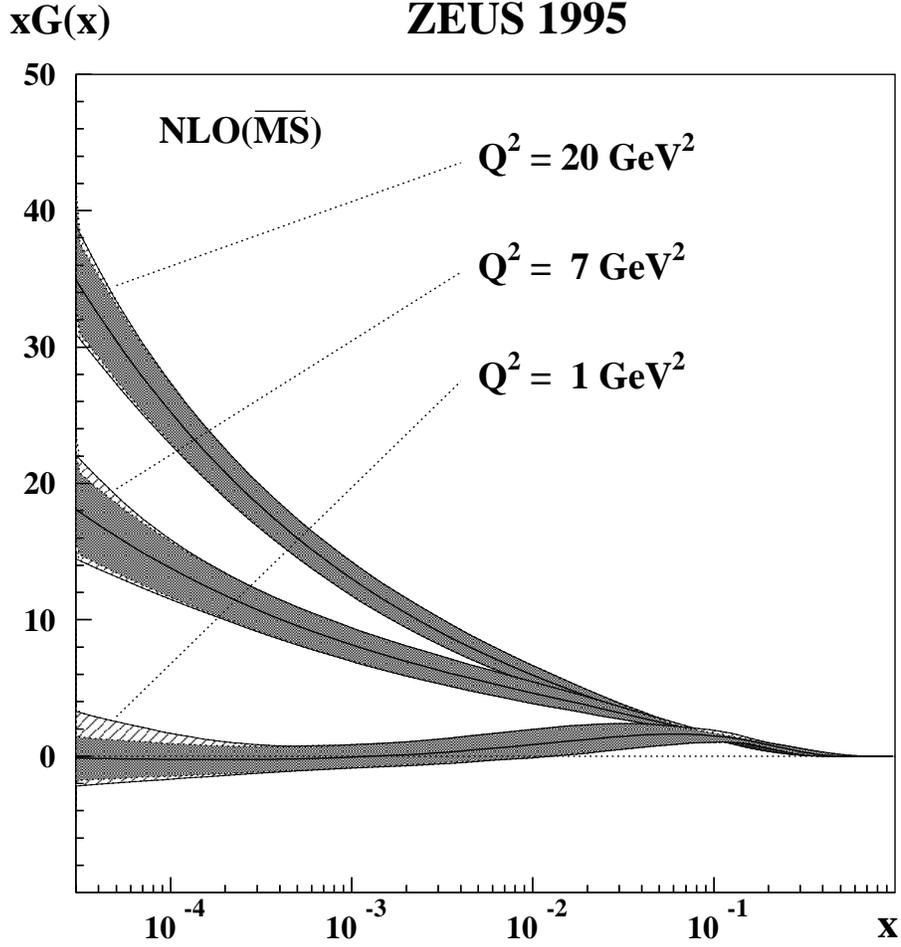,width=12cm}
\end{center}
 \vspace*{2mm}
\caption[fig5]{{\it
ZEUS \protect\cite{ZEUSGLUON} data for gluon structure function
($xG(x,Q^2)$. }}
\end{figure}

The value of the gluon density turns out to be so large that parameter
\begin{equation} \label{KAPPA}
\kappa\,\,\,=\,\,\,\frac{3\,\pi^2 \alpha_S}{2 Q^2}\,\times\,
\frac{xG(x,Q^2)}{\pi\, R^2}\,\,,
\end{equation}
 which determines the value of SC \cite{GLR} \cite{MUQI} \cite{MU90}
\cite{KOVCH}, reaches unity
\begin{equation} \label{Q0X}
 \kappa \,=\,\frac{3\,\pi^2 \alpha_S}{2 Q^2_0(x)}\,\times\,
\frac{xG(x,Q^2_0(x))}{\pi\, R^2}\,\,=\,\,1
\end{equation}
 in HERA kinematic region (see Fig.6 )

\begin{figure}[h]
\begin{center}
\epsfig{file=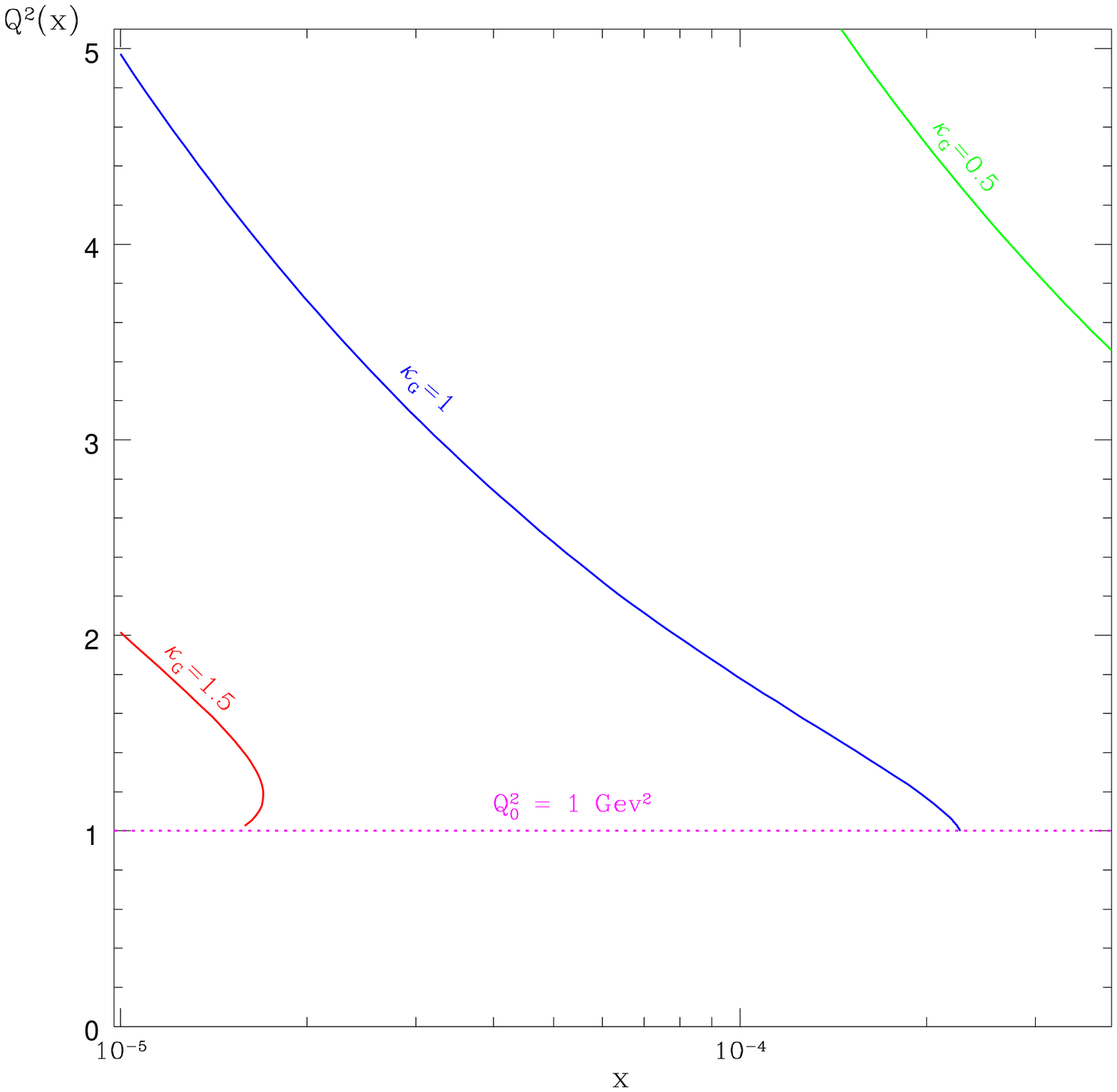,width=12cm}
\end{center}
 \vspace*{2mm}
\caption[fig6]{{\it  Kinematic plot with lines $\kappa$ = 0.5, 1  and 1.5
. }}
\end{figure}

It means that in large kinematic region where $\kappa\,\gsim\,1$ ( to the
left from line $\kappa \,=\,1 $ in Fig.6 ), we expect that the SC should
be large and important for the description of the experimental data.
At first sight such expectations are in clear contradiction with the
experimental data.  Certainly, this fact gave rise to the suspicion ( or
even
mistrust ) that our theoretical approach to the SC is not consistent.
However, the revision and re-analysis  of the SC have been completed
\cite{AGLA} \cite{AGLN} \cite{ML} \cite{KOV} \cite{MK}\cite{KOVCH} with
the  result that $\kappa$ is the parameter which is reponsible for the
value of SC.

  Therefore, we face a puzzling question: {\em where are SC in
DIS?}\, I am happy to tell you that  we now where is the energy
 have two
indications in the HERA
experimental data that the SC are rather large and important.
These indications are:
\begin{enumerate}
\item\,\,\,   $x_P$ -
behaviour of the cross section of the diffractive dissociation
( $\sigma_{DD}$ ) in DIS;

\item\,\,\,    $Q^2$ -
behaviour of $F_2$-slope (  $\frac{\partial F_2(x_B,Q^2)}{\partial \ln
Q^2}$
\,\,).

\end{enumerate}

Here, we would like to discuss both phenomena and their relation to 
the SC.

\subsection{ $ \mathbf{x_P}$ - dependence of   $\mathbf{\sigma_{DD}}$}

~

{\bf Data:} Both collaborations ( H1 and ZEUS ) found that
\begin{equation} \label{DD}
\sigma_{DD}\,\,\, \propto\,\,\,\frac{1}{x^{2\,\Delta_P}}
\end{equation}
where $\Delta_P$ \,=\,$\alpha_P( 0 )
\,\,-\,\,1$. The values of $\alpha_P( 0 )$ are:

\begin{tabular}{l l}
 $\mathbf{\bullet}$\,\,\,\,{\bf H1 \cite{H1DD}\,:} & $\alpha_P( 0
)$\,=\,1.2003
\,$\pm$\,0.020
(stat.)\,$\pm$ \,0.013 (sys.)\,\,;\\

 $\mathbf{\bullet}$\,\,\,\,{\bf ZEUS\cite{ZEUSDD}\,:} &
$\alpha_P( 0  ) $\,\,=\,\,1.127\,\,$\pm$ \,\, 0.009
(stat.)\,\,
$\pm$ \,\,0.012 (sys.)\,\,.\\
\end{tabular}

In Fig.7 the ZEUS data on the Pomeron intercept are plotted together with
the intercept the ``soft" Pomeron \cite{DL}. It is clear that the Pomeron
intercept for   the diffractive processes in DIS is higher than the
intercept of the ``soft" Pomeron.

\begin{figure}
\begin{center}
\epsfig{file=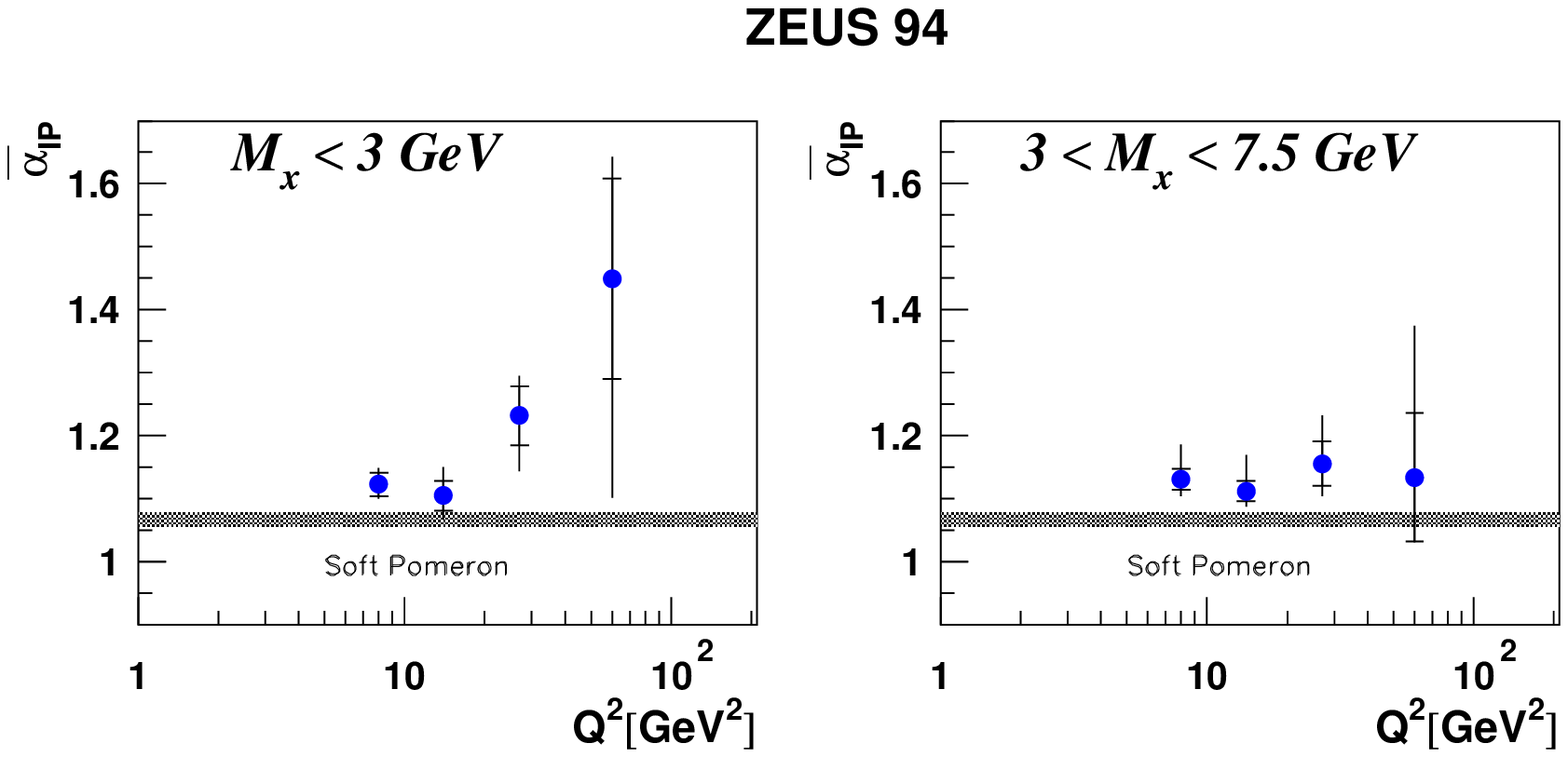,width=170mm}
\end{center}
\vspace*{2mm}
\caption[fig7]{{\it  ZEUS\protect\cite{ZEUSDD} data on the Pomeron
intercept for $\sigma_{DD}$ in DIS. }}
\end{figure}

{\bf Why is it surprising and interesting?}  To answer this question let
us consider the simplest diffractive process - the diffractive production
of the quark - anti quark pair with mass $M$ ( see Fig.8 )

\begin{figure}[h]
\begin{tabular}{l l}
\epsfig{file=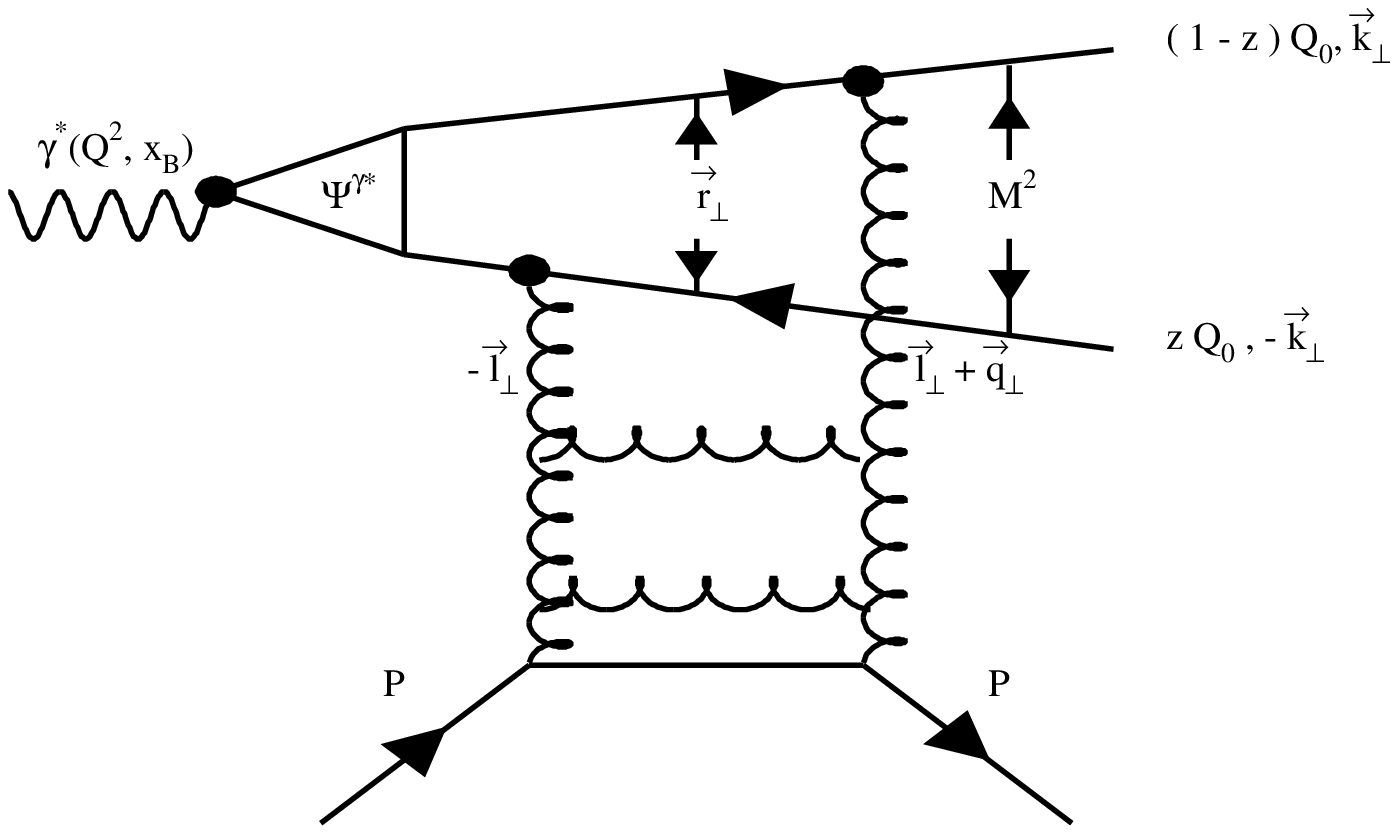,width=8.5cm,height= 7cm} &
\epsfig{file=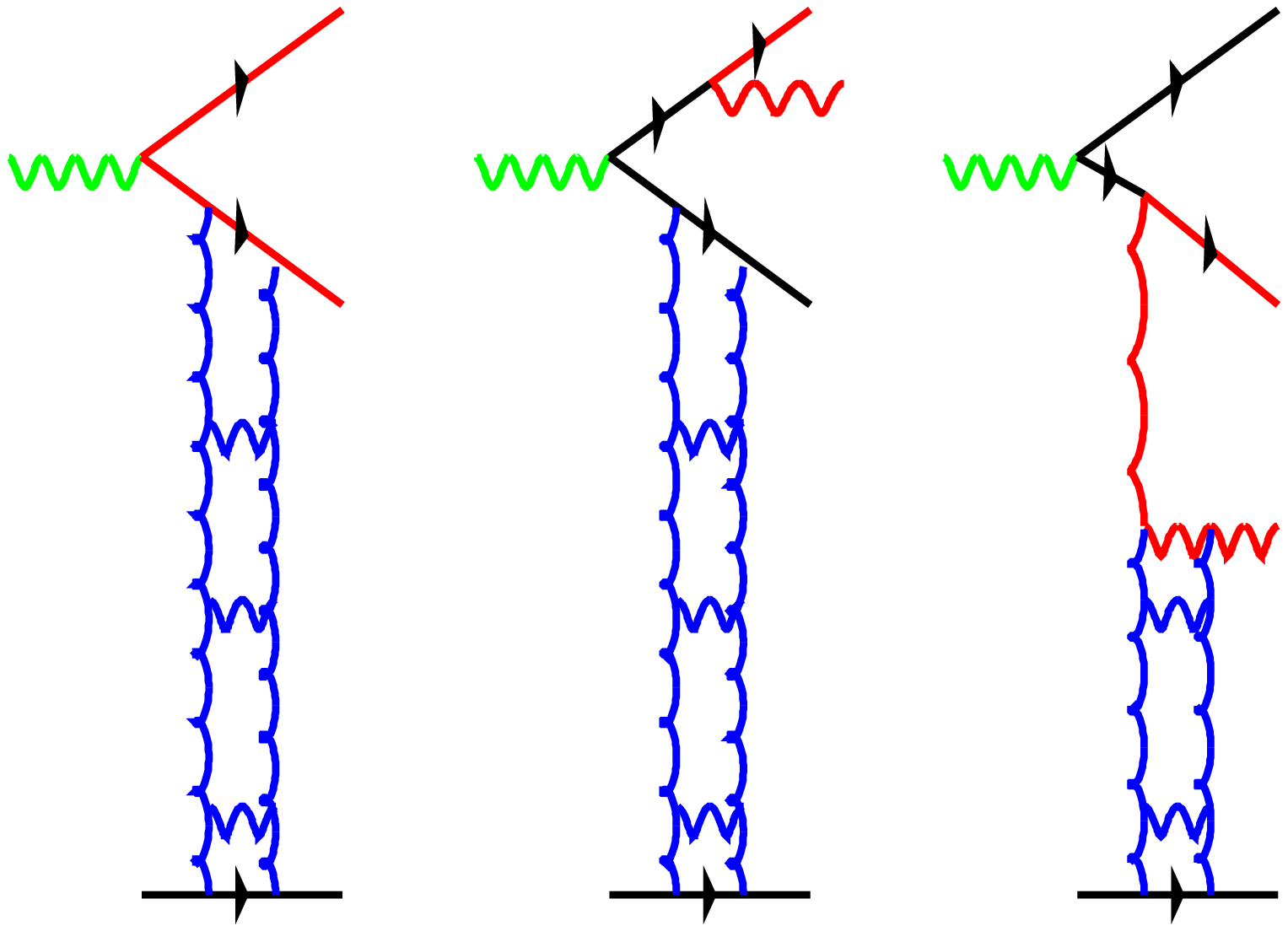,width=8.5cm,height=6cm}\\
Figure 8-a &
Figure 8-b\\
\end{tabular}
 \vspace*{2mm}
\caption[fig8]{{\it
The simplest examples of the diffractive processes: ( a ) the production
of
quark - antiquark pair with mass $M$ and (b) the production of quark -
antiquark pair with one extra gluon.
  }}
\end{figure}

As it was shown ( see Refs. \cite{BAR} \cite{NN} \cite{GLMSM}    )   the
cross
sections of this process are proportional to:

\begin{equation} \label{DD1}
x_P\,\frac{d \sigma^T_{DD}( \gamma^* \,\rightarrow\, q + \bar q )
}{d x_P d
t}\,\,\,\propto\,\,\,\int^{\frac{M^2}{4}}_{Q^2_0}\,\,\frac{d
k^2_{\perp}}{k^2_{\perp}}\,\,\times\,\,\frac{\left(\,\alpha_S\, \,x_P
\,G(x_P,
\frac{k^2_{\perp}}{1 - \beta})\,\right)^2}{k^2_{\perp}}\,\,
\end{equation}
for transverse polarized virtual photon and
\begin{equation} \label{DD2}
x_P\,\frac{d \sigma^L_{DD}( \gamma^* \,\rightarrow\, q + \bar q )
}{d x_P d
t}\,\,\,\propto\,\,\,\int^{\frac{M^2}{4}}_{Q^2_0}\,\,\frac{d
k^2_{\perp}}{Q^2}\,\,\times\,\,\frac{\left(\,\alpha_S\, \,x_P
\,G(x_P,\frac{k^2_{\perp}}{1 - \beta})\,\right)^2}{k^2_{\perp}}
\end{equation}
for the longitudinal polarized photon. In these  formulae we used the
notation ( see Fig.8 ): $Q^2$ is the photon virtuality; $M$ is the
produced mass; $x_P\,\,=\,\,\frac{Q^2 + M^2}{s}$ , where $s$ is the energy
of  virtual photon proton collisions; and $\beta
\,=\,\frac{Q^2}{Q^2\,+\,M^2}$.

The same factor also enters a more complicated process as, for example, a
diffractive production of the quark - antiquark pair and one extra gluon
 (see Fig.8-b ).
Indeed, the cross section of this process is proportional to \cite{GLMSM}:
\begin{equation} \label{DD3}
 x_P\,\frac{d \sigma^T_{DD}( \gamma^* \,\rightarrow\, q \,+ \,\bar q \,
+\, G )
}{d x_P d
t}\,\,\,\propto\,\,\,\,\int^{\frac{M^2}{4}}_{Q^2_0}\,\,\frac{d
k^2_{\perp}}{k^2_{\perp}}\,\,\times\,\ln(M^2/4 k^2_{\perp})
\,\,\times\,\,
\frac{\left(\,\alpha_S\, \,x_P
\,G(x_P,k^2_{\perp})\,\right)^2}{k^2_{\perp}}\,\,.
\end{equation}

One can see that for the transverse polarized photon the diffractive cross
section stems from the low values of the transverse momenta $k_{\perp}$
( see Eq.(~\ref{DD1} ) and  Eq.(~\ref{DD3} ) ) if we use for $xG$ the
solution of the DGLAP equations. However, the situation changes crucially
if we take into  account the SC in expression for $xG$. For estimates of
the value of the SC,  we use the Glauber - Mueller approach \cite{MU90}
for
$xG$, namely \cite{GLMSM}  \cite{GLMN}:
\begin{equation} \label{DD4}
x_P G^{SC}( x_P,k^2_{\perp} )\,\,\,=\,\,\,\frac{2}{\pi^2}
\,\int^1_{x_P}\,\frac{d x'}{x'}
\int^{\infty}_{\frac{1}{k^2_{\perp}}}\,\,\frac{d r^2}{r^4} \int d b^2_t
 \{\,\,1\,\,\,-\,\,\,e^{- \kappa^{DGLAP}(x',r^2) \,S(b_t)}\,\,\}\,\,
\end{equation}
where  $S(b_t)\,\,=\,\,e^{- \frac{b^2_t}{R^2}}$.

From Eq.(~\ref{DD4} ) we can see that in the limit of low $x$
\begin{equation} \label{DD5}
 x_P G(x_P,k^2_{\perp})\,\,\,\,\,
\Longrightarrow\,\,\,\, k^2_{\perp} \,R^2\,\,\int^1_x \,\frac{d x'}{x'}
\,\ln \kappa(x',\frac{1}{k^2_{\perp}})
\end{equation}
if $k^2_{\perp}\,\,\ll\,\,Q^2_0(x)$, where $Q^2_0(x)$ is plotted in Fig.6.
Therefore, the  substitution of  $xG^{SC}$ in  Eq.(~\ref{DD1} ) and
Eq.(~\ref{DD3} ) leads to the situation that the integral over
$k^2_{\perp}$ is not
infrared divergent, and the main contribution comes from the region of
$k^2_{\perp}$ of the order of $Q^2_0(x_P)$. It means that
\begin{equation} \label{DD6}
x_P \,\frac{d \sigma_{DD}}{d x_P dt}
\,\,\,\,\,\Longrightarrow\,\,\,\,\,\frac{1}{Q^2_0(x_P)}\,\,\times\,\,
\left(\,x_P\,G(x_P,Q^2_0(x_P))\,\right)^2\,\,\,\,\propto\,\,\,\frac{1}{x^{2
\lambda_{eff}(Q^2)}_P}
\end{equation}
Eq.(~\ref{DD6} ) allows us to find the typical distances which are
reponsible for the diffraction dissociation in DIS. Comparing the
experimental values of $\lambda_{eff}(Q^2)$ with the ones calculated using
Eq.(~\ref{DD4} ) (see Fig.9 )\footnote{Doing this comparison, we have to
take into account that $\lambda_{eff} (Q^2) \,\,=\,\,\lambda(Q^2) ( Fig. 9
)\, - \,\frac{1}{2}\,\,\frac{d \ln Q^2_0(x)}{d \ln(1/x)}$. $Q^2_0(x)$ in
Fig.6 can be described by simple formula $Q^2_0(x)\,\,=\,\,1
\,GeV^2\,(\,\frac{x_0}{x}\,)^{0.52}$. }
 , one can
conclude that the typical
$k^2_{\perp}$ is not small, but rather  $k^2_{\perp}\,\,\approx \,\,1 -
2\,GeV^2$ \cite{GLMSM}\cite{GLMN},, if we take $x_P \,\approx \,10^{-3}$,
or even larger.

\begin{figure}
\begin{center}
\epsfig{file=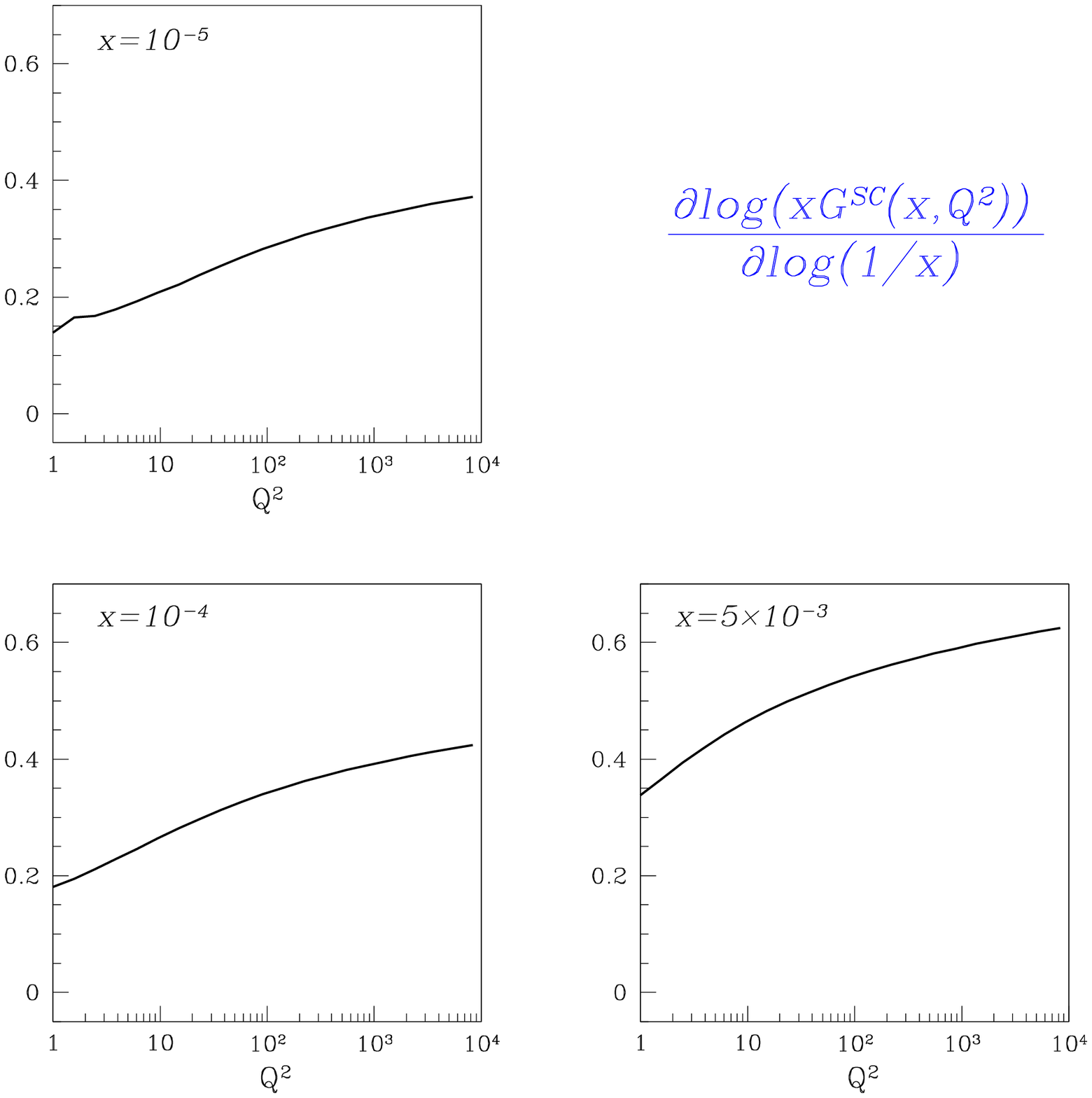,width=130mm}
\end{center}
\vspace*{2mm}
\caption[fig9]{{\it $\lambda(Q^2)
\,=\,\frac{1}{x_P\,G^{SC}(x_P,k^2_{\perp})}\,\, \frac{d
x_P\,G^{SC}(x_P,k^2_{\perp})}{d \ln(1/x_P) }$
calculated using Eq.(~\protect\ref{DD4} ).}}
\end{figure}

It should be stressed that the diffraction dissociation for longitudinally
polarized photon ( see Eq.{~\ref{DD2} ) for quark - anriquark production 
comes from  long distances and its cross section is proportional to
\begin{equation} \label{DDL1}
x_P\,\frac{d \sigma^L_{DD}( \gamma^* \,\rightarrow\, q + \bar q )
}{d x_P d
t}\,\,\,\propto\,  \left( \,\alpha_S \,x_P \,G(x_P,\frac{M^2}{4( 1 -
\beta )})\,\right)^2
\end{equation}

One of the striking feature of the experimental data, is the fact that the
cross section of the diffractive dissociation in DIS and the total DIS
cross section have the same energy dependence. Recently, an explanation of
such a behaviour was suggested based on the SC contributions\cite{GBW1}
\cite{GBW2} \cite{LAKOV}. Indeed, the total cross section and cross
section of diffractive production are equal to:
\begin{eqnarray} 
&
\sigma_{tot} ( \gamma^* p )\,\,\,  =  \,\,\, \int d^2 r_t
\int^1_0 \,d z\,\, | \Psi( Q^2; r_t,z ) |^2 \,\,\,\sigma_{tot} ( r^2_t, x
)\,\,;& 
\label{DDEB1}\\
&
\sigma_{DD} ( \gamma^* p )\,\,\, = \,\,\, \int d^2 r_t
\int^1_0 \,d z \,\,| \Psi( Q^2; r_t,z ) |^2 \,\,\,\{\,\sigma_{tot} (
r^2_t, x
)\,\,-\,\,\sigma_{in}( r^2_t,x)\,\}\,\,;&\label{DDEB2}
\end{eqnarray}
where $\sigma_{tot} ( r^2_r, x )$ and $\sigma_{DD} ( r^2_r, x)$ are total
and inelastic cross section for dipole with distance $r_t$ between quark
and antiquark, respectively.

The solution of the unitarity constraint for dipole scattering gives:
\begin{eqnarray}
&
\sigma_{tot} ( r^2_t, x )\,\,=\,\, 2\,\int\,d^2b_t \,\{\,1\,\,\,-\,\,\,
e^{ - \kappa^{DGLAP}_q \,S(b_t)}\,\}\,\,; & \label{DDEB3}\\
&
\sigma_{in} ( r^2_t, x )\,\,=\,\, \,\int\,d^2b_t \,\{\,1\,\,\,-\,\,\,
e^{ - 2\,\kappa^{DGLAP}_q \,S(b_t)}\,\}\,\,; & \label{DDEB4}
\end{eqnarray}
where $ k^{DGLAP}_q\,\,= \frac{4}{9} \,\,\kappa^{DGLAP}$ ( see 
Eq.(~\ref{KAPPA} ) ). 

After integration over $z$ we get for transverse polarized photon
\cite{AGLFRST}
\begin{equation} \label{DDEB5}
\int^1_0 \,d z \,\,| \Psi_T( Q^2; r_t,z ) |^2
\,\,\,=\,\,\,\frac{4\,\alpha_{em}N_c}{3 \,\pi^2\,Q^2}
\,\times\,\frac{1}{r^4_t}\,\,.
\end{equation}
Using the expilict expression for   $S(b_t)$ and assuming that the
anomalous
dimension of $xG$ is not large $\gamma\,\ll\,1$ ( in this case  we can
replace
$xG(x,k^2_t)$ by $xG(x,Q^2)$ in the integrals )  we can obtain
\cite{LAKOV}:
\begin{eqnarray}
&
\sigma_{tot} ( \gamma^* p )\,\,\,\propto\,\,2\,R^2\,\{\,\ln (\kappa_q )
\,\,+\,\,C\,\,(\,1 \,+\,\kappa_q\,)\,E_1( \kappa_q) \,+\,1\,-\,e^{-
\kappa_q} \,\}\,\,;& \label{DDEB6}\\
&
\sigma_{DD}( \gamma^* p )\,\,\,\propto \,\,\,2\,R^2\,\{\,\ln (\kappa_q )
\,\,+\,\,C\,\,(\,1 \,+\,\kappa_q\,)\,E_1( \kappa_q) \,+\,1\,-\,e^{-
\kappa_q} \,\}\,\,&\\ \nonumber
 &
  -\,\,\,R^2\,\{\,\ln
(2\,\kappa_q )
\,\,+\,\,C\,\,(\,1 \,+\,2\,\kappa_q\,)\,E_1( 2\,\kappa_q) \,+\,1\,-\,e^{-
2\,\kappa_q} \,\}\,\,;&\label{DDEB7}
\end{eqnarray}
where $C$ is the Euler constant and $E_1$ is the interral exponant
\cite{MATHBOOK}.

These formulae indicate  that $\sigma_{tot} \,\rightarrow xG(x,Q^2) $ and 
$\sigma_{DD}\,\rightarrow\,\left( \,xG(x,Q^2 )\,\right)^2$ for small
$\kappa_q\,\ll\,1$, but at $\kappa_q\,\,\gg\,\,1$ we obtain that
\begin{eqnarray}
&
\sigma_{tot} ( \gamma^* p )\,\,\, \longrightarrow \,\,2\,R^2\,\{\ln
(\kappa_q)\,\,+\,\,C\,\}
\,\,;&\label{DDEB8}\\
&
\sigma_{DD}( \gamma^* p ) \,\,\, \longrightarrow \,\,\,2\,R^2\,\{\,\ln
(\kappa_q )
\,\,+\,\,C\,\} \,\,-\,\,
 R^2\,\{\,\ln (2\,\kappa_q )
\,\,+\,\,C\,\} \,\,=\,\, R^2\,\ln( \kappa_q)
\,\,=\,\,\frac{1}{2}\,\sigma_{tot}.&\label{DDEB9}
\end{eqnarray}
This simple calculation shows that indeed at low $x$, where $\kappa_q$ is
expected to be large, $\sigma_{DD} $ and $\sigma_{tot}$ \cite{LAKOV},
should have the
same energy dependence but accuracy of such a statement is controlled by
ratio
$ C/ln (\kappa_q) \,\,\ll\,\,1$. In Refs. \cite{GBW1} and \cite{GBW2} one
can find estimates of how SC reveal themselves in real experimental
situation. It turns out that they are able to describe the experimental
energy behaviour of $\sigma_{DD}$.

\subsection{The effect of screening on $\mathbf{Q^2}$ - dependence of the
$\mathbf{F_2}$ - slope.}

~

{\bf Data:} The experimental data \cite{ZEUSSLOPE} for the $F_2$ - slope
$\frac{d F_2(x,Q^2)}{d \ln Q^2}$
is shown in Fig.10 ( Caldwell plot ). These data give rise to  a hope that
the
matching between
``hard" ( short distance ) and ``soft" (long distance ) processes occurs
at sufficiently large $Q^2$. Indeed, the $F_2$ - slope starts to deviate
from the DGLAP predictions around $Q^2\, \approx\, 5 - 8 \,GeV^2$.

\begin{figure}[h]
\begin{center}
\epsfig{file=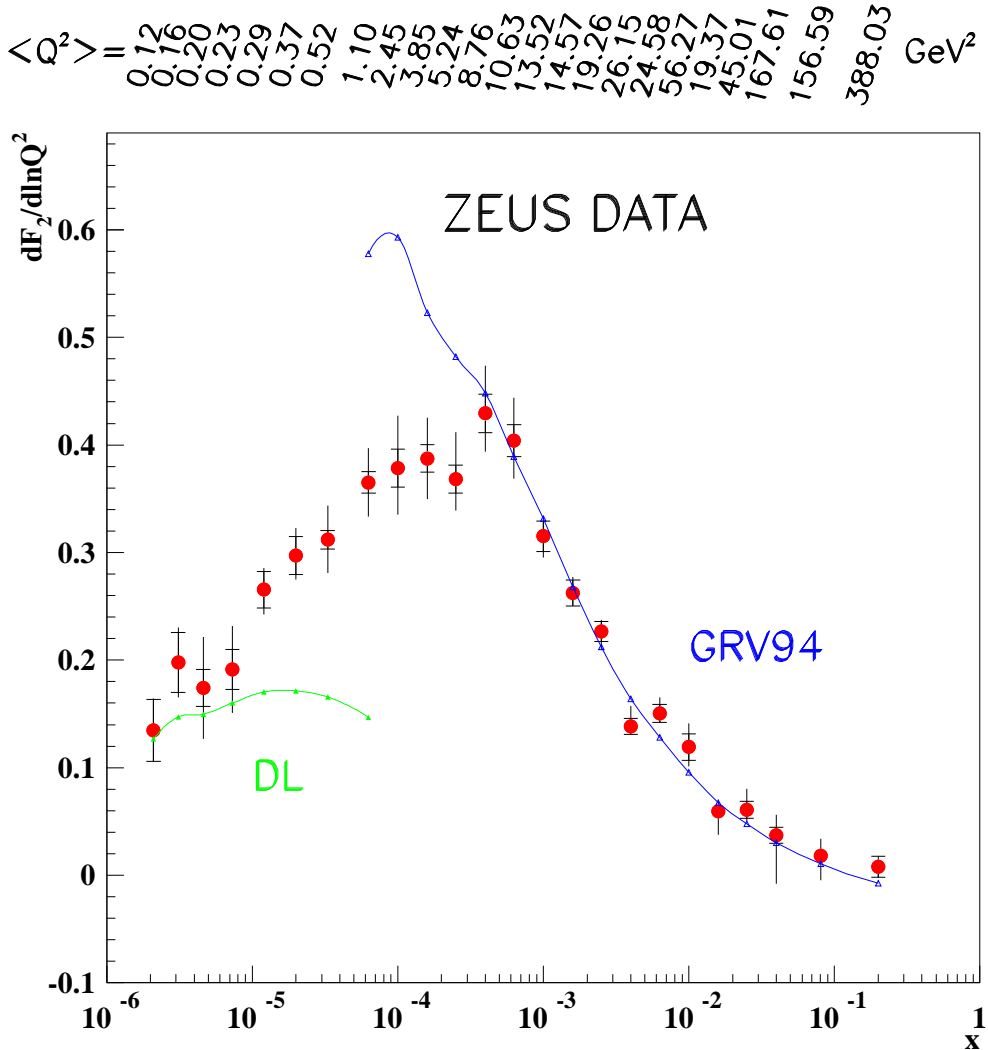,width=130mm}
\end{center}
\vspace*{2mm}
\caption[fig10]{{\it The ZEUS data on the $F_2$ - slope ( Caldwell plot
).}}
\end{figure}

Such a large value of this separation parameter between short and long
distance processes leads to a natural question:``Could the experimental
behaviour of the $F_2$ - slope as a function of $Q^2$ be a manifestation
of the shadowing corrections or, better to say, of the saturation 
of the gluon density in DIS ?".  The answer \cite{MUSL} \cite{GLMSL}
\cite{GLMN}
is: ``Yes".

{\bf $\mathbf{F_2}$ - slope and SC:} In Refs. \cite{GLMSL} \cite{GLMN}
 the influence of the SC on the behaviour was calculated using
Eq.(~\ref{DD4} ) and  similar formula in quark-antiquark sector.
The result is given in Fig.11 where the $F_2$ - slope in ALLM
parameterization \cite{ALLM} is also plotted. Comparison with the
experimental data as
well as with the ALLM  predictions shows that the SC can give a plausible
explanation of the effect.

\begin{figure}
\begin{center}
\epsfig{file=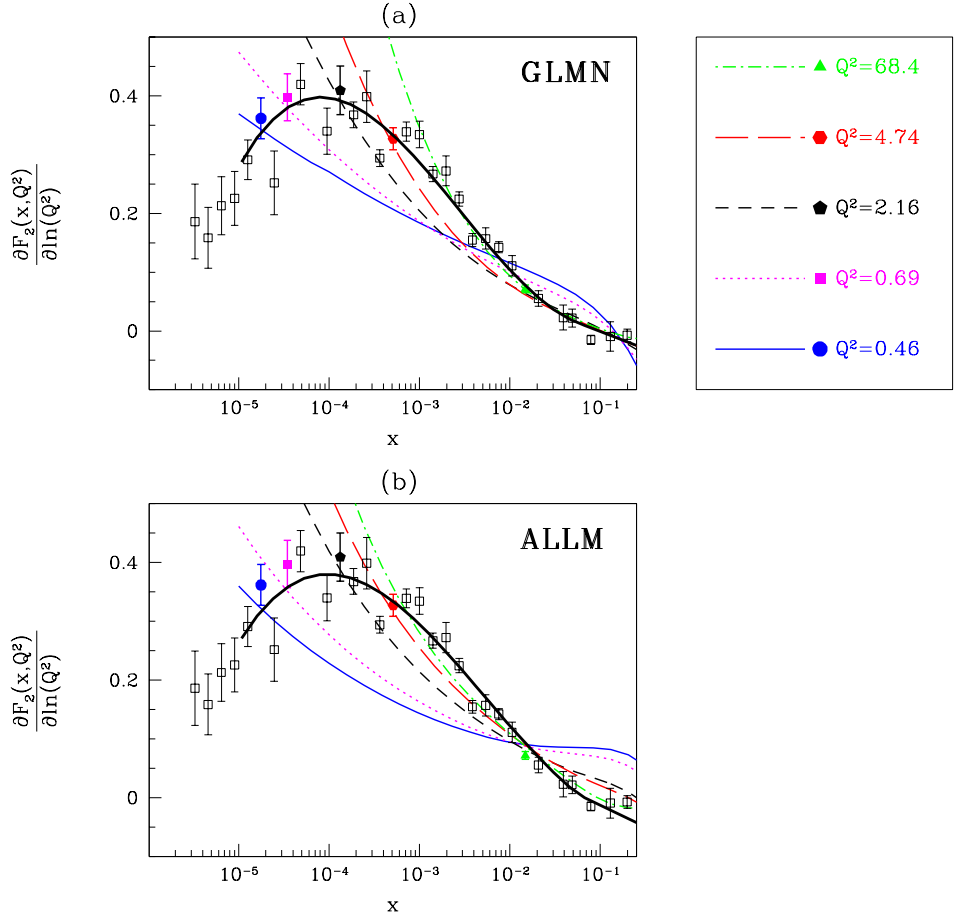,width=130mm}
\end{center}
\vspace*{2mm}
\caption[fig11]{{\it The $F_2$ - slope in QCD calculations taking into
account SC ( GLMN ) \protect\cite{GLMN} and in ALLM'97
parameterization \protect\cite{ALLM}, which describes the matching
between ``soft " and ``hard" interactions in DIS.}}
 \end{figure}

We would like to draw your attention to the fact that at fixed value of
$Q^2$ the SC do not lead to qualitative change of $x$ - dependence of the
$F_2$-slope ( see lines with fixed $Q^2$ in Fig.11 ). However, the
saturation of the gluon density gives $F_2 \,\,\propto\,\,Q^2\,R^2$ at 
$Q^2\,<\,Q^2_0(x)$. Therefore, the $F_2$ - slope shows a decrease at small
$Q^2$. The calculations support the idea that the experimmental selection
of the data on the $F_2$ - slope shows predicted $Q^2$-dependence.
However, it is important to stress that only data on $F_2$ - slope at
fixed energy ( or $x$ ) would clarify how much of the experimentally
observed effect is related to the  $Q^2$ - dependence. It is worthwhile
mentioning
that the ALLM'97 parameterization which describes phenomenologically the
matching between ``hard" and ``soft" interactions  in DIS, predicts 
the same 
behaviour of the $F_2$ - slope as the SC calculations ( see Fig.11 ).

\subsection{Why these two facts are still only the indications of
 SC contributions?}

~

I hope that I have convinced you that the SC provide a natural explanation
of both experimental effects. Even more, the new experimental data give
a possible way to understand and resolve the HERA puzzle, that has been
formulated at the beginning of this section. We have expected such effects
for a long time, and the SC recieved  strong support from the experiment.

However, we cannot  not make a final conclusion because both  the
experimental sets of data on $x_P$-dependence of the diffractive cross
section, and on $Q^2$ - dependence of the $F_2$ - slope can have an
alternative explanation. For example, the MRST parameterization
\cite{MRST} gives the explanation for the $F_2$ - slope behaviour assuming
the strange behaviour of the gluon structure function at the initial
virtuality $Q^2 = Q^2_0$ - decreasing  at low $x$.

Nevertheless, we would like to stress that the SC give a good agreement
with the ALLM'97 parametrization in the wide range of $Q^2$ and $x$.
This   provides  strong support for  SC  which, we hope, will be
confirmed
by the future experimental data.

We need  to introduce  SC in the Monte Carlo codes to be prepared for
the future experiments.

\section{Matching of ``soft" and ``hard" photon - proton interactions}

~
{\bf Gribov's approach:} The rich and high precision data  on deep
inelastic $e p $ scattering
at HERA \cite{H1} \cite{ZEUS}, covering both low and high $Q^2$ regions,
lead to a theoretical problem of matching the non-perturbative (``soft")
and perturbative (``hard ") QCD  domains. This challenging problem has
been under close investigation over the past two decades, starting from
the pioneering paper of Gribov \cite{GRIBOV} ( see also \cite{GVD} ).

Gribov suggested  two stages of the
 $\gamma^* p$ interaction in any QCD description (see Fig.12-a):
\begin{enumerate}
\item\,\,\,The $\gamma^*$ converts into  a hadron system (quark-antiquark
  pair to lowest order) well before the interaction with the target;

\item\,\,\,The quark-antiquark pair (or hadron system) then interacts with
the
target.

\end{enumerate}
These two stages are expressed explicitly in the double dispersion
relation
suggested in Ref. \cite{GRIBOV} ( see Fig.12-b ):

\begin{figure}[h]
\begin{tabular}{l l}
\epsfig{file=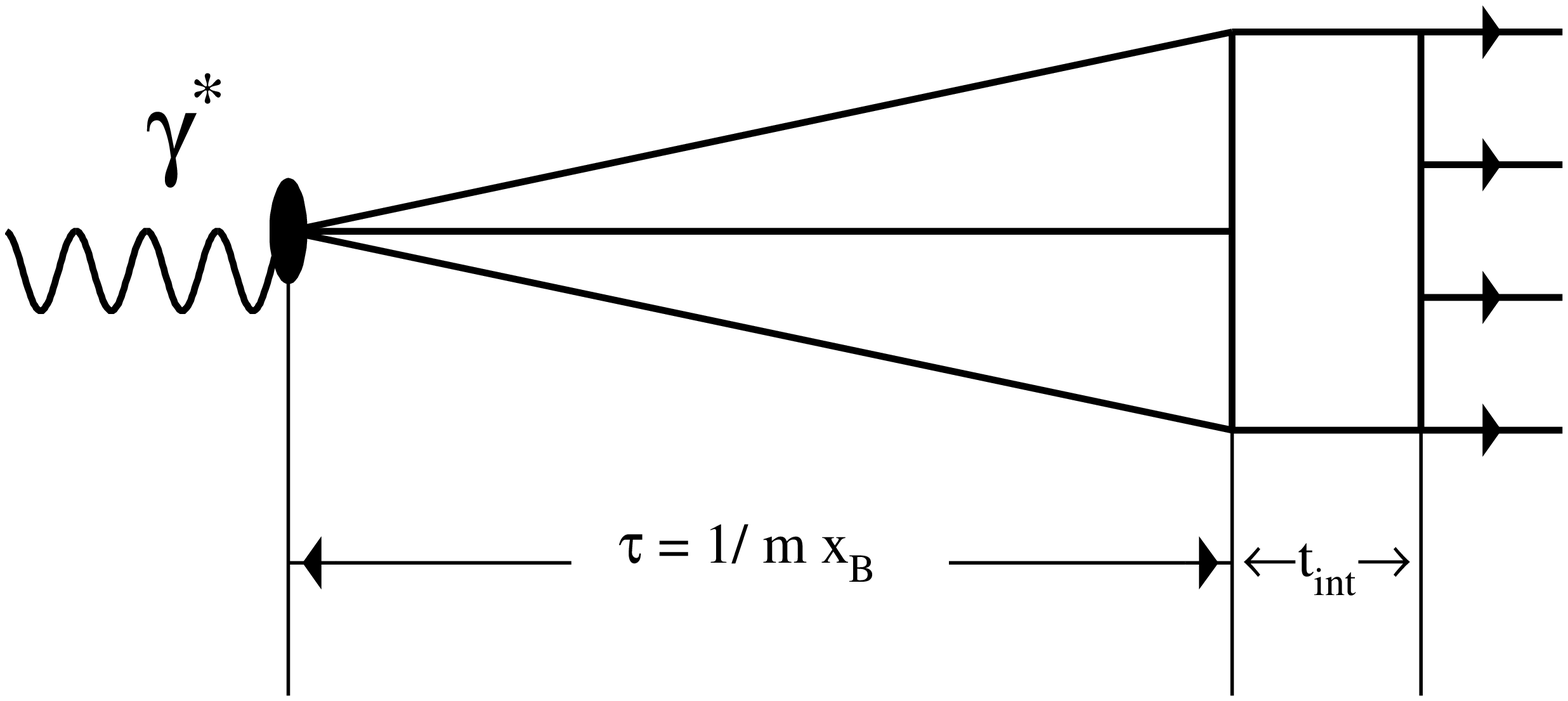,width=8.5cm,height= 7cm} &
\epsfig{file=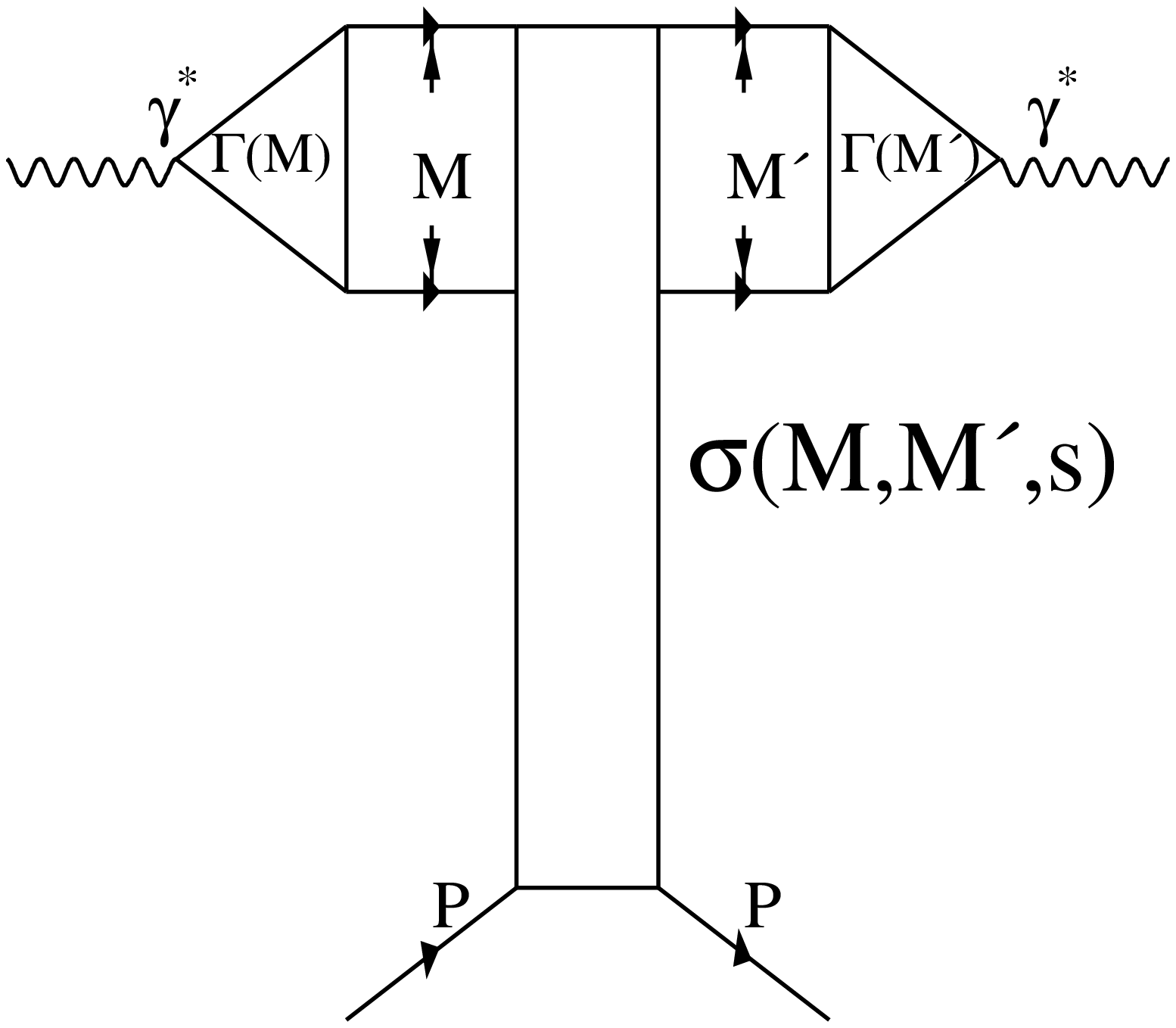,width=8.5cm,height=7cm}\\
Figure 12-a &
Figure 12-b\\
\end{tabular}
 \vspace*{2mm}
\caption[fig12]{{\it
Gribov's approach to photon - hadron interaction: (a) the time structure
of this interaction and (b) the double mass dispertion relation. }}
\end{figure}
\begin{equation}\label{GRIBOVgen}
\sigma(\gamma^* N)=\frac{\alpha_{em}}{3\pi}\int
  \frac{\Gamma(M^2) dM^2}{(Q^2+M^2)}
    \sigma(M^2,M'^2,s)
  \frac{\Gamma(M'^2) dM'^2}{(Q^2+M'^2)},
\end{equation}
where $M$ and $M'$ are the invariant masses of the incoming and outgoing
quark-antiquark pairs, $\sigma(M^2,M'^2,s)$ is the cross section of a
$q \bar q $ interaction with the target, and the vertices $\Gamma^2(M^2)$
and
$\Gamma^2(M'^2)$ are given by  $\Gamma^2(M^2)=R(M^2)$, where $R(M^2)$ is
the
ratio:
\begin{equation}\label{Rdef}
R(M^2)=\frac{\sigma(e^+e^-\to \mbox{hadrons})}{\sigma(e^+e^-\to
\mu^+\mu^-)},
\end{equation}
which has beem measured experimentally.

The key problem in all approaches utilizing  Eq.(~\ref{GRIBOVgen} ) is the
 description of the cross section $\sigma(M^2,M'^2,s)$.
\begin{enumerate}
\item\,\,\,We introduce $M_0 \,\,\approx \,\,1\,GeV$ in the integrals over
$M$ and $M'$, which play the role of a separation parameter.  For $M,\,
M'
\,>\,M_0$ the quark - antiquark pair are produced at short distances (
$r_{\perp}\,\,\propto\,\frac{1}{M}\,\,<\,\,\frac{1}{M_0}$ ), while for
$M,\,
M'\,<\,M_0$ the distance between quark and antiquark is too long
($r_{\perp}\,\,\propto\,\frac{1}{M}\,\,>\,\,\frac{1}{M_0} $), and we
cannot
treat this $q \bar q $ - pair in pQCD. Actually, we cannot even describe
 the
produced hadron state as a $q \bar q $ - pair;

\item\,\,\,For $M,\,M'\,<\,M_0$ we use the Additive Quark Model
\cite{AQM} in which
\begin{eqnarray}
\sigma(M^2,M'^2,s) &=&\sigma^{soft}_N (M^2,s)
\,\delta(M^2\,-\,M'^2) \nonumber\\
 &=& \left(\,\sigma_{tot}(qN
)\,+\,\sigma_{tot}( \bar q  N )\,\right)\,\delta(M^2\,-\,M'^2)
\label{SOFT}
\end{eqnarray}
\item\,\,\,For $M,\, M' \,>\,M_0$ we consider the system with mass $M$
and/or $M'$ as a short distance quark - antiquark pair, and describe
its interaction with the target in pQCD.
The exact formulae for $\sigma^{hard}(M^2,M'^2,s)$ is given  in  Ref.
\cite{GLMPHP}), but the key property of these formulae that
this
interaction can be expressed through the gluon structure function, and it
is
{\em not} diagonal with respect to the masses, contrary to the ``soft"
interaction of a hadron system with small mass.
\end{enumerate}

{\bf Descriptions of the experimental data:} Starting from the paper of
Badelek and Kwiecinski \cite{BAKW} Gribov's ideas have been implemented to
describe the rich and precise experimental data on $\gamma^* p$
interaction in the wide region of $Q^2$ and $x$ \cite{GLMPHP} \cite{MRS}
\cite{GLMNPHP}.It has been  shown that such an approach  is
able to provide a successful description of the experimental data on
photon-proton interaction, over a  wide range of photon virtualities $0\le
Q^2\le 100\,GeV^2$, and energies $ \sqrt{s}\, (= W) \leq  300\,GeV$.
The
key assumption on which this  aproach is based, is that the
non-perturbative
and the perturbative QCD conributions in the Gribov formula can be
separated
by the parameter $M_0$.  The  successful reproduction of the experimental
data \cite{GLMNPHP}
(see Figs. 13 ) shows that this assumption
appears
to be valid. It  lends futher  credence to using the additive quark model
(AQM) to describe the non-perturbative contribution.

\begin{figure}[h]
\begin{tabular}{l l}
\epsfig{file=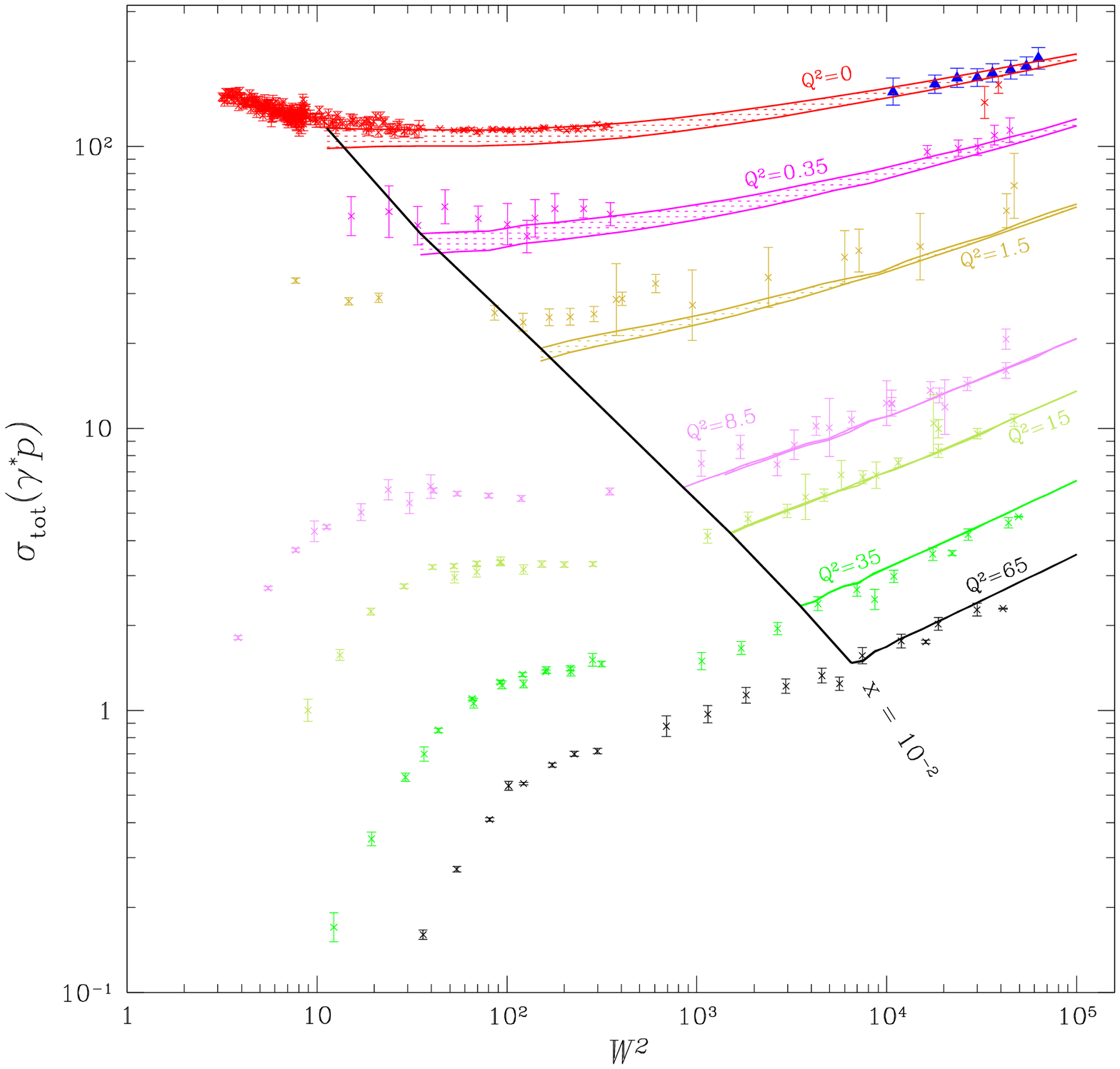,width=8.5cm,height= 7cm} &
\epsfig{file=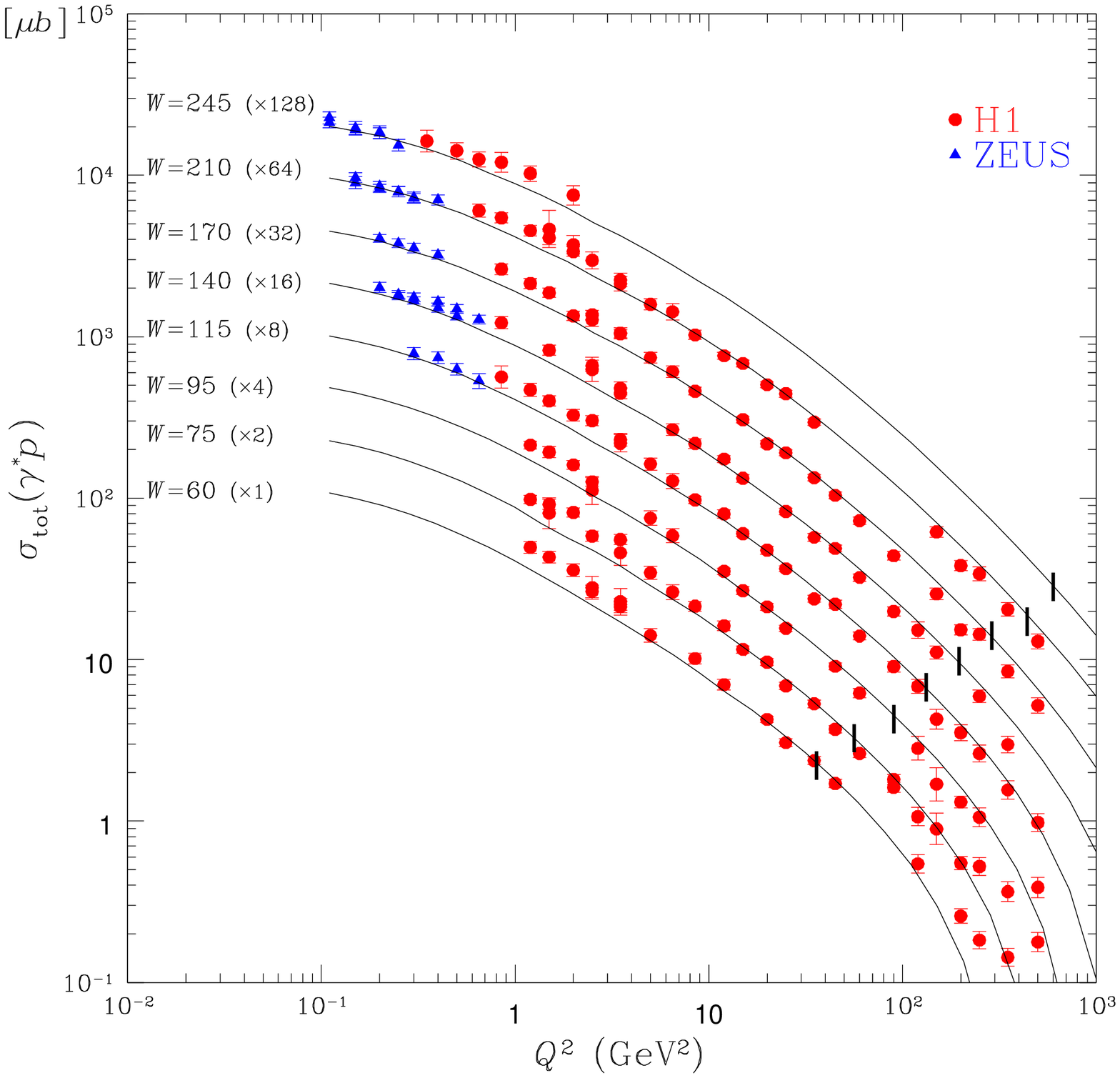,width=8.5cm,height=7cm}\\
Figure 13-a &
Figure 13-b\\
\end{tabular}
 \vspace*{2mm}
\caption[fig13]{{\it  The description of experimental data in  Gribov
approach, given in Ref. \protect\cite{GLMNPHP}.}  }
\end{figure}

A success in phenomenological description gives us hope, that the approach
based on Gribov's ideas can be used for matching of ``soft" and ``hard"
processes for more general observables than the total cross section and,
finally, will lead to selfconsistent Monte Carlo code for all values of
the photon virtualities.

{\bf Golec - Bierat and  W\"{u}sthoff approach:} The natural
generalization
of discussed attempts to describe the matching between long and short
distances physics is to use the SC approach. The full description based on
formulae of Eq.{~\ref{DD4} ) - type has not been developed yet. However,
Golec - Bierat and  W\"{u}sthoff \cite{GBW1} \cite{GBW2} suggested a
simple phenomenological approach which incorporates the main qualitative
and even quantitative feathures of the general SC calculations.

They use for the total cross section of photon - proton interaction the
following formula \cite{GBW1}
\begin{equation} \label{GW1}
\sigma(\gamma^* p )\,\,\,=\,\,\int \,\int d^2 r_{\perp}\, d z \,
| \Psi(Q^2; z, r_{\perp}) |^2\,\,\sigma(x, r_{\perp})\,\,;
\end{equation}
where
\begin{equation} \label{GW2}
\sigma(x, r_{\perp})\,\,\,=\,\,\,\sigma_0\,\,\{\,\,1\,\,\,-\,\,\,e^{-
\,\frac{
r^2_{\perp}}{R^2(x)}}\,\,\}\,\,.
\end{equation}
$R^2(x)$ is a new scale for the DIS,namely, $R^2(x)\,\,=\,\,1/Q^2_0(x)$.
Golec-Bierat and W\"{u}sthoff use the phenomenological parameterization
for $Q^2_0(x) \,\,=\,\,Q^2_0\,\,(x/x_0)^{\lambda}$ instead of calculation
that lead to Fig.6.  With $\sigma_0
\,\,=\,\,23.03\,\,mb$,$\lambda\,\,=\,\,0.288$ and
$x_0\,\,=\,\,3.04\,10^{-4}$.  Golec-Bierat and W\"{u}sthoff successfully
described both the total cross sections of the photon - proton interaction
\cite{GBW1}
as well as the cross sections of the diffractive production in DIS
\cite{GBW2}. Fig.14     show what kind of description can be reached in
 such an approach, as well as the value of the typical momentum
$Q_0(x)$ and
effectivew Pomeron intercept in their model.

\begin{figure}[h]
\begin{tabular}{l l}
\epsfig{file= 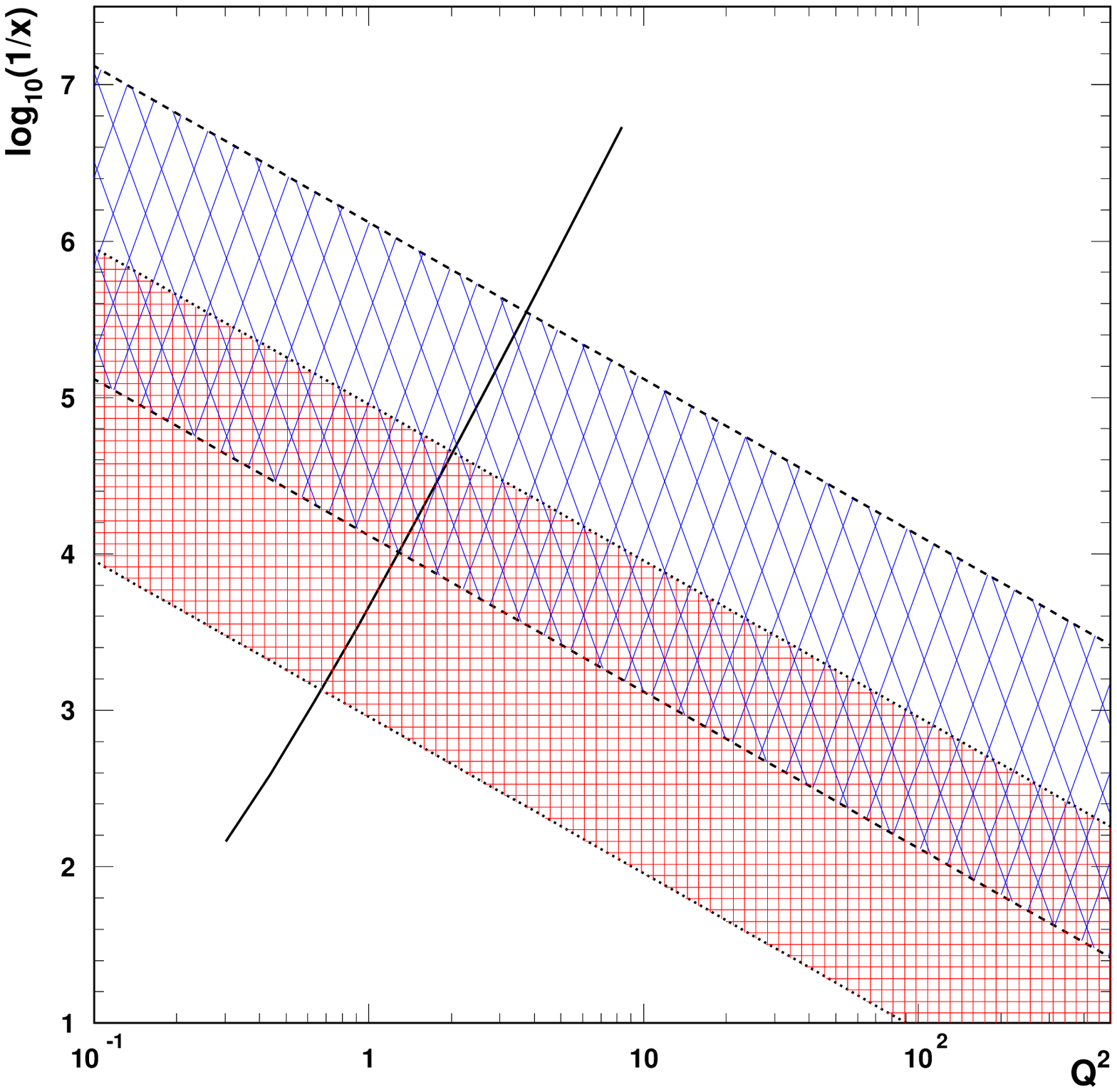,width=8.5cm,height= 7cm} &
\epsfig{file=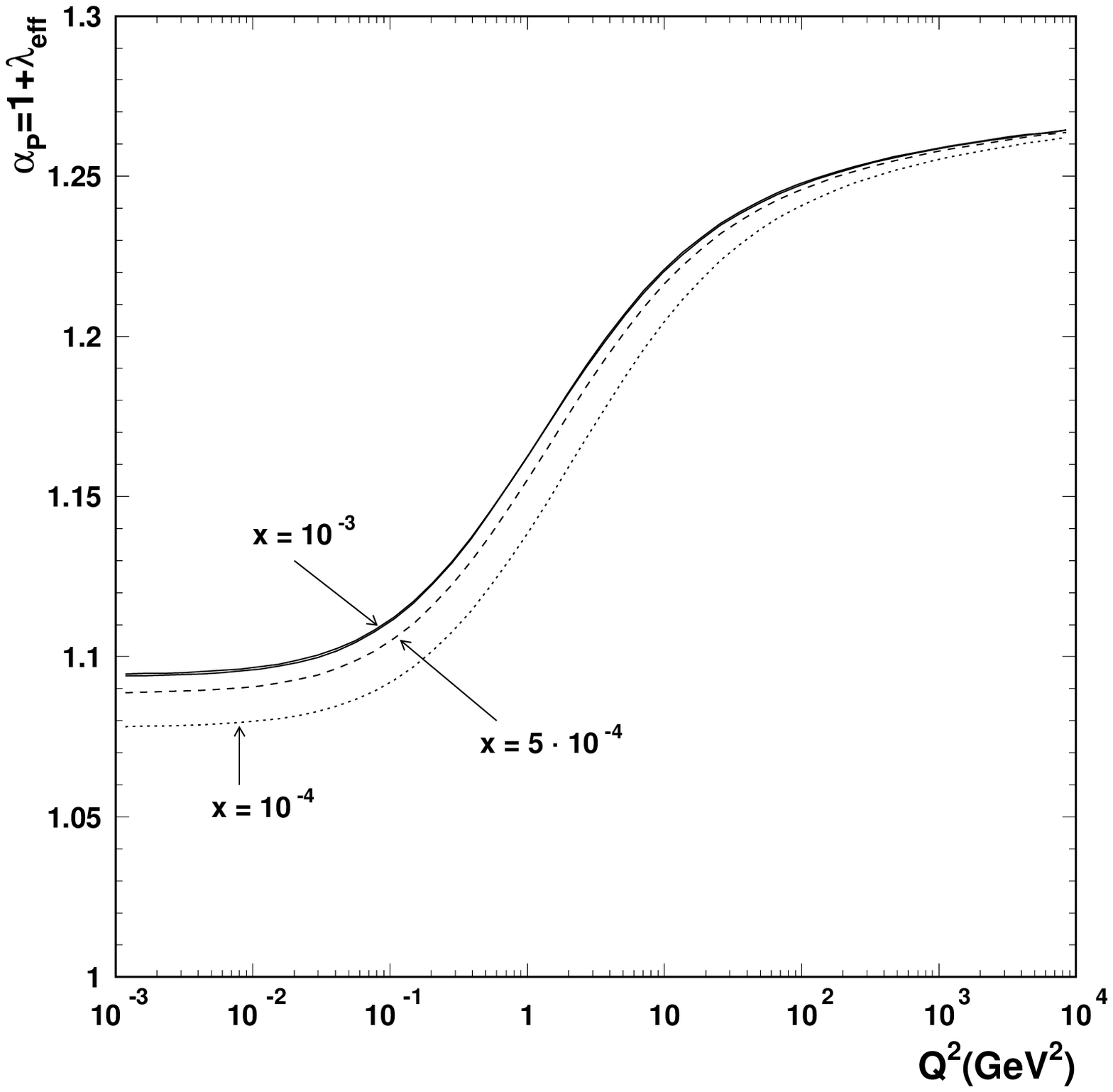,width=8.5cm,height=7cm}\\
Figure 14-a &
Figure 14-b\\
\end{tabular}
\begin{center}
\begin{tabular}{ c}
\epsfig{file= 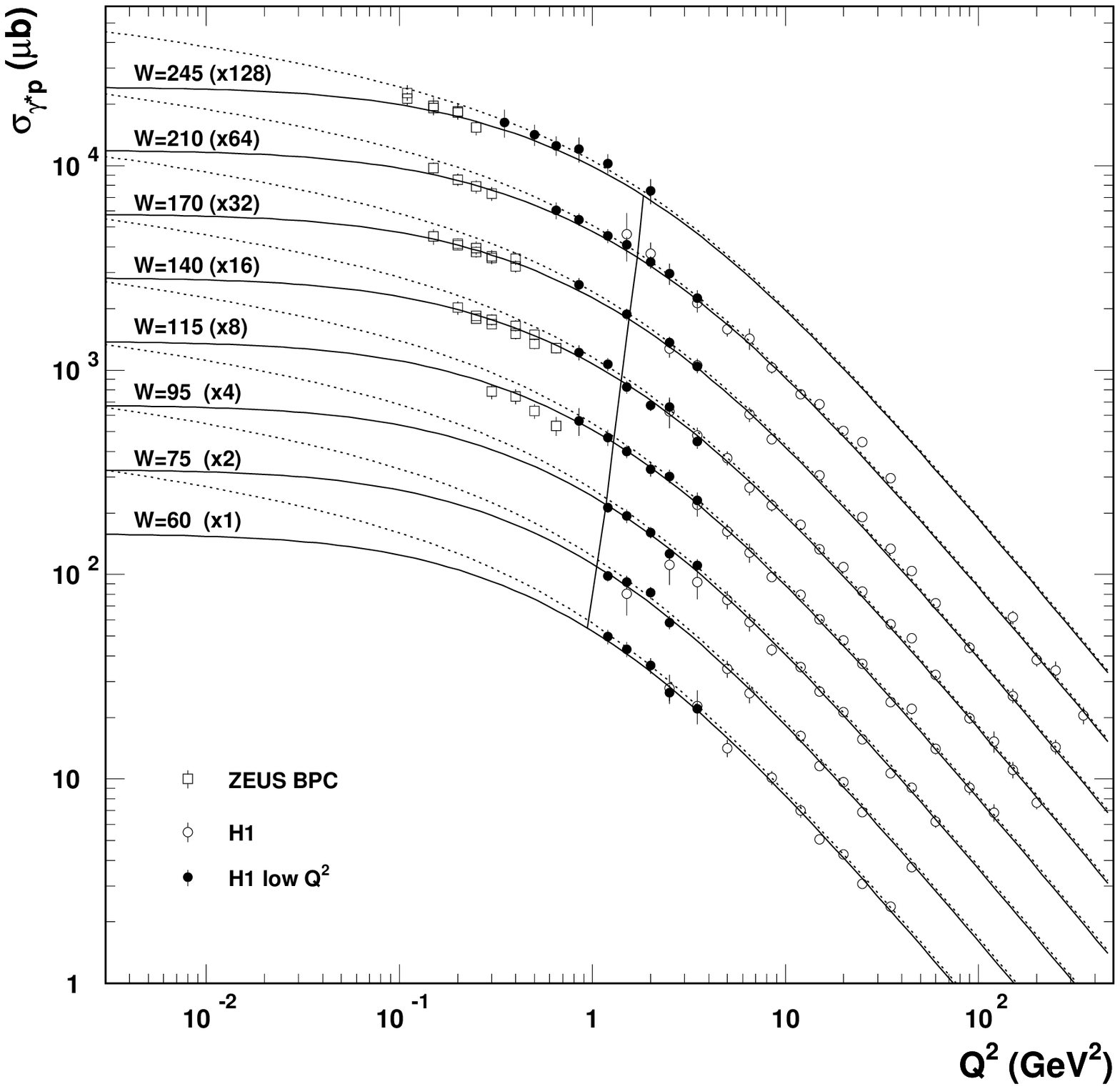,width=10cm}\\
Figure 14 -c\\
\end{tabular}
\end{center}
 \vspace*{2mm}
\caption[fig14]{{\it  Golec-Bierat and W\"{u}sthoff approach:
( a ) the critical line $Q^2 = Q^2_0(x)$; ( b ) the effective Pomeron
intercept and (c) the description of the experimental data} }
\end{figure}

This approach demonstrates that  SC are a possible way of 
describing   the matching of the ``soft" and ``hard" kinematic regions.
On the other hand it gives a practical way to write the matching of pQCD
and np QCD contribution into Monte Carlon codes.

\section{Survival probability of Large Rapidity Gaps ( LRG)}

~

{\bf Definitions and general discussion:} A  LRG
process\cite{LRGKHOZE} \cite{LRGBJ} is a multiparticle production
process in which no      particles are produced in a large window in
rapidity. The typical process of this kind is the production of two
jets with high transverse momenta ($p_{1t}$ and $p_{2 t}$ ) and
large rapidity difference ( $y_1 - y_2\,\,\gg\,\,1 $ ) between them
in which no hadrons are produced (
 see Fig.15 ).

\begin{figure}
\begin{center}
\epsfig{file= 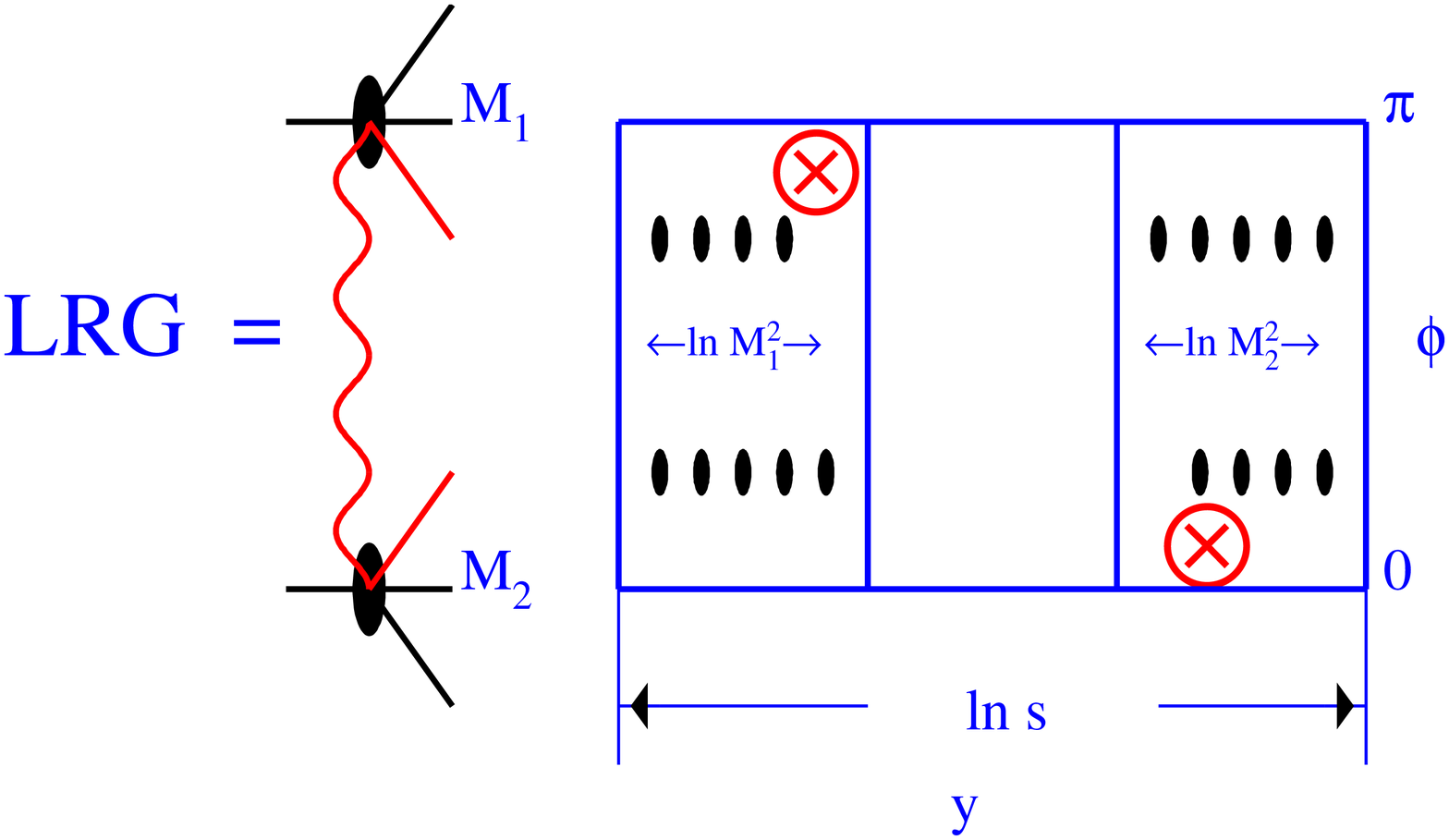,width=15cm,height=7cm}
\end{center}
\vspace*{2mm}
\caption[fig15]{{\it The lego - plot of the typical LRG process of  ``hard" dijets
production with large rapidity gap between them  where no hadrons
are produced}}
\end{figure}

\begin{eqnarray} \label{LRG1}
&
p\,\,+\,\,p\,\,\,\longrightarrow\,\,& \\
&
\,hadrons\,[\,Y -
y_1\,]\,\,+\,\,jet\,[y_1,p_{1t}\,\,+\,\,LRG\,[\,y_1 - y_2\,]
\,\,+\,\,jet\,[\,y_2, p_{2t}\,]\,\,+\,\,hadron\,[\,y_2 - 0\,
]\,\,.&\nonumber
\end{eqnarray}

Bjorken advocated that LRG processes is  a  unique way to measure
high energy asymptotic  at short distances. In our slang,  we call
this asymptotic the exchange of ``hard"  Pomeron, expressing our
hope that we can calculate it in pQCD. The
following observable was suggested \cite{LRGKHOZE} \cite{LRGBJ} to be
measured experimentally ( see Fig.16 ):
\begin{equation}  \label{LRG2}
f_{gap}\;\,\, = \;\,\,  \frac{\sigma ( \;\; dijet\;\; production
\;\; with
\;\; LRG\,\,)}{\sigma_{inclusive} (\;\;  dijet \;\;
production\,\,)}\,\,.
\end{equation}
\begin{figure}[h]
    \begin{center}
\epsfig{file=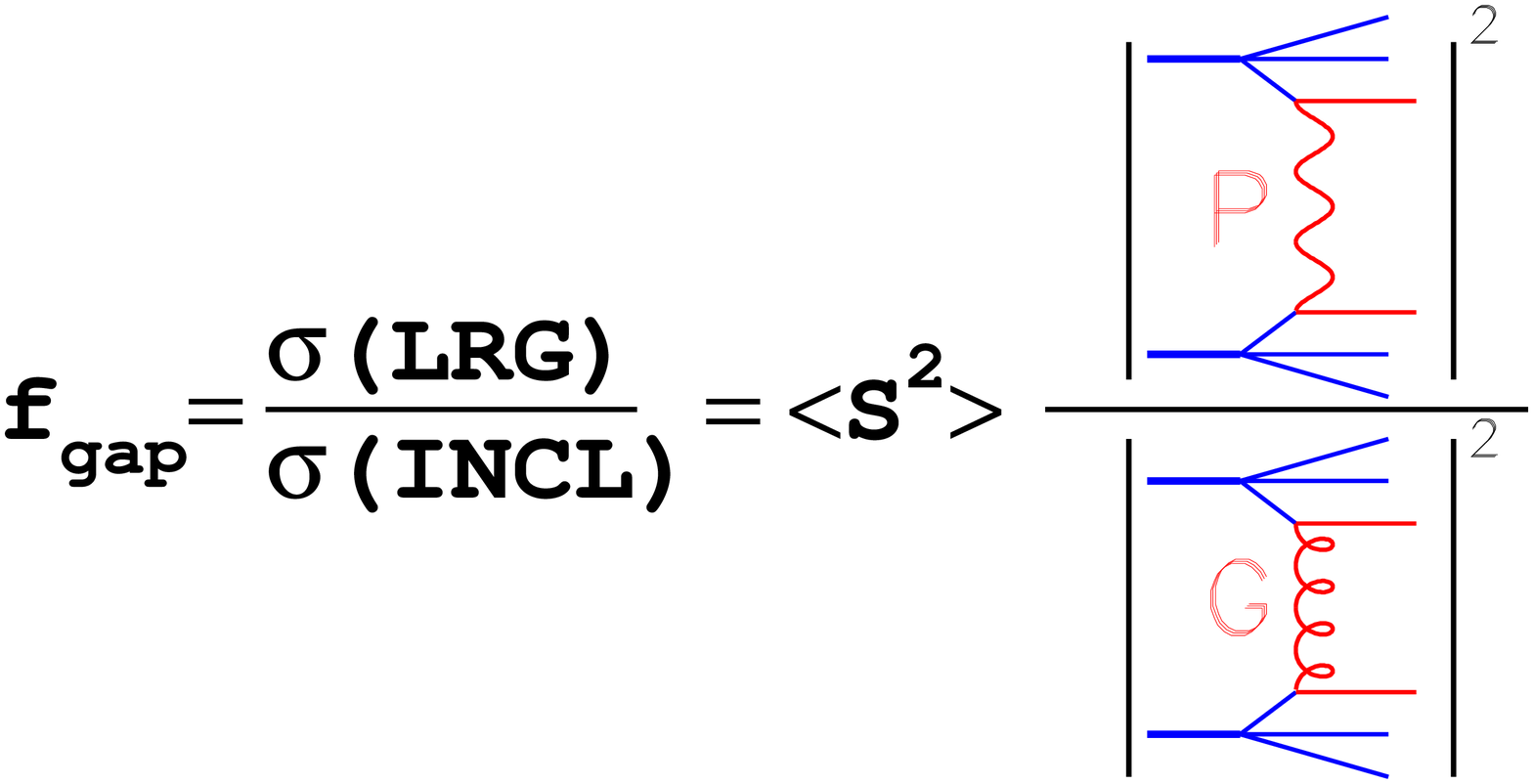, width = 15cm}
    \end{center}
 \vspace*{2mm}
 \caption[fig16]{{\it General expression for  $f_{gap}$.}}
  \end{figure}

Unfortunately, the experimental obsevable $f_{gap}$
 does not measure the contribution of the ``hard" Pomeron but has a factor\cite{LRGBJ}
 $< | S^2 |  >$
  which  we call the survival  probability of LRG and which  is
 the subject of this section. We will try to answer the following questions:
 \begin{enumerate}
\item \,\,\, What is the survival probability of LRG processes\,\,?

\item \,\,\, What is the value of survival probability\,\,?

\item \,\,\, How well can  we estimate  this value \,\,?

\item \,\,\, What is the energy dependence of survival
probability\,\,?

\end{enumerate}

{\bf Data:} We start to discuss the problem of the survival
probability    citing  all experimental data on this
subject given by D0\cite{LRGD0} amd CDF\cite{LRGCDF}.

\begin{center}
\begin{tabular}{ l l l}
   &   D0\cite{LRGD0} & CDF\cite{LRGCDF} \\
 & & \\
$f_{gap}(\,
\sqrt{s}\,\,=\,\,1800\,GeV\,)$ & 0.54\, $\pm$\, 0.17 $\% $ & 1.13
\,$\pm$\,
0.16 $\% $
\\
 & & \\
$f_{gap}(\,
\sqrt{s}\,\,=\,\,640\,GeV\,)$ & 1.85\, $\pm$ \,0.38 $\% $ & 2.7\, $\pm$\,
0.9
$\%$\\
& & \\
$
R_{gap}\,\,=\,\,\frac{f_{gap}(\,\sqrt{s}\,\,=\,\,630\,GeV\,)}
{f_{gap}(\,\sqrt{s}\,\,=\,\,1800\,Gev\,)}$ & 3.4\, $\pm$ \,1.2\,&
2.4\,$\pm$ 0.9\\
&  & \\
$R_S\,\,=\,\,\frac{\langle \mid S \mid^2 \rangle
(\,\sqrt{s}\,\,=\,\,630\,GeV\,)}{\langle \mid S \mid^2 \rangle
(\,\sqrt{s}\,\,=\,\,1800\,GeV\,)} $ &  2.2\,$\pm$\,0.8 &  ???\\

\end{tabular}

\end{center}
We can learn two important properties of $ \langle \mid S \mid^2
\rangle $:
\begin{enumerate}
\item\,\,\,The value of $\langle \mid S \mid^2 \rangle$ is small
 ( $\langle \mid S \mid^2 \rangle
\,\,\lsim\,\,10\,\%$ )  because all estimates \cite{LRGHP} 
show that the ratio of ``hard"  Pomeron
 to  inclusive
  dijet production is about   10 - 20\%\,;

\item \,\,\, $\langle \mid S \mid^2 \rangle$  decreases rapidly  with
energy.
 At least we have to
blame the survival probability for experimentally  observed fast
decreasing of $f_{gap}$,  since the ratio of ``hard" Pomeron to
inclusive dijet production can  only increase in  pQCD.
\end{enumerate}

It should be stressed that only the  first data  on
$f_{gap}$ in DIS \cite{LRGHERA}  have appeared, this  does not allow us to
make any
 conclusions on the energy behaviour of  $\langle \mid S
\mid^2 \rangle$  but tell us that the value of
 the survival probability,
is much  higher ( $\langle \mid S \mid^2 \rangle \,\,\approx\,\,50
- 70 \%$ )   than in hadron - hadron collisions.

{\bf Two source   of survival   probability:} It turns out that the
survival probability $\langle \mid S \mid^2 \rangle$ can be written
as a product of two factor. Each of them has a  different physical
meaning.

\begin{equation} \label{LRG3}
\langle \mid S \mid^2 \rangle\,\,\,=\,\,\,
\langle
\mid S_{bremsstrahlung}(\Delta y = |y_1 - y_2 | )
\mid^2 \rangle\,\,\times\,\,
\langle \mid S_{spectators}(s) \mid^2
\rangle
\end{equation}

\begin{figure}
\begin{center}
\begin{tabular}{l l}
\epsfig{file=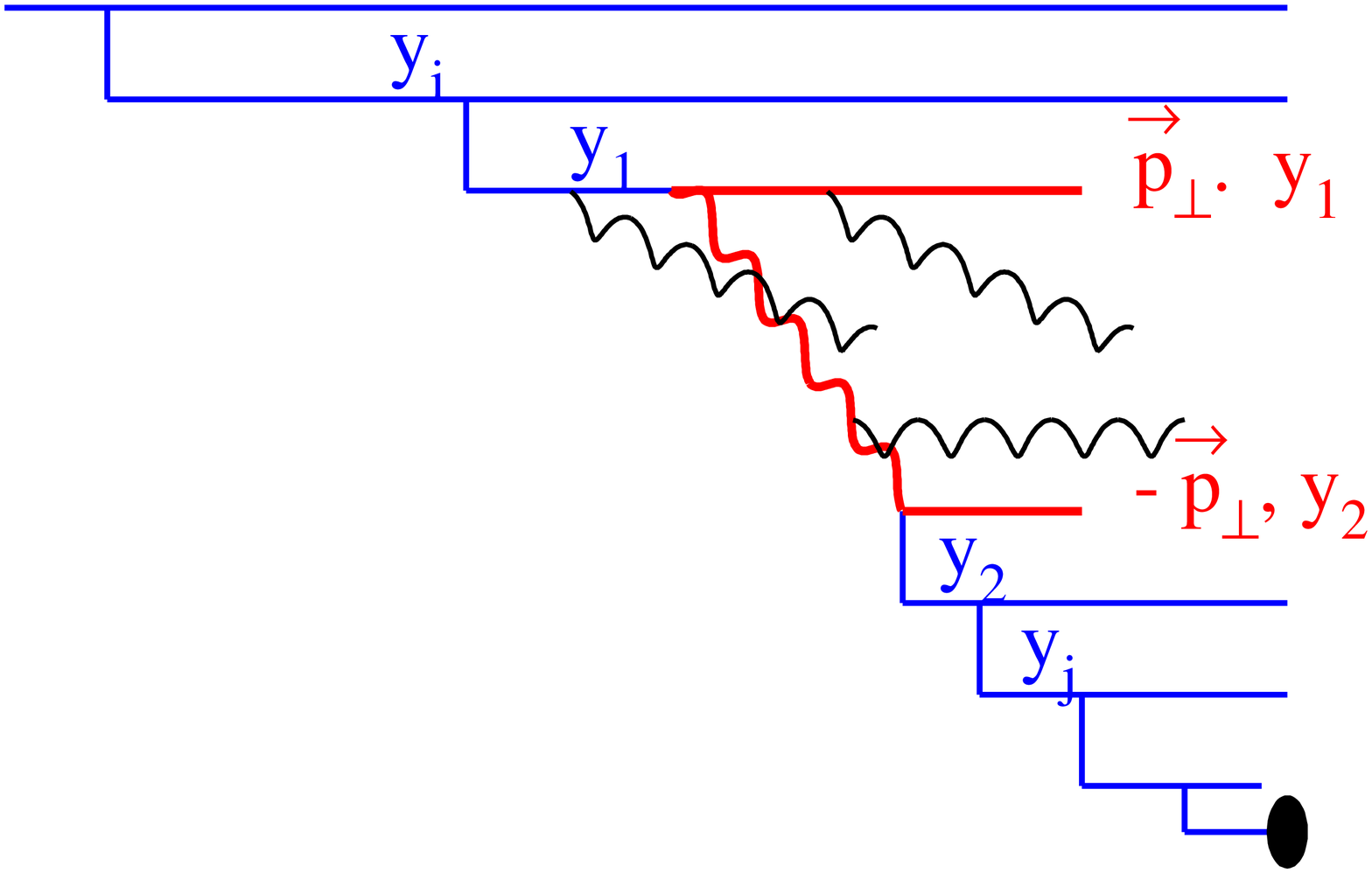,width=7.5cm,height=5cm} &
\epsfig{file=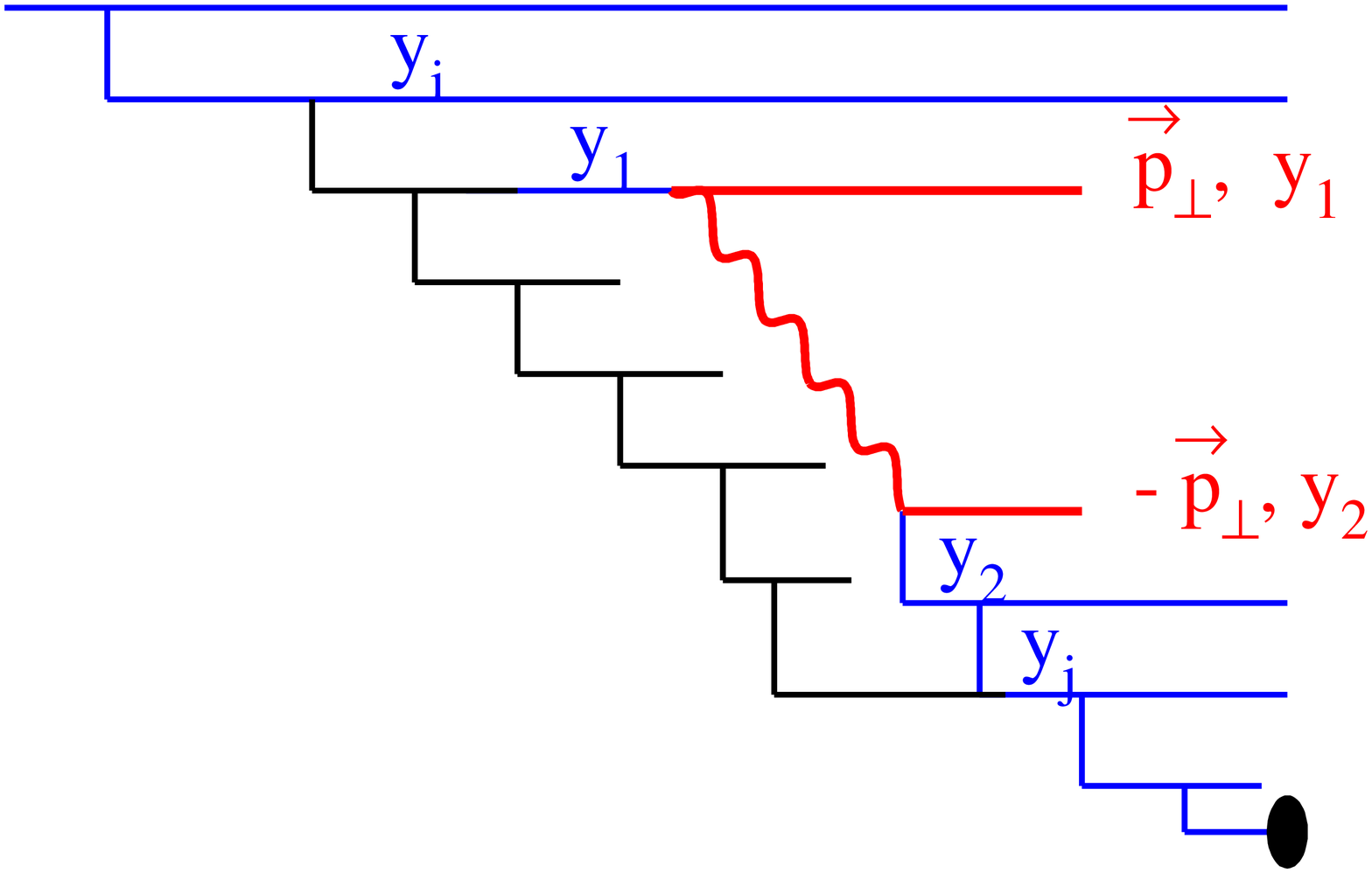,width=7.5cm,height=5cm}\\
Figure 17-a & Figure 17-b\\
\end{tabular}
\end{center}
\vspace*{2mm}
\caption[fig17]{{\it Examples of processes that lead to
$\langle \mid S_{bremsstrahlung} \mid^2 \rangle $. ( Fig.17-a ) and
to   $\langle \mid S_{spectator}| \mid^2 \rangle $ ( Fig. 17-b ).
}}
\end{figure}

The first factor in Eq.(~\ref{LRG3} ) ( $ \langle \mid
S_{bremsstrahlung} \mid^2 \rangle$ ) describes the probability that
the LRG will not be filled by emission of bremsstrahlung gluons
from partons, taking part in the ``hard" interaction (see Fig. 17-a
). It depends mostly on the value of the rapidity gap \cite{LRGBREM}. The
second factor(
$ \langle \mid S_{spectator} \mid^2\rangle
$  ) appears to take into account the probability that no parton
with $x_i\,\,>\,\,x_1$ will have an inelastic interaction with any
parton with $x\,\,<\,\,x_2$ (see Fig. 17-b ). From Fig. 17-b one
can see that this factor depends on the total energy of the
process ( $s$ ) \cite{LRGBJ}
\cite{LRGSOFT}\cite{GLMSP1}\cite{GLMSP2}. Different
physics
behind these two factors lead to a different theoretical status for
their calculations. For calculations of the value of $\langle \mid
S_{bremsstrahlung} \mid^2 \rangle $ the pQCD thechnique could be
and has been developed \cite{LRGBREM}. Only a part of $\langle
\mid S_{spectator} \mid^2 \rangle$ can be controled by pQCD, while the estimates
for  most of  this factor could  only be made in npQCD. In
 pratice, it means that we must rely  on high energy  phenomenology in
our
attempts to provide an  estimate for $\langle \mid
S_{spectator}\mid^2  \rangle $.

Before discussing what we have learned about  $\langle \mid
S_{spectator} \mid^2 \rangle $ we would like to remind the  reader
that all corrections for interactions of two groups of parton  (
see Fig. 17-b )   cancel due to Abramovski-Gribov-Kancheli cutting
rules \cite{AGK}.

{\bf Survival Probability in the Eikonal  Model:}
In order  to  understand the main  features of the survival
probability, let us consider it in the simplest phenomenological
model for the ``soft" interaction - the Eikonal model.

{\em  $\bullet$\,\,\, Main assumptions: }

\parbox{8cm}{
\begin{itemize}
\item\,\, Only the fastest parton interacts with ``wee" partons
\item\,\, Hadrons are correct degrees of freedom at high energy
\end{itemize}}
\parbox{8cm}{
\begin{itemize}
\item\,\, $\frac{\sigma_{SD} + \sigma_{DD}}{\sigma_{el}}\,\,\ll\,\,1$\\

 ~ ~ 
\item\,\,\, Oversimplified  final state
\end{itemize}}

\begin{figure}
\begin{center}
\epsfig{file=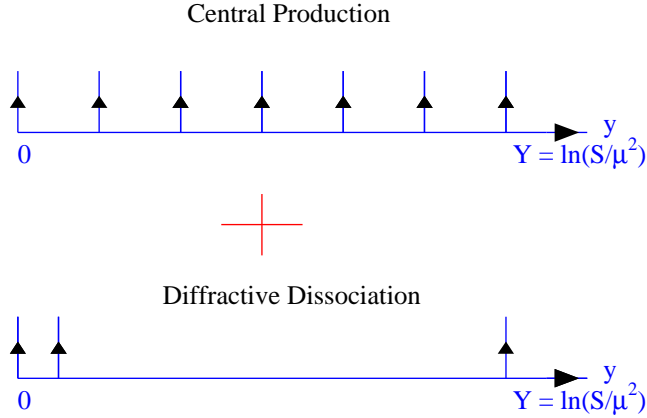,width=10cm}
\vspace*{2mm}
\caption[fig18]{{\it The rapidity structure of the typical event in the
Eikonal model.}}
\end{center}
\end{figure}

\begin{itemize}
\item  \,\,{\em Simple form of unitarity\,:}

\begin{equation} \label{LRG4}
2\, Im a_{el}(s,b)\,\, = \,\,\mid a_{el}(s,b) \mid^{2}\, \,+ \,\,
G_{in}(s,b)\,\,.
\end{equation}

\item\,\,{\em Solution:}
\begin{eqnarray}
& a_{el}\,\,= \,\,i\,[\,1 \,- \,e^{-\frac{ \Omega(s,b)}{2}}\,]\,;&
\label{LRG5}\\
& G_{in}(s,b)\,\, =\,\, 1- e^{- \Omega(s,b)}\,;& \label{LRG6}
\end{eqnarray}
where $\Omega(s,b)$ is an   orbitrary real function.

\item\,\,{\em  Simple parameterization \,:}
\begin{equation} \label{LRG7}
\Omega(s,b)\,\,\,=\,\,\frac{\sigma_0}{\pi
R^{2}(s)}\,\,(\,\frac{s}{s_0}\,)^{\Delta_P}\,\,e^{-\frac{b^{2}}{R^{2}(s)}}\,\,
=\,\,\nu(s)\,\,e^{-\frac{b^{2}}{R^{2}(s)}}
\end{equation}

~

\item\,\,{\em Analytical formulae for ``soft" observable\,:}

\begin{eqnarray}
&
\sigma_{tot} = 2 \int d^{2}b (1 - e^{- \Omega(s,b)/2})
= 2 \pi R^{2}(s)\,[\,ln(\frac{\nu(s)}{2}) + C - Ei(-
\frac{\nu(s)}{2})\,]\,;
 &\label{LRG8}\\
& \sigma_{in} =  \int d^{2}b (1 - e^{- \Omega(s,b)})
=  \pi R^{2}(s)\,[\,ln(\nu(s)) + C - Ei(- \nu(s))\,]\,; & \label{LRG9}\\
&
 \sigma_{el} = \sigma_{tot} - \sigma_{in}
=  \pi R^{2}(s)\,[\,ln(\frac{\nu(s)}{4}) + C + Ei(- \nu(s)) - 2 Ei(-
\frac{\nu(s)}{2})\, ]\,.&\label{LRG10}
\end{eqnarray}

\item\,\,{\em  A simple formula for survival probability\,:}
\begin{equation} \label{LRG11}
 <\mid S_{spectator} \mid^2> = \frac{ \int d^{2}b \Gamma_{H}(b) P(s,b)} { \int
d^{2}b \Gamma_{H}(b)}\,\,,
\end{equation}
 with
$\Gamma_{H}(b) = \frac{1}{\pi
R^{2}_{H}}e^{-\frac{b^{2}}{R^{2}_{H}}}
$

\item\,\,{\em  Analitycal form of $<\mid S_{spectator} \mid^2>$\,:}
\begin{equation} \label{LRGEM}
< \mid S_{spectator} \mid^2> =  \frac{a(s) \gamma[a(s),
\nu(s)]}{[\nu(s)]^{a(s)}}
\end{equation}
with $a(s) = \frac{ R^{2}_{S}(s)}{ R^{2}_{H}} $.

\end{itemize}

{\bf Survival Probability in the Eikonal  Model ( comparison with
experiment ):}

To get more reliable estimates for the value and energy dependance of
$< \mid S_{spectator} \mid^2>$ we tried to extract all parameters directly
from the experimental data \cite{GLMSP1}. The first observation ( see 
Eqs.(~\ref{LRG8} ) - (~\ref{LRG10} ) ) is that the ratio 
\begin{equation} \label{LRG12}
 R_{el} (\nu )\,\,=\,\,   
\frac{\sigma_{el}}{\sigma_{tot}}
\end{equation}
is a function of the only parameter $\nu$. Therefore, we can use the
experimental data on $R_{el}$ to fix $\nu$ at particular value of energy
 ( see Fig.19 ).

To fix the value of $a(s)$ we used the data on J/$\Psi$
production\cite{HERAPSI} which shows two differents slopes in
$t$-dependence for elastic and inelastic processes ( see Fig. 20-a ).
These data can be used to extract the value of $R^2_H = 8 \,GeV^{-2}$ ( 
see
Ref. \cite{GLMSP1} ).

 \begin{figure}
\begin{center}
\begin{tabular}{l l}
\epsfig{file=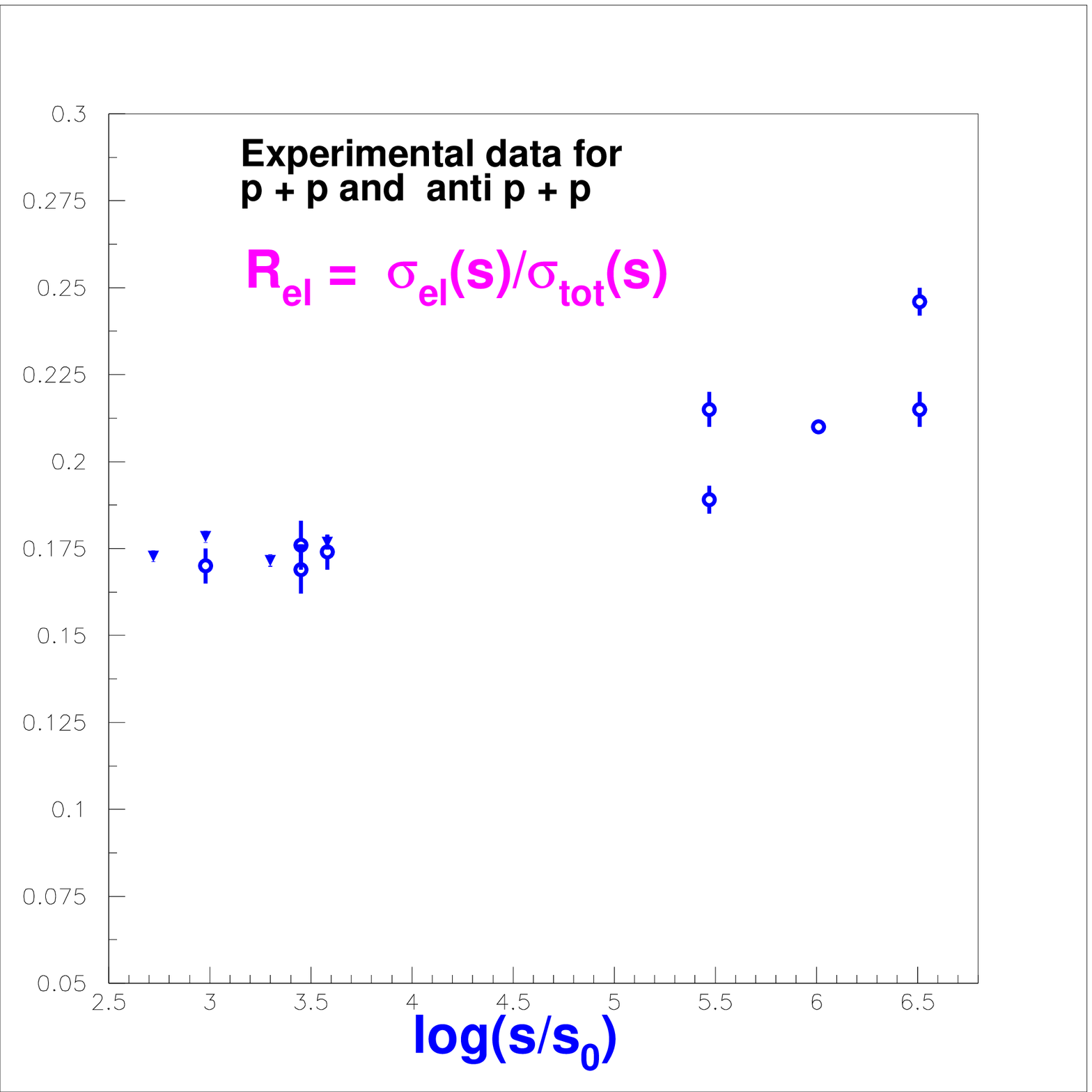,width=7.5cm,height=5cm} &
\epsfig{file=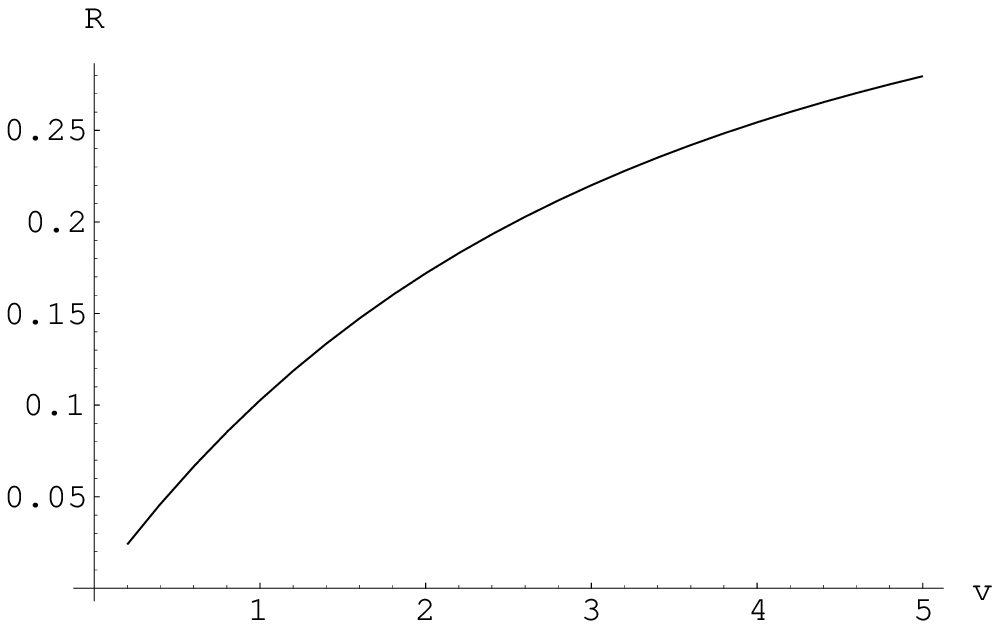,width=7.5cm,height=5cm}\\
Figure 19-a & Figure 19-b\\
\end{tabular}
\end{center}
\vspace*{2mm}
\caption[fig19]{{\it Experimental data ( Fig.19-a ) and the Eikonal model
predictions ( Fig.19-b ) for $R_{el} =
\frac{\sigma_{el}}{\sigma_{tot}}$ . }}
\end{figure}

Using ``soft" phenomenology for $R^2(s)$ \cite{GLMSP1} we obtained
prediction for the survival probability given in Fig.20.
One can see that the Eikonal model reproduces both the experimental value
and the energy dependence of the survival probability.

 \begin{figure}
\begin{center}
\begin{tabular}{l l}
\epsfig{file=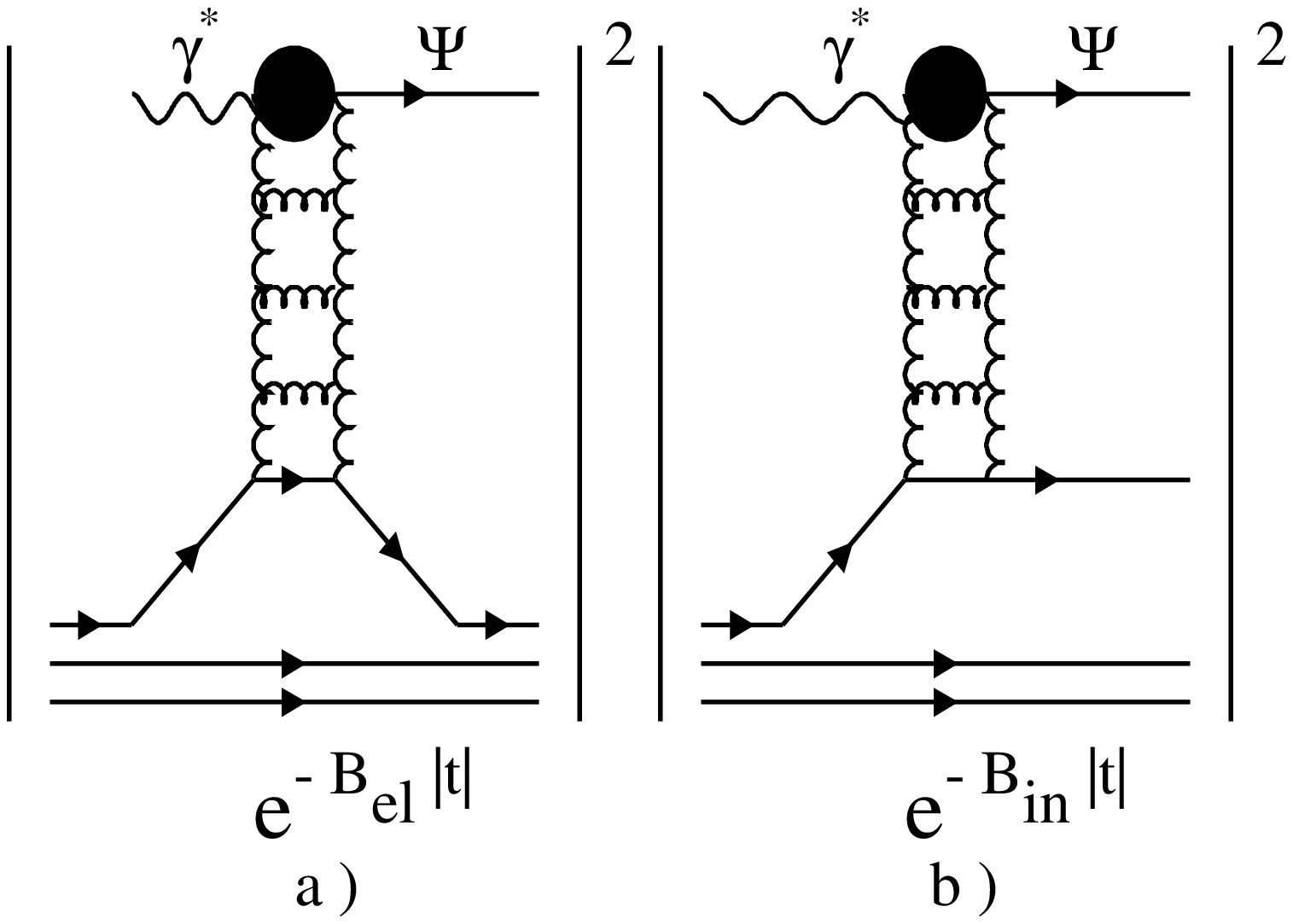,width=8cm} &
\epsfig{file=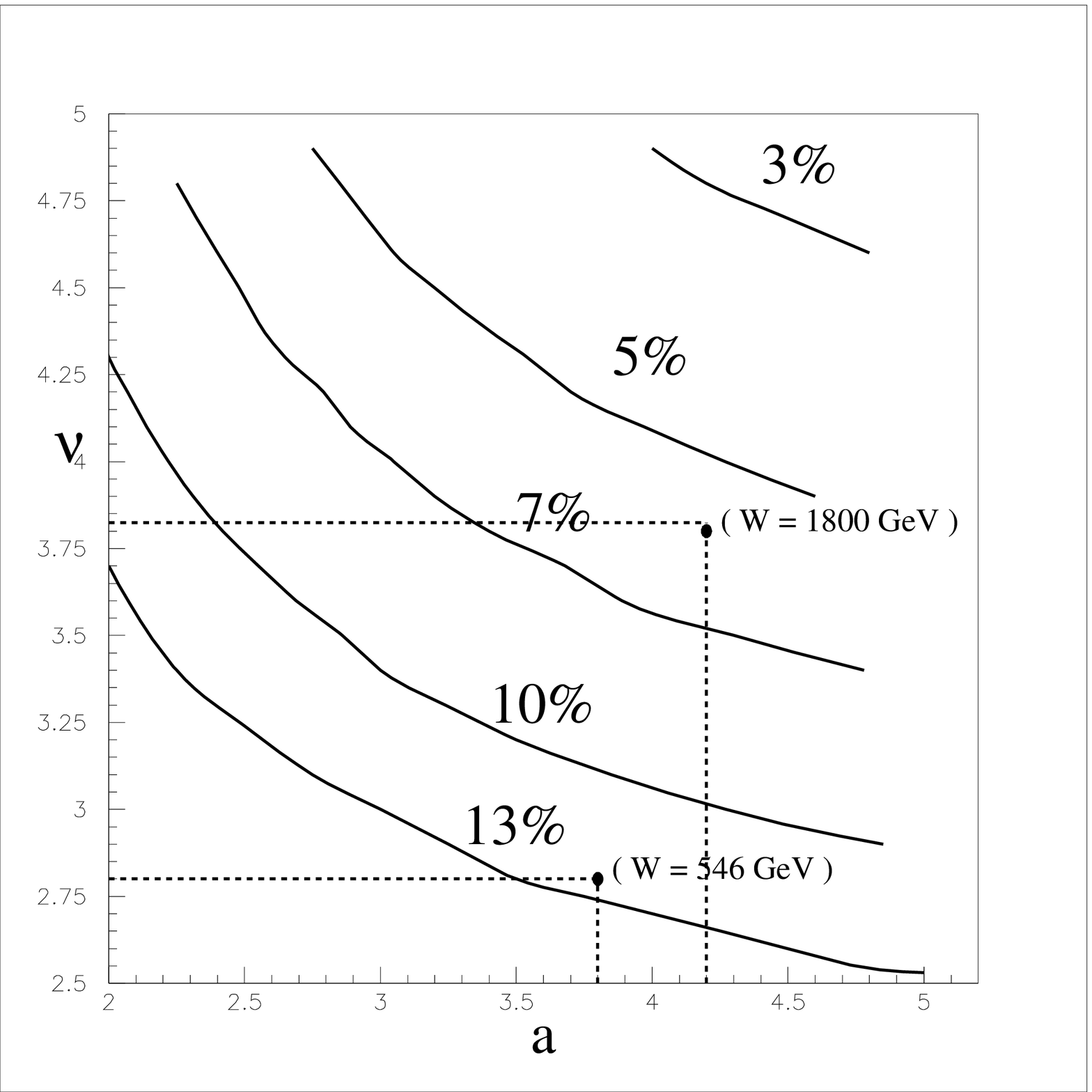,width=8cm}\\
Figure 20-a & Figure 20-b\\
\end{tabular}
\end{center}
\vspace*{2mm}
\caption[fig20]{{\it J/$Psi$ production (Fig.20-a ) and the survival
probability $< \mid S_{spectator}
\mid^2>$  in the
Eikonal model ( Fig.20-b ) }}
\end{figure}

{\bf More general approach to calculation of $\mathbf{\langle \mid
S_{spectator}
\mid^2 \rangle}$.}

 In Ref. \cite{GLMSP2} a more general approach has been developed to
calculation of $< \mid S_{spectator} \mid^2>$ which based on
the following assumptions:
 
\begin{itemize}
\item\,\,\,Only the fastest parton interacts with the target \,\,;

\item\,\,\, Hadrons are not correct degrees of freedom\,\,;
       
\item\,\,\,$\frac{\sigma_{DD}(\,M\,>\,M_0
\,)}{\sigma_{el}}\,\,\ll\,\,1$\,\,;
\item\,\,\,Oversimplified final state: central production ( uniform in
rapidity) and elastic + diffractive production with small masses.
\end{itemize}

One can see that all the  above assumptions are very close to the Eikonal
model,  and the  difference actually is only in the way of  how  
the diffractive production processes are treated.
 In the model of Ref.
\cite{GLMSP2} we assume that the correct degrees of freedom at high energy
are not hadrons but some different states which are described by wave
functions $\Psi_1 $ and $\Psi_2$. Our interaction matrix is diagonal
\begin{equation} \label{3CH1}
< \Psi_n \times \Psi_m | {\mathbf{ T }} | \Psi_{n'} \times \Psi_{m'}
>\,\,\,=\,\,\,A_{n,m} (s,b_t) \,\delta (n
- n')\,\delta( m - m')\,\,,
\end{equation}
and only for $ A_{n,m}$  ( not for hadrons ) unitarity has a form:

\begin{equation} \label{UNIT}
2 \,\,Im\,\,A_{n,m}(s,b_t)\,\,\,=\,\,\,| A^2_{n,m} (s,b_t)|^2
\,\,\,+\,\,G^{in}_{n,m}(s,b_t)\,\,.
\end{equation}

Produced hadrom and a diffractive state have the following wave functions:
\begin{eqnarray}
&
\Psi_{hadron}\,\,\,=\,\,\,\alpha\,\,\Psi_1\,\,\,+\,\,\,\beta\, 
\,\Psi_2\,\,; & \label{3CH2}\\
&
\Psi_D\,\,\,=
\,\,\,-\,\beta\,\,\Psi_1\,\,\,+\,\,\,\alpha\,\,\Psi_2 \,\,.& \label{3CH3}
\end{eqnarray}

Therefore, 
 three channels were taken into account in this model,
namely, elastic scattering,
single and double diffractice production while only elastic rescatterings
were included in the Eikonal model.   

Using the Eikonal - like parameterization ( see Fig.21 ), the three
channel model leads to the prediction for the value and energy dependence
of $ \langle \mid S_{spectator} \mid^2 \rangle $ given in Fig.22.
One can see that this model can reproduce the experimental data including
the energy dependence.

 \begin{figure}[h]
\begin{center}  
\epsfig{file=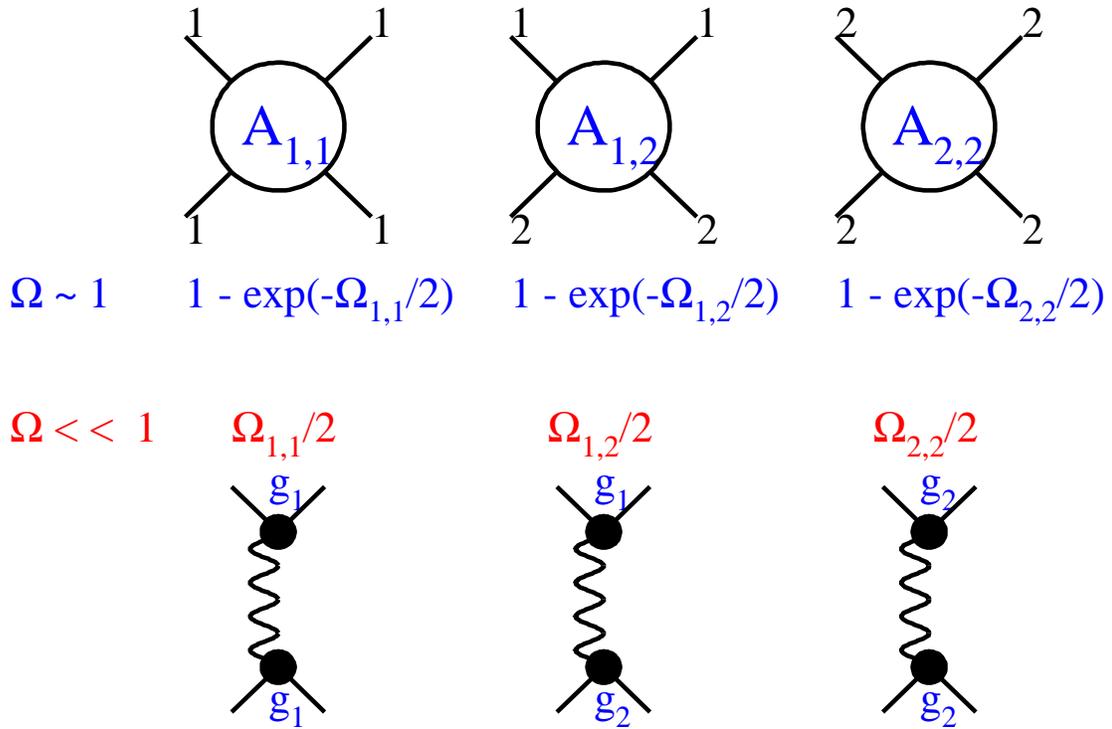,width=17cm}
\end{center} 
\vspace*{2mm}
\caption[fig21]{{\it The Eikonal - like parameterization in the three
channel model. }}
\end{figure}

 \begin{figure}
\begin{center}  
\begin{tabular}{l l}
\epsfig{file=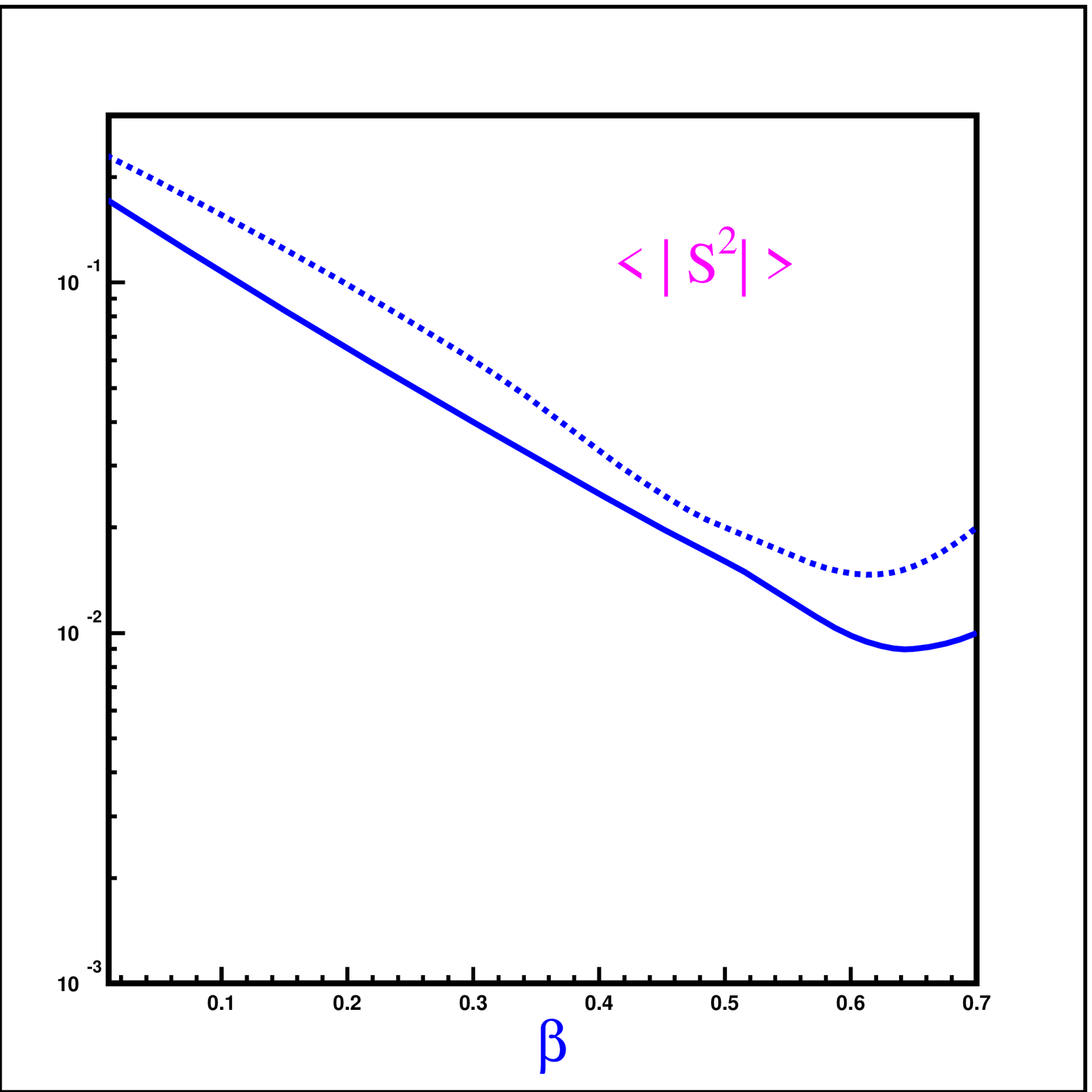,width=8cm} &
\epsfig{file=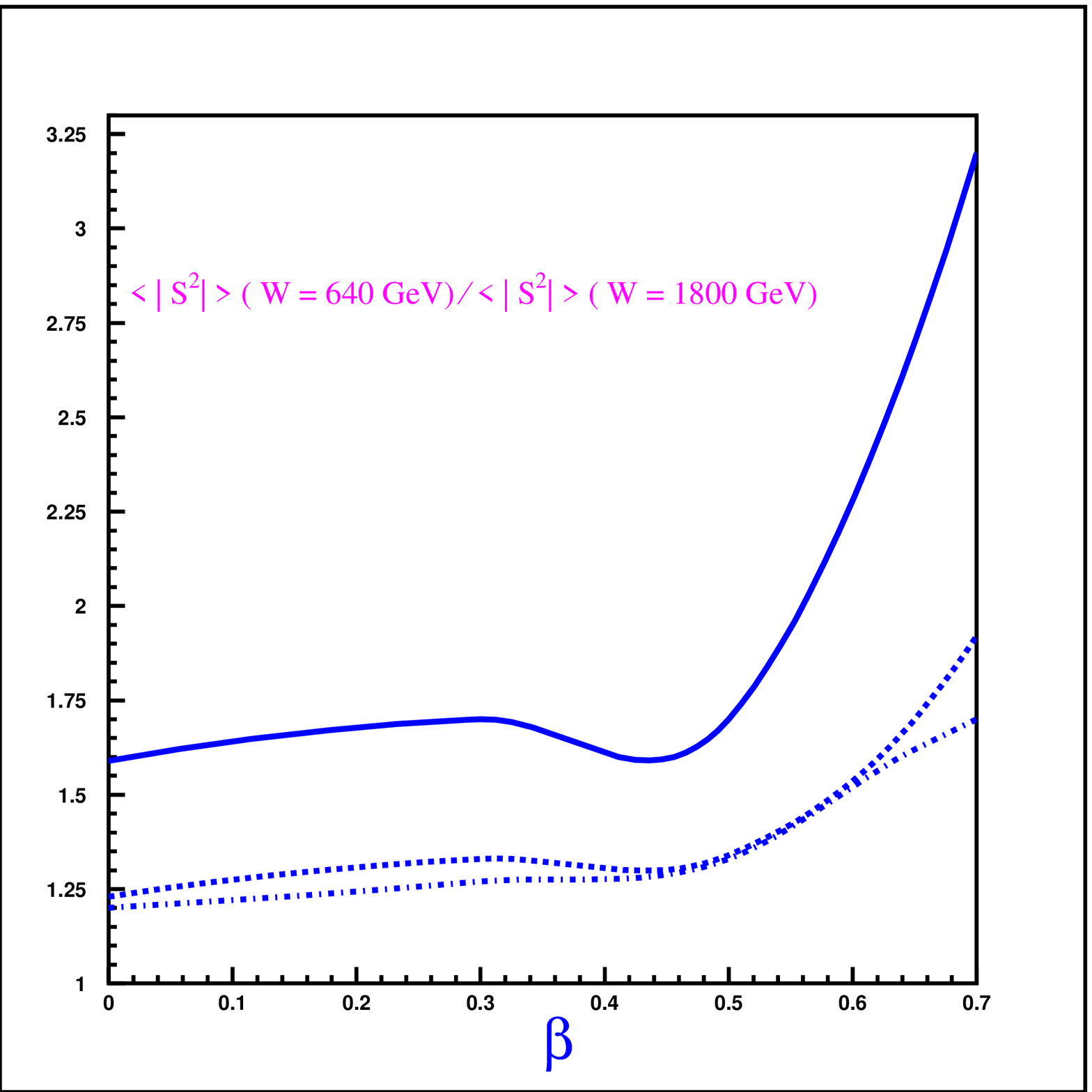,width=8cm}\\
Figure 22-a & Figure 22-b\\
\end{tabular}
\end{center} 
\vspace*{2mm}
\caption[fig22]{{\it The value ( Fig.22-a ) of survival probability
$< \mid S_{spectator}\mid^2>$   at $\sqrt{s} =  W 
= 1800\,GeV$ and the
ratio   $ < \mid S (\, W\, = \, 640\,GeV\, )\mid^2>/< \mid S (\, W\, = \,
1800\,GeV\, )\mid^2 $ ( Fig.22-b )
versus $\beta$ in the three channel model. }}
\end{figure}

Summarizing the experience dealing with three channel model,
 we can conclude that:

\begin{itemize}

\item\,\,\, Accuracy of the three channel model, in principle, is much
better
than in the Eikonal Model, but still we do not know what to do with
diffraction in the region of large mass ( $ M\,\,\approx\,\,s$ )\,\,;

\item\,\,\,The scale of SC is not given by
$
R_D\,\,\,=\,\,\,\frac{\sigma_{el}\,\,+\,\,\sigma_{SD}\,\,\,+\,\,\,\sigma_{DD}}
{\sigma_{tot}}\,\,,
$  but rather by the separate ratios

\begin{center}
\begin{tabular}{l l l }
$R_{el}\,\,=\,\,\frac{\sigma_{el}}{\sigma_{tot}}$ &
$R_{SD}\,\,=\,\,\frac{\sigma_{SD}}{\sigma_{tot}}$ &
$R_{DD}\,\,=\,\,\frac{\sigma_{DD}}{\sigma_{tot}}$\\
\end{tabular}

\end{center}
\item\,\,\, The small value of the survival probability as well as
its
strong energy dependence appear naturally in this approach\,\,;

\item\,\,\, The parameters that have been used are  in agreement with the
more detailed fit of the experimental data\,\,;

\item\,\,\,{\em Questions to experimentalists:}

\begin{enumerate}
\item\,\, What is $R_{el}$ at $\sqrt{s}$ equal to 630\,GeV ?
1800\,GeV ?  Tevatron runII energy ?
   
\item\,\,What is the value for $R_{SD}$  at $\sqrt{s}$ equal to
630\,GeV ?
1800\,GeV ? Tevatron runII energy ?

\item\,\, What is the value for $R_{DD}$ at $\sqrt{s}$ equal to
630\,GeV ? 
1800GeV ?  Tevatron runII energy ?

 \begin{figure}[h]
\begin{center}
\epsfig{file=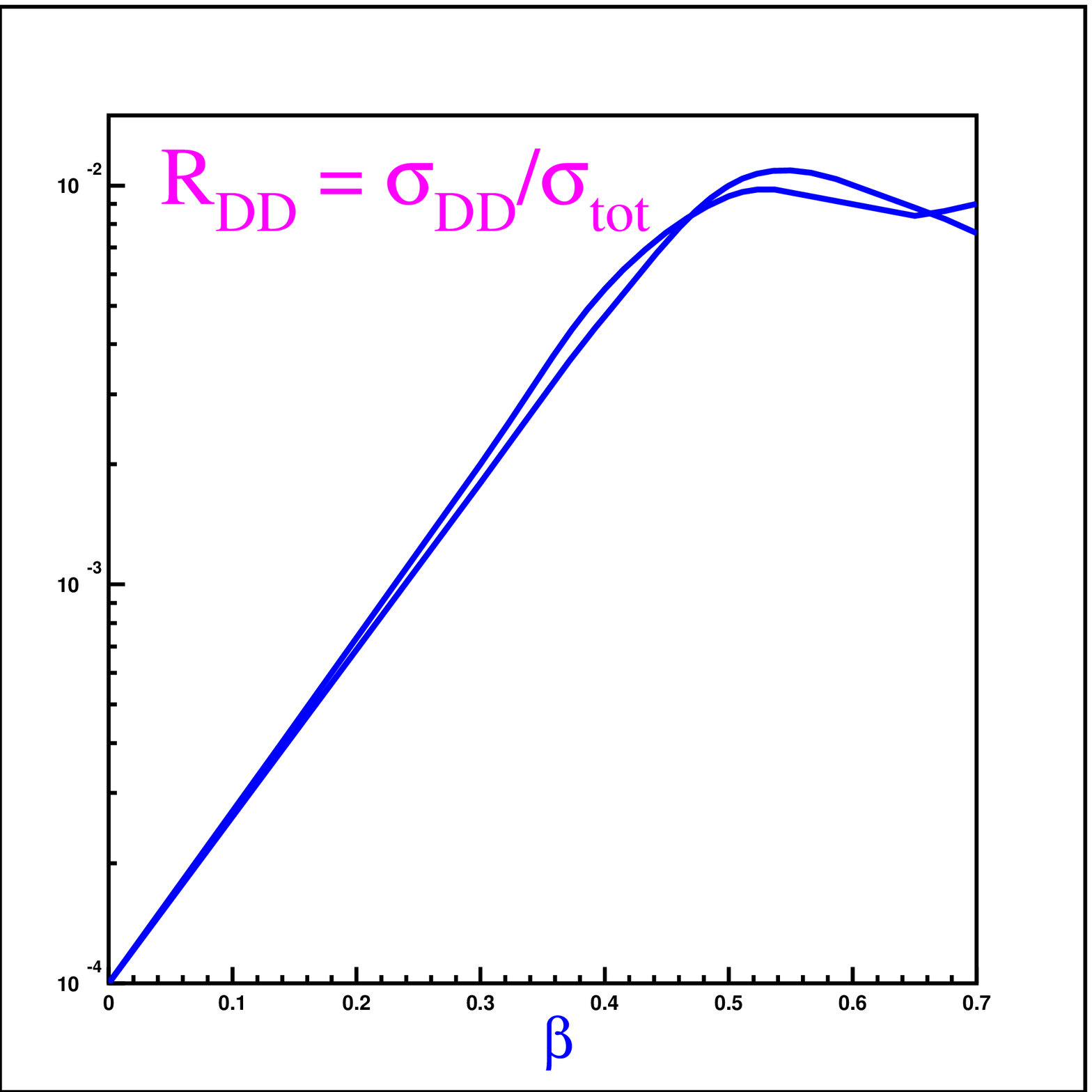,width=10cm}   
\end{center}
\vspace*{2mm}
\caption[fig23]{{\it The three channel model predictions for double
diffractive dissociation ( $R_{DD} = \sigma_{DD}/\sigma_{tot}$. }}
\end{figure}

\item\,\, What is the value of the single diffraction cross section
in the region of large masses  at $\sqrt{s}$ equal to
630\,GeV ?
1800\,GeV ? Tevatron runII energy ?

\end{enumerate}

\end{itemize}

{\bf Q \,\&\, A\,:}

\begin{tabular}{ l l}   
{\bf  Q:} & Have we developed  a theory for 
$\langle \mid  S_{spectator} \mid^2 \rangle$\,? \\
{\bf  A:} & No,  there are only models on the market .\\
 & \\
{\bf  Q:} & Can we give a reliable estimates for the value of  $\langle
\mid
S_{spectator} \mid^2 \rangle$\,\,?\\
{\bf A:} & No,  we have only rough estimates based on the Eikonal - type 
models.\\
\end{tabular}

\begin{tabular}{ l l}
{\bf  Q:} & Can we give a reliable estimates for the energy
behaviour of
  $\langle \mid
S_{spectator} \mid^2 \rangle$\,\,?\\
{\bf A:} & No,  but we understand that
$\langle \mid S_{spectator} \mid^2 \rangle $
  could  decreases steeply
with
energy.\\
 & \\
{\bf  Q:}& Why are you talking about
$\langle \mid 
S_{spectator} \mid^2 \rangle $  if you can do nothing ?\\
{\bf A:}&  Because:\\
\end{tabular}
\begin{flushright}
\begin{minipage}{15cm}{
\begin{itemize}
\item\,\,  Dealing with models we have learned what questions
we should ask experimentalists to improve our calculations;
\item\,\,\, We learned what problems we need to solve
theoretically to provide reliable estimates;
 
\item\,\,\, We understood what kind of questions could be
answered in LRG experiments;

\item\,\,\, We are on the way to an understanding in what
experiments our accuracy is enough, to get   interesting information on
the high energy scattering amplitude at short distances.
\end{itemize}}
\end{minipage}
\end{flushright}
 
{\bf  Survival Probability in DIS: } I hope, that I have explained to you
that we do not have a consistent theoretical approach to the calculation
of   $\langle \mid S_{spectator} \mid^2 \rangle$ for hadron - hadron
collisions.  The natural question to ask is : can  the theoretical
situation  be better for DIS where the ``short" distances can give the
major contribution? 

To answer this question let us consider first two examples of the LRG
processes in DIS: the diffractive dissociation ( DD ) and one-side dijets
production in DD ( see Fig.24 ).

 \begin{figure}[h]
\begin{center}
\begin{tabular}{l l}
\epsfig{file=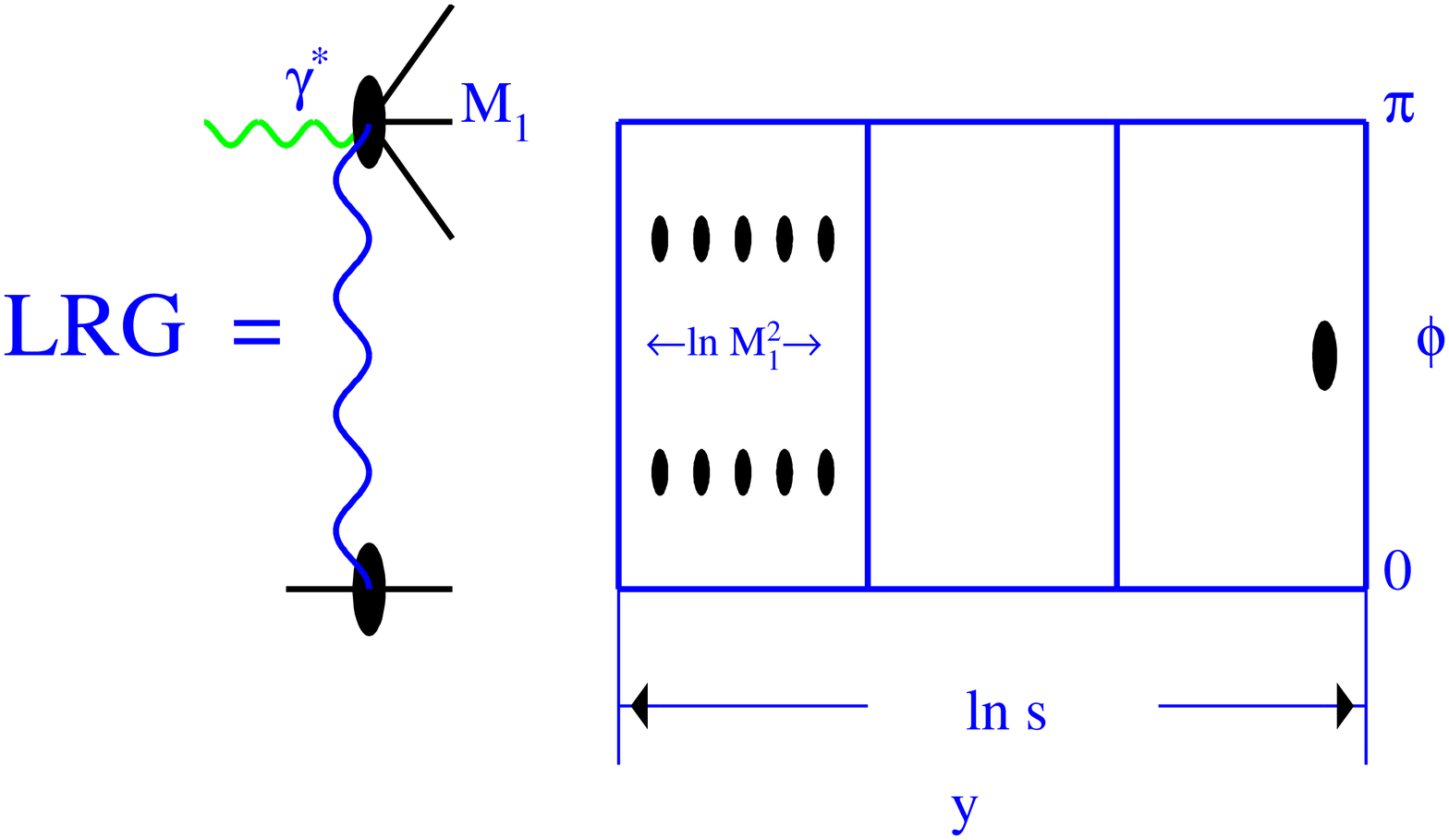,width=8cm} &
\epsfig{file=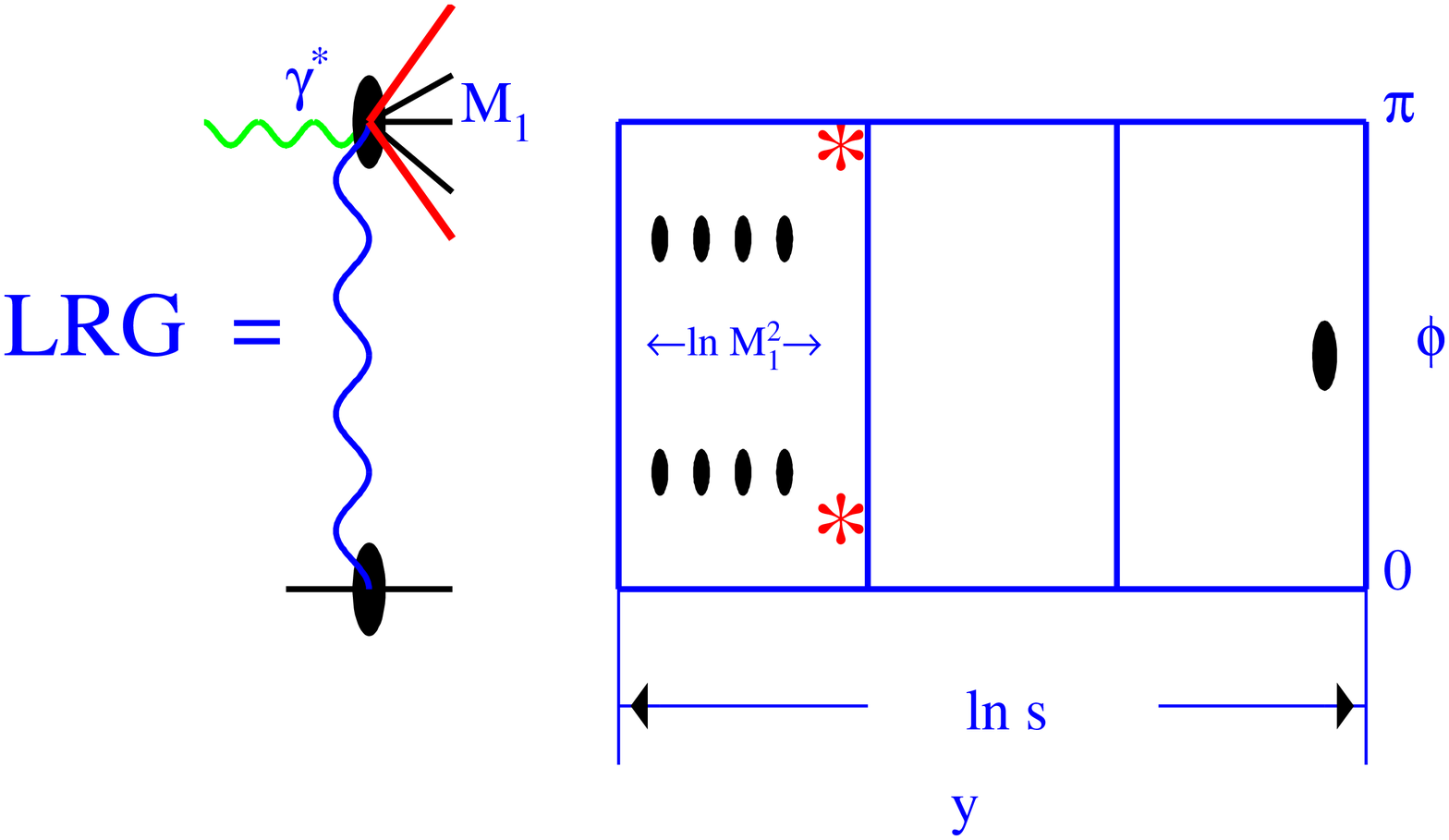,width=8cm}\\
Figure 24-a & Figure 24-b\\
\end{tabular}
\end{center}
\vspace*{2mm}
\caption[fig24]{{\it Lego - plots for DD ( Fig.24-a) and one-side dijets
production in DD ( Fig.24-b ) in DIS.}}
\end{figure}

Fig.25 shows the  parton interaction which can occur for  one-side
dijets
production in DD.

\begin{figure}[h]
\begin{center}

\epsfig{file=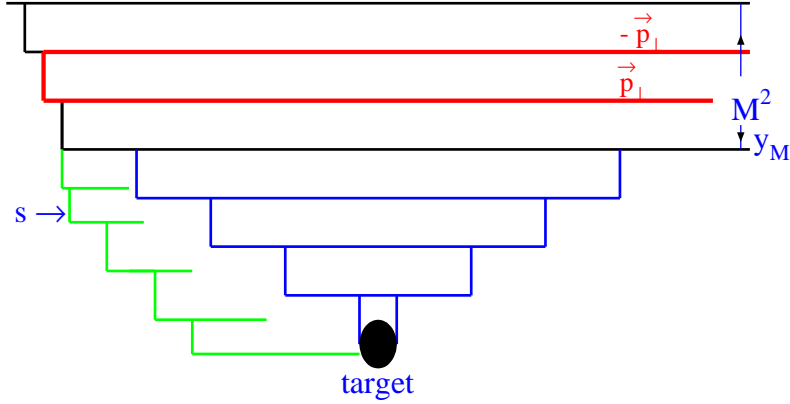,width=12cm} 
\end{center}
\vspace*{2mm}
\caption[fig24]{{\it Picture of parton interactions for  one-side
dijets  
production in DD  in DIS. }}
\end{figure}

Using the main properties of the parton cascade ( mostly the AGK - cutting
rules \cite{AGK} ) we can make three observation on the survival
probability for this process:
\begin{enumerate}
\item\,\,\, Survival probability for one - side dijets   depend
on $Q^2$ and $x_B$, but does not depend on  $p_t$ and $x_1$\,\,;

\item\,\,\, Survival probability is not big in comparison with LRG
survival probability in proton - proton collisions\,\,;

\item\,\,\,For large $Q^2$ the survival probability can be calculated in
pQCD\,\,.

\end{enumerate}

In the Glauber - Mueller approach the cross section of virtual photon -
proton interaction can be written in the form \cite{LERY87} \cite{MU90}:

\begin{equation} \label{DISLRG1}
\sigma(\gamma^* p)\,=\,\int\int\, d^2r_t\, d z\,\,
|\Psi(Q^2;z,r_t)|^2\,\sigma(x,r^2_t)\,\,;
\end{equation}
with
\begin{equation} \label{DISLRG2}
\sigma(x,r^2_t)\,\,=
\,\,2\,\,\int\,\, d^2b_t\,\,[\,\,1\,\,
\,-\,\,\,e^{-\,\,\frac{\Omega( r^2_t,x_B;b_t ) }{2}}\,\,]\,\,;
\end{equation}
where $\Psi$ is the wave function of the virtual photon and
\begin{equation} \label{DISLRG3}
 \Omega( r^2_t,x_B;b_t )\,\,\,=\,\,\, \frac{16 \pi
\alpha_S}{3}\,r^2_t\,\,x_B \,G( x_B,\frac{r^2_t}{4}
)\,\, \frac{1}{R^2}\,e^{- \,\frac{b^2_t}{R^2}}\,\,.
\end{equation}
$R^2 \,=\,2\,B^{DD}(x_B)$ where
$$
\frac{\frac{d \sigma_{DD}}{d t}( t )}{\frac{d \sigma_{DD}}{d t}( t= 0
)}\,\,=\,\,e^{- B^{DD}\,|t|}$$
and this parameter is closely related to non pertutbative physics. At the
moment we can use the experimental data on the slope of the DD processes
in DIS. It should be stressed that factoring out the $b_t$ - dependence 
can be justified in QCD using the factorization theorem \cite{FATEO}.

The formula for the survival probability is a direct generalization of 
Eq.(~\ref{LRG11} ) in the Eikonal model, namely:
\begin{equation} \label{DISLRG4}
\langle \mid S_{spectator}\mid^2 \rangle\,\,\,=\,\,\,\frac{\int\int\int 
d^2r_t\,
d z \,d^2 b_t \,\,|\Psi(Q^2;z,r_t)|^2\,e^{-\,\,\Omega( r^2_t,x_B;b_t
)}\,\,e^{
-\,\,\frac{2 \,b^2_t}{R^2}}}{\int\int\int
d^2r_t\,
d z \,d^2 b_t \,\,|\Psi(Q^2;z,r_t)|^2\,\,\,e^{-\,\,\frac{2 \,b^2_t}{R^2}}}
\end{equation}

The calculations \cite{GLMSM}, using Eqs. (~\ref{DISLRG1} ) -
(~\ref{DISLRG4},
shows that the survival probability is much less in DIS than in hadron -
hadron collisions ( see Fig.26, where $D^2_T \equiv S^2 $ is plotted for
the transverse polarized induced photon reactions with LRG ).

 \begin{figure}[h]
\begin{center}
\begin{tabular}{l l}
\epsfig{file=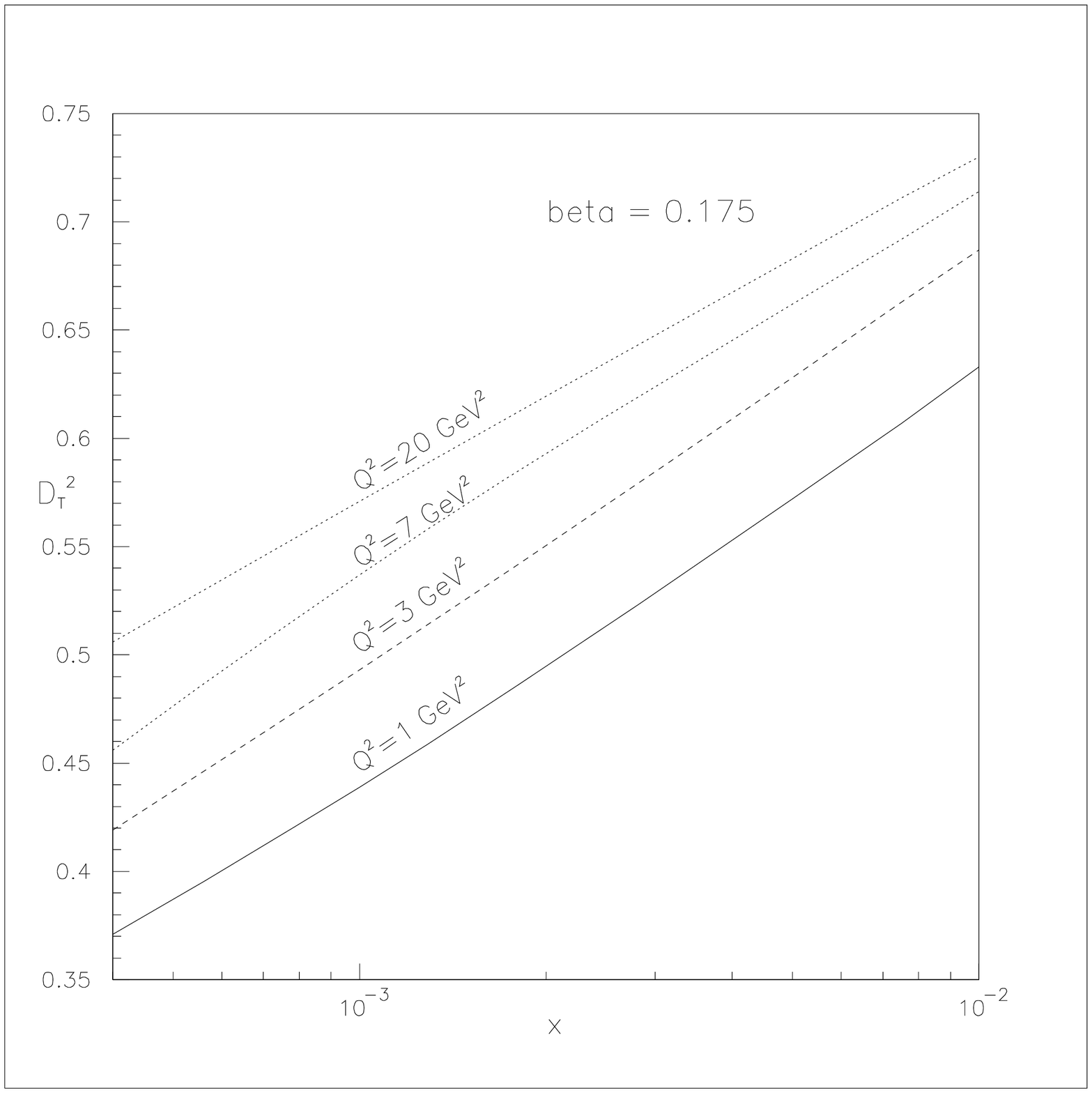,width=8cm} &
\epsfig{file=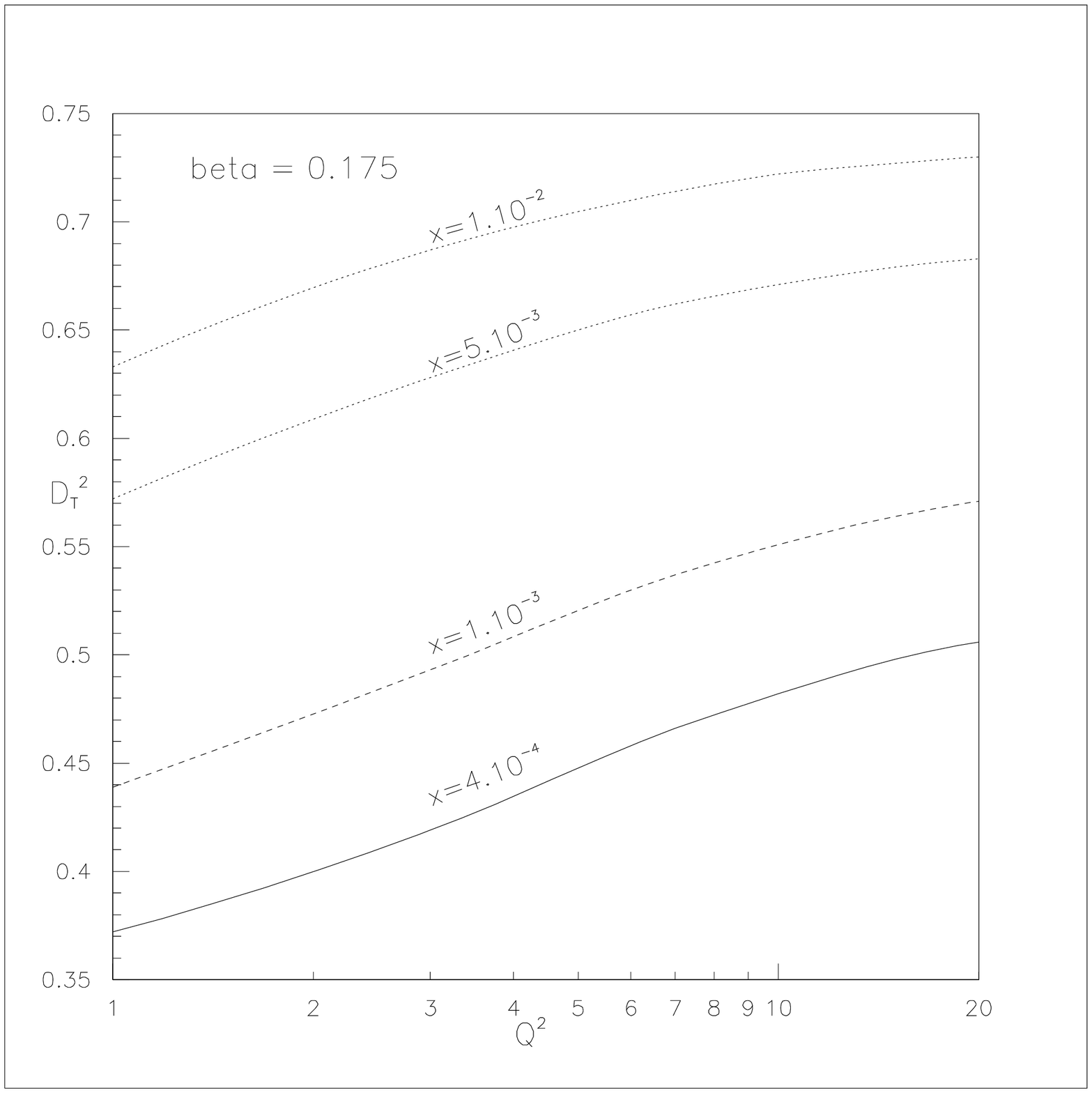,width=8cm}\\
Figure 26-a & Figure 26-b\\
\end{tabular}
\end{center}  
\vspace*{2mm} 
\caption[fig26]{{\it The survival probability $\langle \mid
S_{spectators} \mid^2 \rangle \,\equiv\,D^2_T$  in DIS versus $x$
 ( Fig.26-a ) and $Q^2$ ( Fig.26-b ). }}
\end{figure}

{\bf Multiparticle production in DIS:}
Shadowing corrections or, in other words, the rescatterings of quark -
antiquark pair in a target lead to a new predictions for the processes of
multiparticle productions. Namely, it gives probabilities for production
of several parton showers. For every parton shower we can use the Monte
Carlo code based on the evolution equations, while the probability for
configuration with $n$ - parton showers can be calculated, using 
the following simple formula:
\begin{equation} \label{MP1}
\sigma_n\,\,\,=\,\,\int\,\,d^2r_t d z d^2b_t |
\Psi(Q^2,r_t,z)|^2\,\,\frac{\Omega^n}{n!}\,e^{ - \Omega}  
\end{equation}
Fig.26 shows the first three terms in our decomposition of the
multiparticle production in the parton showers: diffraction dissociation, 
one parton shower  and two parton showers productions. For better
understanding, in Fig.26 the probability of each configuration in
the limit of small $\Omega$ is also given.

\begin{figure}[h]
\begin{center}
\epsfig{file=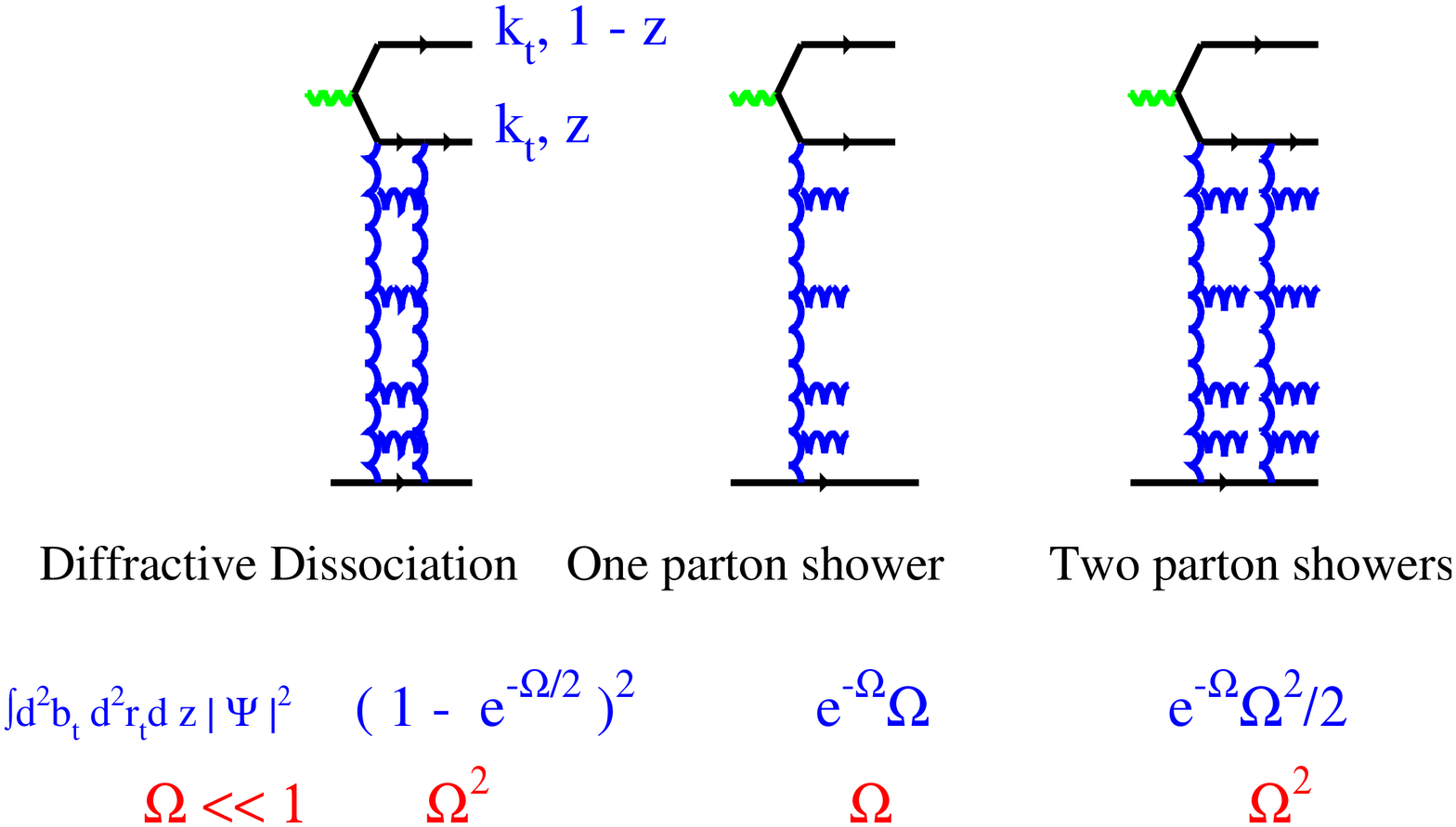,width=17cm} 
\end{center}
\vspace*{2mm}
\caption[fig27]{{\it The first three terms in parton showers
decomposition.}}
\end{figure}

I am firmly believe that such a decomposition will be very useful for
writing  Monte Carlo programs.

\section{Conclusions}
 In this talk I tried to give you a picture of recent progress in low $x$
physics. I chose the subject which, I hope, would be useful for Monte
Carlo experts. Let me summarize in short the present situation in four
subjects that I have discussed  here:
\begin{enumerate}

\item\,\,\,We are on the right track in understanding of the
next-to-leading BFKL Pomeron,  but, at the
moment, we cannot guarantee any calculations of the parameters of the BFKL
Pomeron;

\item\,\,\, First experimental indications of the strong shadowing
corrections have appeared  in energy behaviour of the diffraction
dissociation cross sections in DIS and in the $Q^2$ - behaviour of the
$F_2$ - slope ( Caldwell plot ). We hope, that these data will stimulate
the new experimental systematic search of  SC, as well as the  creation of
the new Monte Carlo codes, that will include the SC and physics, induced
by
them;

\item\,\,\, Considerable progress, based on Gribov's ideas, has been
achieved in  describing  
the matching between the non-perturbative ``soft" processes, and 
the perturbative ``hard" ones in photon - proton interactions. We firmly
believe that this progress will be useful both for future non-perturbative
approaches in QCD and for  creating  new Monte Carlo program, which
take this matching  into account;

\item\,\,\, We can estimate the survival probability in hadron - hadron
collisions only in rather primitive models, which  can reproduce
the basic experimental data on it. The prediction for the survival
probability in DIS are more solid from theoretical point of view, but 
the experimental information is so poor that we cannot compare the main
features of our calculations with the data. We think that time has come to
include in Monte Carlo codes the decomposition of  multiparticle
production process in the series of  multi parton showers production, 
which is  closely  related the the value of the survival probability.
\end{enumerate}

{\bf Acknowlegments:} I am very grateful to E. Gotsman and U. Maor for 
 permanent  useful discussions on the subject which only partly were
reproduced in  our common papers. I would like to thank E.Naftali for
preparing Fig.9 specially for this talk.  I would like also to thank
all participants of the MC  Workshop whose discussions which were useful
and
supportive for me.

\end{document}